\documentclass[nofootinbib]{revtex4-2}

\usepackage{amsmath}
\usepackage{eufrak}
\usepackage[hidelinks]{hyperref}
\usepackage{graphicx}\graphicspath{{Figures/}}
\usepackage{subcaption}
\usepackage{tensor}
\usepackage{amsfonts}
\usepackage{float}

\usepackage{color}
\usepackage{xcolor}

\colorlet{darkgreen}{green!50!black}
\colorlet{brightyellow}{yellow!75!red}
\colorlet{orange}{red!50!yellow}
\colorlet{darkblue}{blue!60!black}
\colorlet{darkred}{red!80!black}

\newcommand{\lp}{\left(}
\newcommand{\rp}{\right)}
\newcommand{\ls}{\left[}
\newcommand{\rs}{\right]}
\newcommand{\lc}{\left\{}
\newcommand{\rc}{\right\}}
\newcommand{\lb}{\left.}
\newcommand{\rb}{\right.}
\newcommand{\nn}{~\nonumber}
\newcommand{\comment}[1]{}
\newcommand{\numberthis}{\addtocounter{equation}{1}\tag{\theequation}}

\begin{document}

\title{Analytic and Approximate Solutions to Color Glass Condensate in the Classical Weak-Field Limit}

\author{S. Robicheaux}
\affiliation{Department of Chemistry and Physics, McNeese State University, Lake Charles, LA~~70605, USA}
\affiliation{Cyclotron Institute and Department of Physics and Astronomy, Texas A\&M University, College Station, TX~~77845, USA }

\author{R.~J.~Fries}
\affiliation{Cyclotron Institute and Department of Physics and Astronomy, Texas A\&M University, College Station, TX~~77845, USA}

\date{\today}

\begin{abstract}
    We discuss two-point functions and the energy momentum tensor of the classical gluon field after the collision of sheets of color charges on the light cone in the weak-field limit. The classical fields created by such a setup is thought to approximate the behavior of the gluon matter created right after the collision of heavy nuclei at large energies. Our discussion is based on a general expression for the gluon distribution in a nucleus, which contains the McLerran-Venugopalan (MV) Model as a special case. We derive the time-dependence of the energy momentum tensor in this general scenario. We show that the large-time behavior is universal, i.e.\ independent of the specific model for the gluon distribution, e.g.\ for energy density, transverse pressure and longitudinal pressure $\epsilon, P_T \sim 1/\tau$ and $P_L \sim 1/\tau^3$, where $\tau$ is longitudinal proper time.
    Subsequently, we focus on two special cases, the MV model and a proposed improved Gaussian (iG) model with improved ultraviolet (UV) and infrared (IR) behavior, the latter inspired by earlier work by Lam and Mahlon. We explicitly discuss the time dependence of the energy momentum tensor in both models. In the case of the MV-model, for infinite colliding nuclei, it is possible to give closed-formed analytic solutions for the energy momentum tensor in terms of special functions. Components of the energy momentum tensor take the form $\sim C (m\tau)^{-n} H(m\tau)$, where $n$ is an integer power, $m$ is the infrared cutoff, $H$ is a linear combination of Meijer-G functions with constant asymptotic value, and $C$ is a known constant. For the iG-model, we obtain reliable series expansions for both small and large times and show that the MV-model is recovered qualitatively in the UV limit. We briefly comment on implications for the angular momentum carried by the gluon field.
\end{abstract}

\maketitle

\section{Introduction}

Color glass condensate (CGC) has been studied as an effective theory of quantum chromodynamics (QCD) in the saturation region of the wave function of hadrons and nuclei \cite{Iancu:2003xm,Gelis:2010nm}. Gluon fields dominate the wave function at small Bjorken-$x$, and up to a saturation scale $Q_s$. The non-abelian interaction of two such fields in collisions of heavy nuclei at large energies seeds many important properties of the matter left behind in the forward light cone after the collision. This initial glasma state \cite{Lappi:2006fp} is thought to locally thermalize into a quark gluon plasma (QGP) \cite{Gale:2012rq,Berges:2013eia,Kurkela:2018vqr,Berges:2020fwq}. 

Large occupation numbers allow for a classical description of key features of color glass and the initial glasma state. This was first discussed in the context of the McLerran-Venugopalan (MV) model \cite{McLerran:1993ni,McLerran:1993ka} in which the nuclei constitute sources of color charges fixed to the light cone which exhibit random local Gaussian fluctuations. The Yang-Mills equations can then be used to compute the gluon distribution functions in the relevant kinematic regime \cite{Jalilian-Marian:1996mkd}. Collisions of nuclei in the MV-model are described by intersecting currents of the pertinent color charges, which create a gluon field in the forward light cone \cite{Kovner:1995ts,Kovner:1995ja}. 

The problem of decoherence of the classical gluon fields and the approach to equilibrium remain topics of active investigation. 
%The application of kinetic theory to the post-glasma phase has been promising \cite{.}.
In this work, we will focus on the classical field description of the glasma phase. It provides a powerful tool to describe the bulk of nuclear collisions at very early times. Classical field calculations have been applied successfully to the phenomenology of bulk observables in nuclear collisions when matched directly to viscous fluid dynamics, either directly or through an intermediate kinetic stage \cite{Schenke:2012hg,Gale:2012rq,Schenke:2020mbo,Heffernan:2023gye}. The classical field phase is considered a good approximation for a few tenths of one fm/$c$ in collisions at the Relativistic Heavy Ion Collider (RHIC) and the Large Hadron Collider (LHC). In this phase, the longitudinal pressure is negative \cite{Fries:2005yc,Chen:2015wia}, as the nuclei slow down and lose a significant amount of their longitudinal momentum as confirmed in experiment \cite{BRAHMS:2003wwg}. A kinetic or fluid dynamic description only becomes reasonable once the longitudinal pressure is non-negative. 

Phenomenological applications mostly rely on numerical solutions of the Yang-Mills equations \cite{Krasnitz:2001qu,Lappi:2003bi,Schenke:2012wb,Schenke:2012hg,Gelfand:2016yho,Ipp:2017lho,McDonald:2023qwc,Ipp:2024ykh}. Some of the later numerical work goes beyond the boost-invariant setup of the original color glass model that we will retain here. For comparisons with data, the IP-Glasma model \cite{Schenke:2012wb,Schenke:2012hg} is deployed, often in conjunction with a subsequent fluid dynamic stage. IP-Glasma combines solutions to the Yang-Mills equations with a prescription for the modeling of the color charge densities taken from the IP-Sat model \cite{Bartels:2002cj,Kowalski:2003hm}. 
Alternatives, like the McDipper model \cite{Garcia-Montero:2023gex}, based on $k_T$-factorized momentum space approach have also been deployed.

Analytic solutions to coliding color glass have been discussed in the literature with some success. Initial energy densities \cite{Lappi:2006hq} and energy flow \cite{Chen:2013ksa} of the gluon field, as well as axial charge densities and fluctuations of these quantities \cite{Lappi:2017skr,Guerrero-Rodriguez:2021ask} have been computed explicitly. A full recursive analytic solution to the Yang-Mills equations has been found in \cite{Chen:2015wia} based on a power series in longitudinal proper time $\tau$. Taking the few lowest orders of this expansion is akin to a ``near-field" approximation for the gluon field. The leading terms have been discussed in several works in the literature \cite{Chen:2015wia,Fujii:2008km,Carrington:2020ssh,Carrington:2021qvi}. However, the convergence of the series for interesting observables, like the energy density $\varepsilon(\tau)$ as a function of time, is somewhat poor. Re-summations have only been successful after further approximations are made to the recursive solution  \cite{Li:2016eqr}.

One case in which the resummation of powers in $\tau$ is straight forward, and leads to \emph{analytic results in transverse momentum space} \cite{Chen:2015wia} is the case of weak fields in the forward light cone. The pertinent solutions had already been found without recursive techniques in the early years of color glass \cite{Kovner:1995ja,Kovner:1995ts}.
However, corresponding \emph{analytic solutions in coordinate space} have been missing. The qualitiative picture is as follows.
Non-abelian interactions between gluons from the two nuclei are necessary to produce a non-trivial gluon field in the forward light cone, which goes beyond a simple superposition of the two nuclear gluon fields. Subsequently, as the nuclei separate, the field strength dilutes quickly, and the fields start to linearize. The weak-field limit is appropriate once the non-abelian terms in the recursive solution are small, i.e.\ $|\partial \mathcal{A}|\gg g|\mathcal{A}\mathcal{A}|$ where $\mathcal{A}$ generically denotes components of the gauge potential $A^\mu$, $\partial$ is a derivative in the transverse coordinate plane, and $g$ is the strong coupling constant. If the linearization is applied right after the initial interaction of fields at time $\tau=0+\epsilon$, i.e.\ at order $g|\mathcal{A}\mathcal{A}|$, where one field comes from each of the two nuclei, we consider this as the \emph{weak field limit}. In that case, resummation of the solution in the forward light cone is possible, and leads to expressions of fields in terms of Bessel functions in transverse momentum space. We will review these solutions in Sec. \ref{sec:two}. We will also discuss situations in which the weak field limit might resemble good approximations to the full solution.

In this work we focus on computing the time dependence of the energy momentum tensor $T^{\mu\nu}(\tau, \mathbf{R})$ of the initial classical gluon field in the weak-field limit on coordinate space, expanding on ideas by Guerrero-Rodriguez and Lappi \cite{Guerrero-Rodriguez:2021ask}. We assume a rather general ansatz for the gluon two-point function $\gamma(\mathbf{x}_\perp, \mathbf{y}_\perp)$ in a nucleus, which includes the MV-model as a special case. We discuss solutions both for constant average charge densities $\mu$ and for realistic nuclear profiles. For the latter, we employ the generalization of the MV-formalism, first discussed in \cite{Chen:2015wia}, using an expansion in gradients of $\mu$. We answer the question, whether a closed analytic solution in closed form in coordinate space is possible, at least for the simplest scenario. The answer is yes %for constant $\mu$ 
in the MV model. The energy momentum tensor is given in terms of Meijer-G functions. . 

Unfortunately, the MV-model suffers from UV-singularities which need to be regularized.  
We propose a modified model, the improved Gaussian (iG) model for $\gamma(\mathbf{x}_\perp, \mathbf{y}_\perp)$ which is UV-regularized through coarse graining for the color charges, and has improved IR behavior by enforcing global color neutrality. Solutions for the energy momentum tensor for this model are not in closed form, but the remaining integral can be carried out numerically or by series expansion.  
Compare to Ref.\ \cite{Guerrero-Rodriguez:2021ask} our work focuses on closed-form solution for the MV-model in coordinate space, and generalizes the discussion to non-constant $\mu$, which introduces flow and anisotropy effects in the transverse plane.

The paper is organized as follows. In the next section we review the setup to compute classical gluon fields in nuclear collisions, and we discuss the restriction to the weak field limit. In Sec. \ref{sec:three} we introduce two-point correlation functions of the gluon field in the forward light cone, and express them in terms of the gluon correlation functions %$G(\mathbf{x}_\perp,\mathbf{y}_\perp)$ 
in the single nuclei. In Sec. \ref{sec:EM_tensor} we utilize these results to obtain expectation values of the energy momentum tensor, focusing on the MV-Model. The iG-model, with UV-regularization and better IR behavior is introduced in Sec. \ref{sec:AGaussianModel}. The analytic expressions for the energy momentum tensor in the MV-model, as well as our final results using the improved model are presented in Sec. \ref{sec:ASTE}. We also discuss connections between models and phenomenological consequences. We end with a brief epilogue on angular momentum in Sec. \ref{sec:AngularMomentum}.
%\remark{}{Finish up outlook.}\remark{SR}{Done.}

\section{Classical Gluon Fields in Nuclear Collision}\label{sec:two}

Nuclei probed at increasingly high energies exhibit  steep growth of the number of gluons at small momentum fractions $x=p_g/P$ with respect to a single nucleon momentum $P$. This growth in gluon number eventually has to saturate, with the resulting density defining a scale $1/Q_s^2$, where $Q_s$ is the saturation momentum. In this limit, gluons can be effectively described in a quasi-classical picture. Partons with large $x\lesssim 1$ are dilute and act as color sources. Gluons at small $x \ll 1$ are generated by these sources, as described by the classical Yang-Mills equations \cite{McLerran:1993ka,McLerran:1993ni,Jalilian-Marian:1996mkd}. The large momenta of the sources render them essentially frozen during time scales of interest, and they can thus be chosen to be constant along the light cone. The cutoff between sources and fields is subject to a renormalization equation \cite{Jalilian-Marian:1997jhx,Jalilian-Marian:1997ubg,Iancu:2001ad}.

The same picture can be applied to describe gluon fields generated in the collision of two high energy nuclei \cite{Kovner:1995ts, Kovner:1995ja}. The sources are the large-$x$ partons in both nuclei, forming intersecting color charge currents. This choice of sources, if assumed to be on the light cone, is boost-invariant.
%a system that is inherently boost invariant and can be solved in 2+1 dimensions. see also \cite{Kovchegov:1996ty,Kovchegov:1998bi}. 
In this section we review the setup of this effective classical theory and discuss its weak-field limit.

\subsection{Computing Glasma Fields}

The color charges representing the large-$x$ components of the two nuclear wave functions are propagating along the $+$ and $-$ directions along the light cone. The relevant color current can thus be written as
\begin{subequations}\label{eq:current}
\begin{align}
  J^+ (x) &= \delta(x^-) \rho_1(\mathbf{x}_\perp) \, , \\
  J^-_2(x) &= \delta(x^+) \rho_2(\mathbf{x}_\perp) \, , \\
  J^i_{1,2} (x) &= 0 \, ,
\end{align}
\end{subequations}

\noindent with $i=1,2$, where $\rho_1(\mathbf{x}_\perp)$ and $\rho_2(\mathbf{x}_\perp)$ are the transverse densities of color charge in nucleus $1$ and $2$, respectively. Light cone coordinates are defined as $x^\pm = (x^0\pm x^3)/\sqrt{2}$. The current also satisfies the continuity equation $[D_\mu J^\mu]=0$ if we choose an axial gauge with
\begin{equation}~\label{eq:Fock-Sch}
  x^+ A^- + x^- A^+ = 0 \, .
\end{equation}

\noindent The gluon field here is defined as $F^{\mu\nu} = i[D^\mu,D^\nu]/g$, with covariant derivative $D^\mu =\partial^\mu -ig A^\mu$, and is generated by the color current through the Yang-Mills equation
\begin{equation}\label{eq:ym}
  \left[ D_\mu, F^{\mu\nu}\right] = J^\nu  \, .
\end{equation}

\noindent The Yang-Mills equations respect the boost symmetry of the sources, making the generated gluon field boost-invariant. The global structure of the gauge potential in space-time coordinates in our axial gauge can be written as \cite{Kovner:1995ts}
\begin{subequations}\label{eq:ASpacetimeStructure}
\begin{align}
    A^+ &= \Theta(x^+)\Theta(x^-)x^+A(\tau,\mathbf{x}_\perp) \\
    A^- &= -\Theta(x^+)\Theta(x^-)x^-A(\tau,\mathbf{x}_\perp) \\
    A^i &= \Theta(x^-)\Theta(-x^+)A^i_1(\mathbf{x}_\perp) + \Theta(x^+)\Theta(-x^-)A^i_2(\mathbf{x}_\perp) + \Theta(x^-)\Theta(x^+)A^i_\perp(\tau, \mathbf{x}_\perp) \quad (i=1,2) .
\end{align}
\end{subequations}

\noindent We use Milne coordinates with the proper time $\tau$ and space-time rapidity $\eta=\frac{1}{2} \log\frac{t+z}{t-z}$ in the forward light cone. The fields $A_1^i(\mathbf{x}_\perp)$, $A_2^i(\mathbf{x}_\perp)$ of the individual colliding nuclei can be obtained by solving their Yang-Mills equations with the appropriate charge densities $\rho_{1,2}$, respectively. The Yang-Mills equations lead to matching conditions at $\tau=0$ given by \cite{,Chen:2015wia}
\begin{subequations}\label{eq:a0}
\begin{align}
  A_\perp^i(0,\mathbf{x}_\perp) &= A_1^i(\mathbf{x}_\perp) + A_2^i(\mathbf{x}_\perp)   \\
  A(0,\mathbf{x}_\perp) & = -\frac{ig}{2} \delta^{ij} \left[ A_1^i(\mathbf{x}_\perp), A_2^j(\mathbf{x}_\perp) \right]  .
\end{align}
\end{subequations}

\noindent These fields on the light cone are the initial condition for the series expansion
\begin{equation}
  A^{\mu} (\tau,\mathbf{x}_\perp) = \sum_{n=0}^\infty \tau^n
  A^{\mu}_{(n)}(\mathbf{x}_\perp) 
\end{equation}
in the forward light cone.

We have omitted the dependence of the field on the space-time rapidity $\eta$ which is fully determined by boost-invariance. The field strength tensor $F^{\mu\nu}$ and the energy-momentum tensor can be expanded in $\tau$ in a similar way. We note that for fixed charge densities $\rho_{1,2}$ the fields $A$ and $A_\perp^i$ are finite at $\tau=0$.

The Yang-Mills equations imply equations for the fields $A$ and $A_\perp^i$ which are are known to have a recursive solution \cite{Chen:2015wia} in terms of their series coefficients in $\tau$. While the 0th order is determined by the initial conditions in Eq.\ (\ref{eq:a0}), the following even terms are given by
\begin{subequations}\label{eq:ExactGlasmaSolution}
\begin{align}
    A_{(n)} &= \frac{1}{n(n+2)}\sum\limits_{k+l+m=n-2}\ls D^i_{(k)}, \ls D^i_{(l)}, A_{(m)} \rs\rs  \\
    A^i_{\perp(n)} &= \frac{1}{n^2}\lp \sum\limits_{k+l=n-2}\ls D^j_{(k)}, F^{ji}_{(l)} \rs + ig \sum\limits_{k+l+m=n-4}\ls A_{(k)}, \ls D^i_{(l)}, A_{(m)} \rs\rs \rp 
\end{align}
\end{subequations}

\noindent while odd terms vanish. Note that $[D^i_{(n)},\cdot]$ for $n>0$ is simply the commutator with the $n$th term in the expansion of $-igA^i_\perp$.

We decompose the field strength tensor $F^{\mu\nu}$ into the usual chromo-electric and chromo-magnetic fields. They can be expressed through the gauge potential in our axial gauge as
\begin{subequations}\label{eq:ebVectors}
\begin{align}\label{eq:e3def}
    E_z &= -\frac{1}{\tau} \frac{\partial}{\partial\tau} \tau^2 A  \\\label{eq:b3def}
    B_z &= \partial^2 A_\perp^1 - \partial^1 A_\perp^2 -ig \left[ A_\perp^2, A_\perp^1\right] \\\label{eq:eidef}
    E_i &= -\cosh\eta \, \frac{\partial}{\partial \tau} A_\perp^i + \sinh\eta \, \tau [D^i, A] \\\label{eq:bidef}
    B_i &= \epsilon^{ij} \left( \sinh\eta\,  \frac{\partial}{\partial \tau} A_\perp^j - \cosh\eta \, \tau [D^j , A] \right)   \, .
\end{align}
\end{subequations}

Since boost-invariance is slightly less trivial for these 3-vectors we have kept the explicit dependence on the space-time rapidity in these expressions. The initial conditions from Eq.\ \eqref{eq:a0} imply vanishing transverse electric and magnetic fields at $\tau=0$ in the forward light cone, while longitudinal fields are given by
\begin{subequations}\label{eq:eb0}
\begin{align}\label{eq:e0}
  E_0 \equiv  E_{z,(0)} &= ig \delta^{ij} \left[ A_1^i, A_2^j \right] \, ,\\\label{eq:b0}
  B_0\equiv  B_{z,(0)} &= ig \epsilon^{ij} \left[ A_1^i, A_2^j \right] \, .
\end{align}
\end{subequations}

Further details that can be learned from the lowest orders in the recursion formula can be found in Refs.\ \cite{Chen:2013ksa,Chen:2015wia,Carrington:2021qvi}.

\subsection{The McLerran-Venugopalan Model}\label{sec:Intro_ssec:CGMV}

The MV model approximates the gluon field in heavy nuclei with the classical field determined by the Yang-Mills equations as discussed above. In addition, it makes assumptions about the statistical distribution of color charges $\rho(x)$. 
For an observable $O$ that depends on $\rho$ we can then compute $\langle O\rangle = \int W[\rho] O[\rho]d[\rho]$.

The MV-model assumes that the individual color charges making up the density $\rho(x)$ are uncorrelated with each other. Then the central limit theorem states that the weight $W[\rho]$ will be Gaussian. This Gaussian is completely defined by its first two moments 
\begin{align}\label{eq:MV}
    \langle \rho_{\underline{a}} \rangle_{\mathrm{MV}} &= 0 & \langle \rho_{\underline{a}}(x)\rho_{\underline{b}}(y) \rangle_{\mathrm{MV}} &= \frac{g^2}{N_c^2-1}\delta_{\underline{ab}}\delta^2(\mathbf{y_\perp}-\mathbf{x}_\perp)\delta(x^- - y^-)\mu\left(\frac{\mathbf{x}_\perp+\mathbf{y}_\perp}{2} \right)h(x^-).
\end{align}

The one-point function enforces that color averages to zero for each point in the transverse plane for the ensemble as a whole. We have written the expectation values by expanding $\rho=\rho_{\underline{a}} t_{\underline{a}}$ in terms of Gell-Mann matrices $t_{\underline{a}}$, $\underline{a}=1,\ldots,N_c^2-1$. Generally, underlined lowercase letters here denote SU($N_c$) indices. The two-point function encodes the lack of correlation of color charges through $\delta$-functions and introduces a variance $g^2 \mu/(N_c^2-1)$ for the charge density. In the proper MV-model this average charge density is held constant. A dependence on the transverse coordinates can be added and is, e.g., discussed in Ref.\ \cite{Chen:2015wia}. $h$ is a function normalized to one which parametrizes the thickness of the color charge around the light cone. It can be assumed to be a $\delta$-function here, since we will not discuss subtleties arising from gauge transformations %in 2+1 dimensions 
\cite{Jalilian-Marian:1996mkd}.

\subsection{The Weak Field Limit}

For the rest of the paper we consider the limit of weak fields in which we only keep the leading non-abelian term for fields in the forward light cone. The simplest choice that is consistent in power counting is to keep only those contributions to the gauge potential, that lead to terms of order $(g \mathcal{A}^2)$ for the physical fields $F^{\mu\nu}$. $\mathcal A$ generically represents components $A_{1,2}^i$ of the fields of the individual nuclei. To motivate this choice, recall that the initial chromo-magnetic and electric fields on the light cone are of this order. The recursion relations in Eq.\ (\ref{eq:ExactGlasmaSolution}) will add additional terms of equal or higher order. %$\sim g\mathcal{A}$. 
On the other hand, fields of lowest order $\sim \mathcal{A}$ do not emerge from the collision, but are a trivial superposition of the fields of the two nuclei before the collision. They do not lead to physical electric or magnetic fields in the forward light cone.

To be more precise, in the weak field limit we can omit higher orders in $g\mathcal{A}$ from the recursion (\ref{eq:ExactGlasmaSolution}) to arrive at
\begin{subequations}
\begin{align}\label{eq:arec1}
   A_{(n)} &= \frac{1}{n(n+2)} \partial^i \partial^i A_{(n-2)} \\\label{eq:arec2}
   A^i_{\perp, (n)} &= \frac{1}{n} \epsilon^{ij} \partial^j B^3_{(n-2)}
\end{align}
\end{subequations}

\noindent for all even $n>0$. The recursions for the longitudinal electric and magnetic fields can likewise be written down easily,
\begin{subequations}
\begin{align}\label{eq:ezrec1}
  E_{z,(n)} &= \frac{1}{n^2} \partial^i\partial^i E_{z,(n-2)} \\\label{eq:bzrec1}
  B_{z,(n)} &= \frac{1}{n^2} \partial^i\partial^i B_{z,(n-2)} \,  \, . 
\end{align}
\end{subequations}

\noindent One can drop non-abelian terms in the calculation of Eq.\ \eqref{eq:ebVectors}, except for the zeroth order longitudinal magnetic field for which we need to retain the term $ig[A^2_{\perp(0)},A^1_{\perp(0)}]$ to collect all terms up to order $g\mathcal{A}^2$.

The fields are straight forward to re-sum if we transition to transverse Fourier space. In the following the $\sim$ sign over fields will denote a Fourier transformation with respect to the two transverse coordinates, and $\mathbf{k}_\perp$ denotes transverse momentum vectors. The re-summed power series in time are
\begin{subequations}
\begin{align}\label{eq:GlasmaPotReSum}
    \tilde{A}(\tau,\mathbf{k}_\perp) &= \tilde{A}_{(0)}(\mathbf{k}_\perp) + \tilde{A}_{(0)}(\mathbf{k}_\perp)\sum\limits_{n\geq 2, \mathrm{even}} \frac{2}{n!!^2(n+2)}(ik_\perp\tau)^n = 2\tilde{A}_{(0)}\frac{J_1(k_\perp\tau)}{k_\perp\tau} \\\nn
    \tilde{A}^i_\perp(\tau,\mathbf{k}_\perp) &= \tilde{A}^i_{\perp, (0)}(\mathbf{k}_\perp) + \tilde{B}_0(\mathbf{k}_\perp)\frac{i \epsilon^{ij} k_\perp^j}{k_\perp^2}\sum\limits_{n\geq 2, \mathrm{even}} \frac{1}{n!!^2}(ik_\perp \tau)^{n} \\ \label{eq:GlasmaPotReSum2}
    &= \tilde{A}^i_{\perp, (0)} + \epsilon^{ij}\frac{ik_\perp^j}{k_\perp^2}\tilde{B}_0(\mathbf{k}_\perp)\ls J_0(k_\perp\tau) - 1 \rs 
\end{align}
\end{subequations}

\noindent using Bessel functions of the first kind. Note that the result for the transverse gauge differs from solutions in the literature 
\cite{Chen:2015wia} which ascribes a time dependence to the superposition of the nuclear fields $A^i_\perp$. However, the latter field is constant. Any additional time dependence would have to come from interactions and thus introduce additional powers $\sim g\mathcal{A}$. Eq. \eqref{eq:GlasmaPotReSum2} contains all terms up to order $g\mathcal{A}^2$. The reader can check that the solutions above satisfy the Yang-Mills equations to the chosen order.

The longitudinal electric and magnetic fields can be swiftly computed to be
\begin{subequations}\label{eq:ezbzres2}
\begin{align}\label{eq:ezres2}
  \tilde E_z(\mathbf{k}_\perp,\tau) &= \tilde E_0(\mathbf{k}_\perp) J_0(k_\perp\tau) \\\label{eq:bzres2}
  \tilde B_z(\mathbf{k}_\perp,\tau) &= \tilde B_0(\mathbf{k}_\perp) J_0(k_\perp\tau)  \, .
\end{align}
\end{subequations}

\noindent Note that $\tilde E_0$ and $\tilde B_0$ are the Fourier transformed initial conditions on the forward light cone from Eq.\ \eqref{eq:eb0}, and $k_\perp=|\mathbf{k}_\perp|$. The transverse electric and magnetic fields require slightly more work and re-sum to 
\begin{subequations}\label{eq:eibires2}
\begin{align}\label{eq:eires2}
    \tilde{E}_i(\tau)   &= \left(  \cosh\eta \, \epsilon^{ij} \mathbf{k}^j_\perp \tilde B_0 +  \sinh\eta \, \mathbf{k}^i_\perp \tilde{E}_0  \right) \frac{i}{k_\perp} J_1(k_\perp \tau) \\\label{eq:bires2}
    \tilde{B}_i(\tau)  &= \left(  \sinh\eta \, \mathbf{k}^i_\perp \tilde{B}_0 -  \cosh\eta \, \epsilon^{ij} \mathbf{k}^j_\perp \tilde{E}_0  \right) \frac{i}{k_\perp} J_1(k_\perp \tau)   
\end{align}
\end{subequations}

\noindent where we have omitted arguments $\mathbf{k}_\perp$ in the notation for brevity, but have kept the explicit dependence on space-time rapidity exhibited by transverse fields. The origin of these fields at small times can be traced back to Ampere's Law, Faraday's Law and Gauss' Law \cite{Chen:2013ksa}. Eqs. \eqref{eq:ezbzres2} and \eqref{eq:eibires2} can also be found in Ref. \cite{Guerrero-Rodriguez:2021ask} up to a minus sign difference in \eqref{eq:eires2}. 

\section{Correlation Functions in the Weak-Field Glasma}\label{sec:three}

In the following we will deal with correlation functions of the type
\begin{equation}
   \mathcal{C}^{\mu_1\nu_1,\mu_2\nu_2} =   \left\langle F^{\mu_1\nu_1}_{\underline{a}} (\tau, \eta, \mathbf{x}_\perp ) F^{\mu_2\nu_2}_{\underline{a}} (\tau', \eta', \mathbf{x}'_\perp)     \right\rangle
\end{equation}

\noindent where the $\langle \ldots \rangle$ indicates averaging over $\rho_1$ and $\rho_2$.  These correlation functions are gauge-invariant in the weak field limit, as the restriction to order $\mathcal{O}(g\mathcal{A}^2)$ permits us to drop the gauge link connecting the two fields. These correlation functions encode important physics. The energy momentum tensor $T^{\mu\nu}$, angular momentum density $M^{\mu\nu\lambda}$, and the Chern-Simons charge density $\mathbf{E} \cdot \mathbf{B}$ can be expressed through these correlation functions in the UV limit $(\tau',\eta',\mathbf{x}'_\perp) \to (\tau,\eta,\mathbf{x}_\perp)$. The momentum broadening coefficient $\hat q$ for partons traversing the glasma can be obtained by integration over the path between the two space-time points. In the following we will set $\eta'=\eta=0$ for simplicity, as the boost invariance of the system will allow a reconstruction of the correlation function for arbitrary values.

Let us first focus on $\langle E_zE_z\rangle =\mathcal{C}^{03,03}$ as an example of the correlation function of two longitudinal fields. We apply Fourier transformations to both transverse coordinates and use the known time evolutions \eqref{eq:ezbzres2} to obtain
\begin{equation}
  \label{eq:ezez}
  \left\langle E_z(\tau,\mathbf{x}_\perp) E_z(\tau',\mathbf{y}_\perp) \right\rangle= \int \frac{d^2 p_\perp}{(2\pi)^2}\frac{d^2 k_\perp}{(2\pi)^2} 
  e^{i\mathbf{p}_\perp \cdot \mathbf{x}_\perp+i \mathbf{k}_\perp \cdot \mathbf{y}_\perp}  
    J_0 (p_\perp \tau) J_0(k_\perp \tau') \left\langle \tilde E_0 (\mathbf{p}_\perp) 
    \tilde E_0  (\mathbf{k}_\perp) \right\rangle \, .
\end{equation}

\noindent We now write the initial fields in terms of the fields in the colliding nuclei. We utilize the momentum space versions of Eq.\ (\ref{eq:eb0}) replacing products with convolutions over transverse momenta. We resolve the commutators by writing the fields in terms of their  SU(3) components and Gell-Mann matrices, $A^i_{1,2} = A^{i}_{1,2,\underline{a}}t_{\underline{a}}$. Taking the commutators and applying the implicit trace yields
\begin{equation}
  \label{eq:00corr}
  \left\langle \tilde E_0 (\mathbf{p}_\perp) 
    \tilde E_0  (\mathbf{k}_\perp) \right\rangle = g^2 f^{\underline{abe}}f^{\underline{cde} }
     \delta^{ij}\delta^{kl}  
    \int \frac{d^2 p'_\perp}{(2\pi)^2}\frac{d^2 k'_\perp}{(2\pi)^2}  \left\langle \tilde A_{1,\underline{a}}^{i}(\mathbf{p}'_\perp) 
    \tilde A_{1,\underline{c}}^{k}(\mathbf{k}'_\perp) \right\rangle 
   \left\langle \tilde A_{2,\underline{b}}^{j}(\mathbf{p}_\perp - \mathbf{p}'_\perp) \tilde A_{2,\underline{d}}^{l}(\mathbf{k}_\perp - \mathbf{k}'_\perp) 
   \right\rangle   \, ,
\end{equation}

\noindent where the $f^{\underline{abc}}$ are the SU(3) structure constants. All approaches to model the correlation function of gluons in single nuclei, including the MV-model, assume the absence of correlations between directions in SU(3) color space. We can therefore write
\begin{equation}
    \label{eq:AAansatz}
    \left\langle  A_{k,\underline{a}}^{i}(\mathbf{x}_\perp) 
    A_{k,\underline{b}}^{j}(\mathbf{y}_\perp) \right\rangle = \frac{g^2}{N_c^2-1} \delta^{\underline{ab}} G_k^{ij}(\mathbf{x}_\perp,\mathbf{y}_\perp)
\end{equation}

\noindent where we have taken out some additional factors for later convenience, and $G_k$ is the gluon correlation function for nucleus $k$. Using Fourier transformations on the two-gluon correlation functions allows us to substitute them in Eq.\ (\ref{eq:00corr}). 

Subsequently, one can take the $d^2p_\perp'$ and  $d^2k_\perp'$ integrals which forces matching coordinates in the correlation functions of both nuclei to be identical,
\begin{equation}
  \left\langle \tilde E_0 (\mathbf{p}_\perp) 
    \tilde E_0  (\mathbf{k}_\perp)  \right\rangle = g^6 \frac{N_c}{N_c^2-1} 
      \delta^{ij}\delta^{kl}
	\int d^2x'_\perp d^2 y'_\perp  e^{-i\mathbf{x}'_\perp\cdot\mathbf{p}_\perp -i\mathbf{y}'_\perp\cdot\mathbf{k}_\perp }
  G_1^{ik}(\mathbf{x}'_\perp,\mathbf{y}'_\perp) G_2^{jl}(\mathbf{x}'_\perp,\mathbf{y}'_\perp)    \, .
\end{equation}

\noindent One can now explicitly take the momentum integrals over Bessel functions in Eq.\ (\ref{eq:ezez}), utilizing \cite{Arfken}
\begin{equation}\label{eq:momentum_1}
    \int \frac{d^2 p_\perp}{(2\pi)^2}  e^{i\mathbf{p}_\perp \cdot (\mathbf{x}_\perp-\mathbf{x}'_\perp) }
        J_0(p_\perp \tau)    = \int_0^\infty  \frac{ d p_\perp}{2\pi } p_\perp J_0\left (p_\perp |\mathbf{x}_\perp-\mathbf{x}'_\perp|\right) J_0(p_\perp \tau)
    = \frac{1}{2\pi\tau} \delta(\tau- |\mathbf{x}_\perp-\mathbf{x}'_\perp|)   \, .
\end{equation}

\noindent One arrives at a final result of this section
\begin{multline}\label{eq:ll2}
   \left\langle E_z(\tau,\mathbf{x}_\perp) E_z(\tau',\mathbf{y}_\perp) \right\rangle =  \frac{N_c}{N_c^2-1}   \frac{g^6}{(2\pi)^2 \tau \tau'} \delta^{ij}\delta^{kl} \\ \times \int d^2 x'_\perp d^2 y'_\perp  \delta(\tau- |\mathbf{x}_\perp-\mathbf{x}'_\perp|) \delta(\tau'- |\mathbf{y}_\perp-\mathbf{y}'_\perp|)   G_1^{ik}(\mathbf{x}'_\perp,\mathbf{y}'_\perp) G_2^{jl}(\mathbf{x}'_\perp,\mathbf{y}'_\perp) .
\end{multline}

\noindent It is easy to check that for the analogous correlation functions involving longitudinal magnetic fields one simply needs to replace the Kronecker-$\delta$ symbols with Levi Civita tensors in (\ref{eq:ll2}): $\langle E_zB_z\rangle \sim \delta^{ij}\epsilon^{kl}$, $\langle B_zE_z\rangle \sim \epsilon^{ij}\delta^{kl}$, $\langle B_zB_z\rangle \sim \epsilon^{ij}\epsilon^{kl}$. Note that Eqs.\ (\ref{eq:e0}) and (\ref{eq:b0}) allow us to rephrase all expressions also in terms of the initial longitudinal fields at $\tau=0$, e.g. for the correlation function of electric fields
\begin{equation}
  \label{eq:ll2alt}
   \left\langle E_z(\tau,\mathbf{x}_\perp) E_z(\tau',\mathbf{y}_\perp) \right\rangle =    \frac{1}{(2\pi)^2 \tau \tau'} 
  \int d^2 x'_\perp d^2 y'_\perp  \delta(\tau- |\mathbf{x}_\perp-\mathbf{x}'_\perp|)
  \delta(\tau'- |\mathbf{y}_\perp-\mathbf{y}'_\perp|)  
  \left\langle E_0 (\mathbf{x}'_\perp) E_0(\mathbf{y}'_\perp) \right\rangle
   \, .
\end{equation}

As an example of a correlation function of transverse fields we consider $\langle E_mE_n\rangle =\mathcal{C}^{0m,0n}$, ($m,n=1,2$). Proceeding as in the case of two longitudinal fields we obtain
\begin{multline}
   \langle E_m(\tau,\mathbf{x}_\perp) E_n(\tau',\mathbf{y}_\perp) \rangle =  g^6 \frac{N_c}{N_c^2-1} 
   \varepsilon^{mt}\varepsilon^{ij}\varepsilon^{ns}\varepsilon^{kl}  
   \int d^2 x'_\perp d^2 y'_\perp G_1^{ik}(\mathbf{x}'_\perp,\mathbf{y}'_\perp) G_2^{jl}(\mathbf{x}'_\perp,\mathbf{y}'_\perp)  \\ \times  \left(
   \int \frac{d^2 p_\perp}{(2\pi)^2}  \left( -\nabla^t_{\mathbf{x}'_\perp} \right) e^{i\mathbf{p}_\perp\cdot (\mathbf{x}_\perp-\mathbf{x}'_\perp) }
        \frac{1}{p_\perp} J_1(p_\perp \tau)  \right)\left( 
    \int \frac{d^2k_\perp}{(2\pi)^2} \left( -\nabla^s_{\mathbf{y}'_\perp} \right) e^{i\mathbf{k}_\perp\cdot (\mathbf{y}_\perp-\mathbf{y}'_\perp)} \frac{1}{k_\perp}  J_1 (k_\perp\tau') \right) \, .
\end{multline}
The new Bessel integral appearing in this case is \cite{Lin:2014}
\begin{equation}\label{eq:momentum_2}
    \int \frac{d^2 p_\perp}{(2\pi)^2}  e^{i\mathbf{p}_\perp\cdot (\mathbf{x}_\perp-\mathbf{x}'_\perp) } \frac{1}{p_\perp}
        J_1(p_\perp \tau)    = \int_0^\infty  \frac{d p_\perp}{2\pi } J_0\left (p_\perp |\mathbf{x}_\perp-\mathbf{x}'_\perp|\right) J_1(p_\perp \tau)
    = \frac{1}{2\pi\tau} \Theta(\tau- |\mathbf{x}_\perp-\mathbf{x}'_\perp|)   \, ,
\end{equation}
where $\Theta$ denotes the Heaviside step function. Taking these integrals simplifies the transverse-transverse case to
\begin{multline}
  \label{eq:tt2}
   \langle E_m(\tau,\mathbf{x}_\perp) E_n(\tau',\mathbf{y}_\perp) \rangle = 
   \frac{N_c}{N_c^2-1}  \frac{g^6}{(2\pi)^2 \tau \tau'}  
 \varepsilon^{mt}\varepsilon^{ij}\varepsilon^{ns}\varepsilon^{kl}
    \\ \times  
  \int d^2 x'_\perp d^2 y'_\perp  \delta(\tau- |\mathbf{x}_\perp-\mathbf{x}'_\perp|)
  \delta(\tau'- |\mathbf{y}_\perp-\mathbf{y}'_\perp|)  \frac{-{\left( \mathbf{x}_\perp-\mathbf{x}'_\perp\right)}^t }{\left| {\mathbf{x}_\perp-\mathbf{x}'_\perp}\right|}  \frac{-{\left( \mathbf{y}_\perp-\mathbf{y}'_\perp\right)}^s }{\left| {\mathbf{y}_\perp-\mathbf{y}'_\perp}\right|} G_1^{ik}(\mathbf{x}'_\perp,\mathbf{y}'_\perp) G_2^{jl}(\mathbf{x}'_\perp,\mathbf{y}'_\perp)
    \, .
\end{multline}
Different combinations of transverse fields simply alter the second and fourth Levi-Civita tensors in this equation. One can check that these tensors need to be replaced in the following way, $\langle E_mB_n\rangle \sim -\varepsilon^{ij}\delta^{kl}$, $\langle B_mE_n\rangle \sim -\delta^{ij}\epsilon^{kl}$, $\langle B_mB_n\rangle \sim \delta^{ij}\delta^{kl}$, with the rest of Eq.\ (\ref{eq:tt2}) remaining unchanged.

The correlation function between one transverse and one longitudinal field can be computed analogously. We simply present the result ($m=1,2$)
\begin{multline}
	\label{eq:tl2}
   \langle E_m(\tau,\mathbf{x}_\perp) E_z(\tau',\mathbf{y}_\perp) \rangle= 
   \frac{N_c}{N_c^2-1}  \frac{g^6}{(2\pi)^2 \tau \tau'}   
    \epsilon^{mn}\epsilon^{ij} 
   \delta^{kl} \\ \times  
  \int d^2 x'_\perp d^2 y'_\perp  \delta(\tau- |\mathbf{x}_\perp-\mathbf{x}'_\perp|)
  \delta(\tau'- |\mathbf{y}_\perp-\mathbf{y}_\perp'|)  
   \frac{-{\left( \mathbf{x}_\perp-\mathbf{x}'_\perp\right)}^n }{\left| {\mathbf{x}_\perp-\mathbf{x}'_\perp}\right|}  G_1^{ik}(\mathbf{x}'_\perp,\mathbf{y}'_\perp) G_2^{jl}(\mathbf{x}'_\perp,\mathbf{y}'_\perp)
    \, .
\end{multline}
For other combinations the last two tensors need to be replaced as given by the following scheme: $\langle E_mB_z\rangle \sim \varepsilon^{ij}\varepsilon^{kl}$, $\langle B_mE_z\rangle \sim -\delta^{ij}\delta^{kl}$, $\langle B_mB_z\rangle \sim -\delta^{ij}\varepsilon^{kl}$.

\begin{figure}[tb]
  \begin{center}
    \includegraphics[width=6cm]{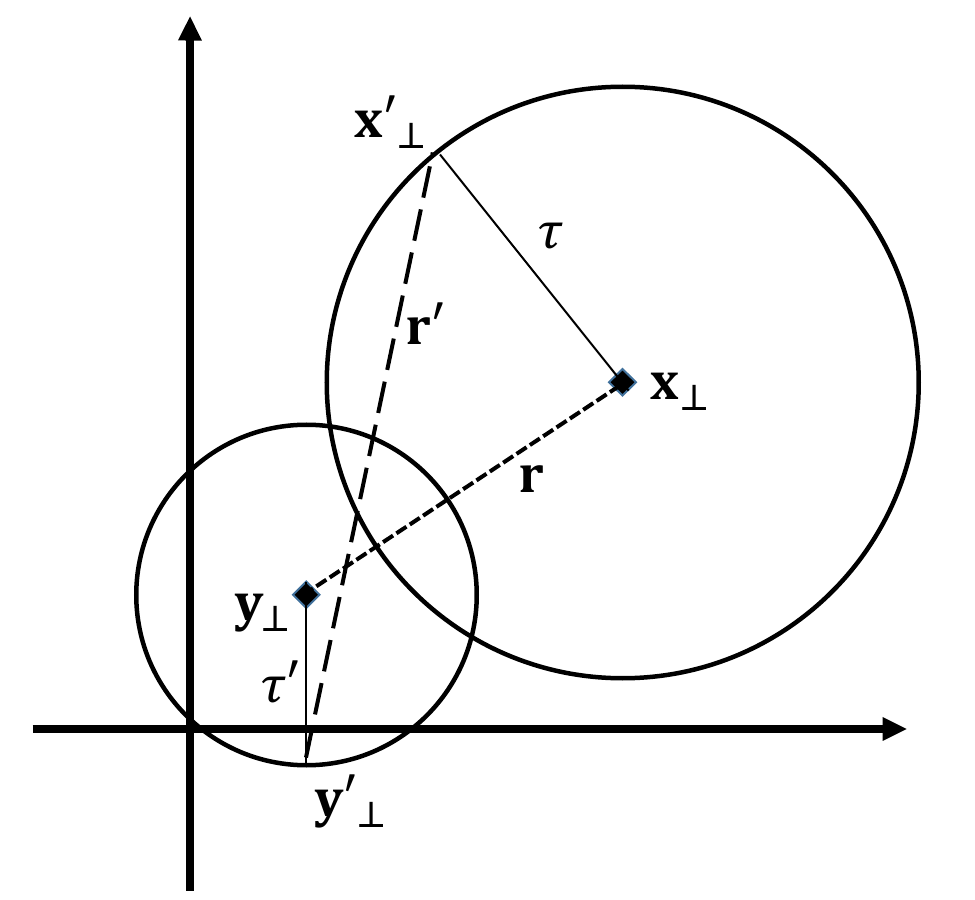}
  \caption{\label{fig:geometry} The locations of points at $\tau=0$ contributing to fields at times $\tau$ and $\tau'$ are circles determined by the backward light cones. Only points on both circles can contribute to the correlation function.}
  \end{center}
\end{figure}

The formulas in Eqs.\ (\ref{eq:ll2}), (\ref{eq:tt2}) and (\ref{eq:tl2}) are the main outcome of this section. We have written the general two-point correlation function of fields in terms of  integrals over non-oscillatory functions. These could fairly easily be evaluated numerically, see Fig.\ \ref{fig:geometry}, if a model for the gluon distributions $G_{1,2}$ is specified.
Our result in the limit $(\tau',\mathbf{y}_\perp)\to (\tau,\mathbf{x}_\perp)$ is consistent with the one-point functions discussed by Guerrero-Rodriguez and Lappi \cite{Guerrero-Rodriguez:2021ask}\footnote{Note the difference in definition. In Ref. \cite{Guerrero-Rodriguez:2021ask}, one-point functions are quadratic in fields (at one space-time point), and two-point functions are quartic (involving two space-time points). In our work, two-point functions are quadratic in fields.}. The space-time picture discussed in their work can be generalized to our correlation functions as shown in Fig.\ \ref{fig:geometry}. The field at time $\tau$ and position $\mathbf{x}_\perp$ in the transverse plane is determined by the fields at time $\tau=0$ at positions $\mathbf{x}'_\perp$ that are on the backward light cone. The $\delta$-functions in (\ref{eq:ll2}), (\ref{eq:tt2}) and (\ref{eq:tl2})  thus determine circles of radii $\tau$ and $\tau'$, respectively. The integrals in Eqs.\ (\ref{eq:ll2}), (\ref{eq:tt2}) and (\ref{eq:tl2}) sample the contributions over these circles. For future convenience we introduce center and relative coordinates $\mathbf{R}=(\mathbf{y}_\perp+\mathbf{x}_\perp)/2$ and $\mathbf{r}=\mathbf{y}_\perp-\mathbf{x}_\perp$, and analogously $\mathbf{R}'$ and $\mathbf{r}'$ for the primed coordinates.

\section{The Energy Momentum Tensor}\label{sec:EM_tensor}

\subsection{The Gluon Correlation Function in the MV-Model and its Generalization}%\label\mathbf{R}_\perp
\label{sec:gluecorr}

To make further progress we have to specify the gluon correlation functions $G_k(\mathbf{x}'_\perp,\mathbf{y}'_\perp)$ for each of the two nuclei. We utilize the usual strategy and first find the 2-point function in a covariant gauge, where $A_{\mathrm{cov}}^\mu(\mathbf{x}_\perp) = \delta^{+\mu}\alpha(\mathbf{x}_\perp)$, and subsequently apply a gauge transformation to the axial gauge. In covariant gauge, the Yang-Mills equations for a single nucleus reduces to a two-dimensional Poisson equation
\begin{align}\label{eq:Poisson}
\triangle\alpha &= -\rho.
\end{align}

The solution can be found using the Green's function, $ \mathcal{G}(\mathbf{x}_\perp)=-\ln(x^2_\perp/\Lambda^2)/4\pi$ with arbitrary length scale $\Lambda$. This function is divergent for low momenta, thus we utilize an infrared regulator $m = 2e^{-\gamma_E}/\Lambda$ which can be thought of as giving the gluon a small but finite mass $\approx 1$ fm$^{-1}$. In Fourier space we use $\tilde{\mathcal{G}}(\mathbf{k}_\perp) = 1/k_\perp^2 \rightarrow \tilde G(\mathbf{k}_\perp)=1/(k_\perp^2 + m^2)$. In coordinate space, this transforms to a modified Bessel function of the second kind, $G(\mathbf{x}_\perp) = K_0(mx_\perp)/2\pi$. This Green's function is equivalent to the previous logarithmic solution in the limit $m\rightarrow 0$. 

Now the covariant gauge two-point function can be found from, e.g., \eqref{eq:MV}. In general we can write
\begin{align}\label{eq:TwoPointAlpha}
\left< \alpha_{\underline{a}}(x)\alpha_{\underline{b}}(y) \right> = \frac{g^2}{N_c^2 - 1} \delta_{\underline{a}\underline{b}}h(x^-)\delta(x^- - y^-) \gamma(\mathbf{x}_\perp, \mathbf{y}_\perp)  \, .
\end{align}
In the MV-model the transverse profile function $\gamma$ is given by the Green's function and the transverse part of the charge correlation function (\ref{eq:MV})
\begin{align}
\gamma_{\mathrm{MV}}(\mathbf{x}_\perp, \mathbf{y}_\perp) &= \int d^2z_\perp d^2z'_\perp G(\mathbf{x}_\perp-\mathbf{z}_\perp)G(\mathbf{y}_\perp-\mathbf{z}_\perp')\mu\lp \frac{\mathbf{z}_\perp+\mathbf{z}_\perp'}{2} \rp \delta^{(2)}(\mathbf{z}_\perp-\mathbf{z}_\perp') \, .
\end{align}
Recall that the MV-model assumes uncorrelated color charges, which is reflected by the presence of the transverse $\delta$-function in the equation above. To allow for the discussion of more general models, we replace the transverse $\delta$-function in (\ref{eq:MV}) with a more general, rotationally invariant, function $\mathcal{D}(\mathbf{z}_\perp-\mathbf{z}_\perp')=\mathcal{D}(|\mathbf{z}_\perp-\mathbf{z}_\perp'|)$.  We can still imagine this function to peak around the origin, but the exact shape is not relevant in this section. After some suitable change of variables to relative and center coordinates we have
\begin{align}
\label{eq:gamma}
\gamma(\mathbf{x}_\perp, \mathbf{y}_\perp)&= \int d^2u_\perp d^2v_\perp G(\mathbf{u}_\perp-\mathbf{v}_\perp/2)G(\mathbf{u}_\perp+\mathbf{v}_\perp/2)\mu( \mathbf{R} - \mathbf{u}_\perp) \mathcal{D}(\mathbf{v}_\perp-\mathbf{r}).
\end{align}
The integrals in Eq.\ (\ref{eq:gamma}) are easy to evaluate in the case of the original MV-model, with constant $\mu$ and $\mathcal{D}_{\mathrm{MV}}=\delta^{(2)}$. In Ref.  \cite{Chen:2015wia}, in was proposed that a non-constant charge variance $\mu$ can be taken into account through a gradient expansion.
Indeed, we can obtain closed form expressions for the functions $\gamma$ by expanding $\mu( \mathbf{R} - \mathbf{u}_\perp)$ around $\mathbf{u}_\perp = 0$,  more precisely $\mu( \mathbf{R} - \mathbf{u}_\perp) = \mu(\mathbf{R}) - u^i_\perp\nabla^i\mu(\mathbf{R}) + \tfrac{1}{2}u^i_\perp u^j_\perp \nabla^{ij}\mu(\mathbf{R}) - \cdots$. We note that all odd terms in the expansion vanish due to parity of the integrand.
Such a gradient expansion will be valid as long as the gradients are not too large on the length scale $\sim 1/m$. Otherwise, each term in the expansion can be computed in Fourier space,
\begin{align*}
\int d^2u_\perp G(\mathbf{u}_\perp-\mathbf{v}_\perp/2)&G(\mathbf{u}_\perp+\mathbf{v}_\perp/2)\mathbf{u}^{i_1}_\perp\cdots \mathbf{u}^{i_n}_\perp \\ 
&= \int (-i)^n\frac{d^2p_\perp}{(2\pi)^2} e^{i \mathbf{p}_\perp\cdot\mathbf{v}_\perp}\frac{1}{p^2_\perp + m^2}(\nabla_z)^{i_1\cdots i_n}\left[ e^{i \mathbf{z}_\perp\cdot\mathbf{v}_\perp/2}\frac{1}{(\mathbf{z}_\perp+\mathbf{p}_\perp)^2 + m^2} \right]_{\mathbf{z}_\perp\rightarrow 0},
\end{align*}

\noindent where the right-hand side can be identified as the inverse Fourier transform of Bessel functions after taking the derivatives,
\begin{align}
    \int\frac{d^2p_\perp}{(2\pi)^2}e^{i\mathbf{p}_\perp\cdot\mathbf{v}_\perp}(p^2_\perp+m^2)^{-1-\ell} &= \frac{(mv_\perp)^{\ell}K_{\ell}(mv_\perp)}{2\pi(\ell !)(2m^2)^{\ell}}.
\end{align}
We can now cast $\gamma$ in the following general form 
\begin{align}\label{eq:gammaf}
\gamma(\mathbf{x}_\perp,\mathbf{y}_\perp) &= \mu(\mathbf{R})g_{0,1}(r) + \nabla^{ij}\mu(\mathbf{R})\lp \delta^{ij}g_{2,1}(r) + \lp\frac{r^ir^j}{r^2}-\frac{1}{2}\delta^{ij}\rp g_{2,2}(r) \rp + O\lp\nabla^{ijkl}\mu(\mathbf{R})\rp
\end{align}
where we have omitted terms that are quartic or higher in gradients of $\mu$, and we have introduced three coefficient functions $g_{0,1}$, $g_{2,1}$, and $g_{2,2}$. Since $\mathcal{D}$ is rotationally invariant, the coefficient functions $g$ only depend on the magnitude of the relative coordinate $r=|\mathbf{r}|$. Their precise form can be computed from the choice of $\mathcal{D}$.  
For the rest of this section we will work in terms of these generic functions  in order to make exchanging models simple. For the MV-model ($\mathcal{D}_{\mathrm{MV}}=\delta^{(2)}$), we can read off the values for these functions from Ref. \cite{Chen:2015wia}:
\begin{subequations}\label{eq:gMV}
\begin{align}
    g_{0,1,\mathrm{MV}}(r) &= \frac{r}{4\pi m}K_1(mr) \label{eq:g01mv} \\   
    g_{2,1,\mathrm{MV}}(r) &= \frac{r^3}{192\pi m}K_3(mr) \\
    g_{2,2,\mathrm{MV}}(r) &= \frac{r^3}{96\pi m}K_1(mr). \label{eq:g22mv}
\end{align}
\end{subequations}
We will discuss another possible choice in the next section.

To turn the two-point function \eqref{eq:TwoPointAlpha} into its corresponding function in light cone gauge we utilize the gauge transformation via Wilson lines as discussed in 
Refs. \cite{Jalilian-Marian:1996mkd,Chen:2015wia}. The result is well known and reads
\begin{equation}\label{eq:expFFAA}
\left< A^i_{\underline{a}}(\mathbf{x}_\perp)A^j_{\underline{b}}(\mathbf{y}_\perp) \right> =\frac{g^2}{N_c^2 - 1}\delta_{\underline{ab}}
U\left( \mathbf{x}_\perp, \mathbf{y}_\perp \right)
\nabla _{x_\perp}^i\nabla_{y_\perp}^j\gamma\left(\mathbf{x}_\perp, \mathbf{y}_\perp\right)
\end{equation}
where 
\begin{equation}
\label{eq:Uexp}
U(\mathbf{x}_\perp, \mathbf{y}_\perp) = \frac{2(N_c^2-1)}{g^4 N_c \Gamma ( \mathbf{x}_\perp,\mathbf{y}_\perp)} 
\left[ \exp\left( \frac{g^4 N_c}{2(N_c^2-1)} \Gamma ( \mathbf{x}_\perp,\mathbf{y}_\perp) \right) -1 \right]  \, ,
\end{equation}  
and $\Gamma ( \mathbf{x}_\perp,\mathbf{y}_\perp)=2\gamma ( \mathbf{x}_\perp,\mathbf{y}_\perp)-\gamma ( \mathbf{x}_\perp,\mathbf{x}_\perp)-\gamma ( \mathbf{y}_\perp,\mathbf{y}_\perp)$ is a version of the gluon profile function with the UV-sensitive terms removed.
The limit of infinitely thin nuclei can be taken once the gauge transformation is complete, leaving only transverse coordinates as arguments.

Computing the derivatives of $\gamma$ for our general ansatz is tedious but straight forward. They can be written in terms of the center and relative derivatives, ${\nabla}_{x_\perp,y_\perp} = \tfrac{1}{2}{\nabla}_R \mp {\nabla}_r$. and work out to be
\begin{align}\label{eq:ddgamma}
\nabla_{x_\perp}^i\nabla_{y_\perp}^j\gamma(\mathbf{x}_\perp, \mathbf{y}_\perp) &= \mu(\mathbf{R})\left\{\left[ \frac{r^ir^j}{r^2} - \delta^{ij} \right]\frac{1}{r}g_{0,1}'(r) - \frac{r^ir^j}{r^2}g_{0,1}''(r)\right\} + \frac{1}{2}\lp \nabla^i\mu(\mathbf{R}) \frac{r^j}{r} - \nabla^j\mu(\mathbf{R}) \frac{r^i}{r} \rp g_{0,1}'(r) \\~\nonumber
&+\frac{1}{4}\nabla^{ij}\mu(\mathbf{R}) g_{0,1}(r) + \nabla^{lm}\mu(\mathbf{R})\left\{ \left[ \frac{r^ir^j}{r^2} - \delta^{ij} \right]\delta^{lm}\frac{1}{r}g_{2,1}'(r) - \frac{r^ir^j}{r^2}\delta^{lm}g_{2,1}''(r)  - 2\left[\vphantom{\frac{r^ir^m}{r^2}} \delta^{li}\delta^{mj}\right.\right. \\~\nonumber
&\left. -\delta^{ij}\frac{r^lr^m}{r^2}\right]\frac{1}{r^2}g_{2,2}(r) - 4\left[ -\delta^{li}\frac{r^jr^m}{r^2} -\delta^{mj}\frac{r^ir^l}{r^2}  + 2\frac{r^ir^jr^lr^m}{r^4} \right]\lp\frac{1}{r^2}g_{2,2}(r) - \frac{1}{2r}g_{2,2}'(r)\rp \\~\nonumber
&- \lp \frac{r^lr^m}{r^2}-\frac{1}{2}\delta^{lm} \rp\frac{r^ir^j}{r^2}g_{2,2}''(r)+\left.\lp \frac{r^lr^m}{r^2}-\frac{1}{2}\delta^{lm} \rp\lp  \frac{r^ir^j}{r^2}-\delta^{ij}\rp\frac{1}{r}g_{2,2}'(r)\right\}     \, ,
%+ O\lp\nabla^{klm}\mu(\mathbf{R})\rp
\end{align}
where primes indicate derivatives with respect to $r$. Comparing Eq.\ (\ref{eq:expFFAA}) to our ansatz (\ref{eq:AAansatz}) we conclude that $G^{ij}_k=U_k\nabla_{x_\perp}^i\nabla_{y_\perp}^j\gamma_k$
where $k$ now denotes the nucleus. This is the main result of this subsection.
Note that we can expand the exponential in Eq.\ (\ref{eq:Uexp}) if we can argue that $g^4\Gamma$ is small, and thus $U= 1+\mathcal{O}(g^4\Gamma)$. This is the case at very short distances $\mathbf{y}_\perp \to \mathbf{x}_\perp$, where $\Gamma$ vanishes due to the subtraction of the UV-dominant terms. This argument was, e.g., used in Ref. \cite{Chen:2015wia} where the small-time expansion requires all quantities to be evaluated at $\tau=0$, leading to
\begin{equation} G^{ij}_k(\mathbf{x}_\perp,\mathbf{y}_\perp)=\nabla_{x_\perp}^i\nabla_{y_\perp}^j\gamma_k(\mathbf{x}_\perp,\mathbf{y}_\perp)  \, .
\end{equation}
This approximation can also be used in the weak-field limit as long as $g^4\Gamma\sim g^4\mu$ is chosen sufficiently small.

\subsection{Correlations of Initial Fields}

We are now ready to return to Eq.\ (\ref{eq:ll2}). We first work out the contractions of transverse indices to leading order gradients in $\mu_1$ and $\mu_2$
\begin{subequations}
\begin{align}\label{eq:expEEpBB}
\delta^{ij} \delta^{kl} G_1^{ik}(\mathbf{x}_\perp,\mathbf{y}_\perp) G_2^{jl}(\mathbf{x}_\perp,\mathbf{y}_\perp)  &= %g^6\frac{N_c}{N_c^2-1}
 \mu_1(\mathbf{R})\mu_2(\mathbf{R})\lc \frac{1}{r^2}  {g_{0,1}'}^2 + {g_{0,1}''}^2 \rc  U_1(r)U_2(r)\\
\epsilon^{ij} \epsilon^{kl} G_1^{ik}(\mathbf{x}_\perp,\mathbf{y}_\perp) G_2^{jl}(\mathbf{x}_\perp,\mathbf{y}_\perp)  &= %g^6\frac{N_c}{N_c^2-1}
\mu_1(\mathbf{R})\mu_2(\mathbf{R}) \frac{2}{r} g_{0,1}' g_{0,1}''  
U_1(\mathbf{R},r)U_2(\mathbf{R},r)
\\
\delta^{ij} \epsilon^{kl} G_1^{ik}(\mathbf{x}_\perp,\mathbf{y}_\perp) G_2^{jl}(\mathbf{x}_\perp,\mathbf{y}_\perp) &= \lp \mu_1(\mathbf{R}) \frac{\mathbf{r}\times{\nabla}_R}{r}\mu_2(\mathbf{R}) - \mu_2(\mathbf{R})\frac{\mathbf{r}\times {\nabla}_R}{r} \mu_1(\mathbf{R})\rp \\\nn &\times \frac{g_{0,1}'}{2}\lc\frac{g_{0,1}'}{r} + g_{0,1}''\rc U_1(\mathbf{R},r)U_2(\mathbf{R},r)  \label{eq:EBinit} \\
\epsilon^{ij} \delta^{kl} G_1^{ik}(\mathbf{x}_\perp,\mathbf{y}_\perp) G_2^{jl}(\mathbf{x}_\perp,\mathbf{y}_\perp) &= -\delta^{ij} \epsilon^{kl} G_1^{ik}(\mathbf{x}_\perp,\mathbf{y}_\perp) G_2^{jl}(\mathbf{x}_\perp,\mathbf{y}_\perp) .
\end{align}
\end{subequations}

Here, the $U_k(\mathbf{R},r)$ are the expression from (\ref{eq:Uexp}) evaluated using only the first term $\sim \mu_k$ in $\gamma_k$ as we currently look only at terms that are leading in an expansion in gradients.
According to Eq.\ (\ref{eq:ll2alt}) the expressions above also give the correlation functions of initial longitudinal fields at $\tau=0$, up to a factor $g^6N_c/(N_c^2-1)$. The four expressions listed correspond to $\langle E_0 E_0\rangle$, 
 $\langle B_0 B_0\rangle$,  $\langle E_0 B_0\rangle$, and  $\langle B_0 E_0\rangle$, respectively. 

\begin{figure}[tb]
 \begin{center}
    \includegraphics[width=8cm]{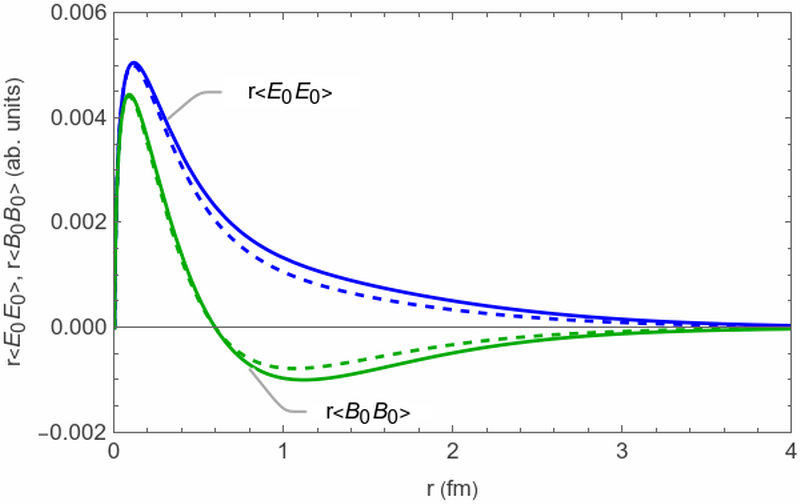}
 \end{center}
  \caption{\label{fig:E0E0} The initial correlation functions $r\langle E_0E_0\rangle$ and $r\langle B_0B_0\rangle$ in the glasma, divided by the factor $\mu^2 g^6 N_c/(N_c^2-1)$,  as functions of the relative distance $r$ of the fields. Calculations use the MV-model with symmetric constant source densities $\mu_1=\mu_2=\mu$ and $m=1$ fm$^{-1}$. The common factor has been divided out to allow shape comparisons between different values of $g$ and $\mu$. The factor $r$ tames the logarithmic singularity at $r=0$ for better presentation. Three scenarios are shown: (i) The correlators with $U$ forced to be 1 (solid lines), (ii) The full form including $U$ using a weak-field scenario $g=0.1$, $\mu=1$ fm$^{-2}$, and (iii) the full form using an almost physical scenario with a realistic $\mu=19.4$ fm$^{-2}$ but $g$ fixed at unity (dashed lines). The weak-field result is indistinguishable from the simplified result with $U=1$.}
\end{figure}

 The shape of the initial field correlation functions and the importance of the factors $U_k(\mathbf{R},r)$ are explored in Fig.\ \ref{fig:E0E0} for the case of the MV-model with constant charge density fluctuations $\mu_1=\mu_2=\mu$. The initial field correlation functions thus are only functions of the relative coordinate $r$. A factor $g^6\mu^2 N_c/(N_c^2-1)$ is scaled out to allow for easy comparison of different values of $g$ and $\mu$. The correlation functions have singularities $\sim \log r$ and $\sim \log^2 r$ at the origin and a weight factor $r$ is thus introduced for plotting. The UV-singularity is a consequence of the choice in the MV-model to allow color charge fluctuations on infinitely small length scales to be uncorrelated ($\mathcal{D}_{\mathrm{MV}}=\delta^{(2)}$), which can lead to arbitrarily large fields on short distance scales. This will be our motivation to introduce a model with UV-regulated charge fluctuations in Sec.\ \ref{sec:AGaussianModel}.  Fig.\ \ref{fig:E0E0} shows $r\langle E_0 E_0\rangle$ and
 $r\langle B_0 B_0\rangle$ (modulo the factors discussed above) since the mixed electric-magnetic correlation functions vanish in absence of gradients of $\mu$. This is consistent with vanishing initial Chern-Simons charge $\langle \mathbf{E}\cdot \mathbf{B}\rangle $ that has been noted by several authors \cite{Lappi:2006fp}. 

Fig.\ \ref{fig:E0E0} shows computations in three scenarios: a weak-field case ($g=0.1$, $\mu=1$ fm$^{-2}$), an almost physical case where the source densities are realistic (see Appendix \ref{sec:app:IPGlasma}) but the coupling moderate ($g=1$, $\mu=19.4$ fm$^{-2}$), and a simplified calculation in which we set $U(r)=1$. The weak coupling results are indistinguishable from the simplified calculation, while the almost physical case shows a roughly 27\% suppression of the correlation function, but no significant change in its shape. This is a good indication that $U=1$ is an acceptable approximation as long as $g$ is not too large, and we will adopt this approximation for our weak field scenario.

\subsection{The Components of the Energy Momentum Tensor}
\label{sec:emtensor}

We are now ready to tackle the remaining integrals in Eqs.\ (\ref{eq:ll2}), (\ref{eq:tt2}) and (\ref{eq:tl2}). For this and the following steps we will restrict ourselves to the special case $\mathbf{ r}=0$ and $\tau'=\tau$. I.e., our correlation functions are expectation values of expressions quadratic in the fields, e.g.\ $\left\langle E_z(\tau,\mathbf{x}_\perp) E_z(\tau',\mathbf{y}_\perp) \right\rangle \to \langle E_z^2(\tau, \mathbf{R})\rangle$. It turns out that for sufficiently smooth color density variances $\mu_k$ all integrals except one can be taken analytically without specifying the details of the gluon distribution (\ref{eq:gammaf}). We will also enforce the weak coupling limit by setting the $U_k=1$.
We report the full results up to second order gradients in Appendix \ref{sec:app:emtensor} and restrict ourselves here to the leading terms in the gradient expansion. We discuss the case of the longitudinal electric field as a blueprint.

To evaluate the $\delta$-functions we carry out one more coordinate transformation to $\mathbf{q}=\mathbf{R}-\mathbf{R}'+\mathbf{r}'/2$.  Recall that the initial field correlation functions have to be taken at the ``primed" coordinates. 
Then Eq.\ (\ref{eq:ll2alt}) becomes
\begin{multline}
  \label{eq:ll2step}
   \left\langle E_z^2(\tau, \mathbf{R}) \right\rangle =    \frac{1}{(2\pi\tau)^2} 
  \int d^2 R' \int d^2 r' \,  \delta(\tau- |\mathbf{R}-\mathbf{R}'+\mathbf{r}'/2|)
  \delta(\tau- |\mathbf{R}-\mathbf{R}'-\mathbf{r}'/2|)  
  \left\langle E_0  E_0 \right\rangle (\mathbf{R}',r') \\ 
 = \frac{1}{(2\pi\tau)^2} 
  \int d^2 r' \int d^2 q \,  \delta(\tau- q)
  \delta(\tau- |\mathbf{q} + \mathbf{r}'|)  
  \left\langle E_0  E_0 \right\rangle (\mathbf{R},r') + \mathcal{O}(\mu_2 \nabla \mu_1, \mu_1 \nabla \mu_2)
   \, .
\end{multline}
In the second step we have used the assumption that the $\mu_k$ should be reasonably smooth and expanded the factors $\mu_1(\mathbf{R}')$ and $\mu_2(\mathbf{R}')$ that appear in $\langle E_0  E_0\rangle (\mathbf{R}',r')$
as
\begin{equation}
   \label{eq:secondmuexp}
   \mu_k(\mathbf{R}') = \mu_k(\mathbf{R})
    + (-\mathbf{q}+\mathbf{r}'/2)^i \nabla^i\mu_k(\mathbf{R}) + \ldots 
\end{equation}
Without this expansion it is impossible to continue without specific input for the $\mu_k$. Here, we aspire to compute results complete  to second order in gradients.  In the following, we will discuss the leading terms explicitly, and quote the remaining terms up to second order in Appendix \ref{sec:app:emtensor}. The result of the expansion is that the initial field correlation function is now computed with the charge density variances at positions $\mathbf{R}$ while the difference coordinate is taken to be the integration variable $\mathbf{r}'$.

We are able to proceed and arrive at the leading term
\begin{multline}
  \label{eq:ll2step2}
   \left\langle E_z^2(\tau, \mathbf{R}) \right\rangle    
   = \frac{2}{(2\pi)^2} 
  \int  d^2 r' \int_0^{2\pi} d\phi\,
  \delta\left(2 r'\tau \cos(\alpha-\phi) - {r'}^2 \right) 
  \left\langle E_0  E_0 \right\rangle (\mathbf{R},r') \\ 
   = \frac{1}{\pi^2 } 
  \int  d^2 r' \frac{\Theta(2\tau - r')}{r'\sqrt{4\tau^2-r'^2}}
  \left\langle E_0  E_0 \right\rangle (\mathbf{R},r')
   \, .
\end{multline}
where $\alpha$ and $\phi$ are the angle of the vectors $\mathbf{r}'$ and $\mathbf{q}$, respectively. The $\delta$-function remaining in the first line
%encodes the intersection of the two circles indicated in Fig.\ \reffig:geometry. 
allows for two values of $\phi$ to contribute to the integral as long as $0<r'<2\tau$. The $\phi$-integral is zero for $r'>2\tau$. %$\Theta$ is the Heaviside step function.
We can easily take the remaining azimuthal angle integral and obtain
\begin{multline}
  \label{eq:ll2step3}
   \left\langle E_z^2(\tau, \mathbf{R}) \right\rangle    
   = \frac{2}{\pi } 
  \int_0^{2\tau}  dr' \, \frac{1}{\sqrt{4\tau^2-r'^2}}
  \left\langle E_0  E_0 \right\rangle (\mathbf{R},r')  \\ 
  = \frac{g^6 N_c}{(N_c^2-1)}\mu_1(\mathbf{R}) \mu_2(\mathbf{R}) \frac{2}{\pi}
  \int_0^{2\tau} \frac{ dr'}{\sqrt{4\tau^2-r'^2}} \left(\frac{1}{r'^2}  {g_{0,1}'(r')}^2 + {g_{0,1}''(r')}^2 \right)
 % U_1(\mathbf{R},r')U_2(\mathbf{R},r')
   \, .
\end{multline}
where we have used the result from Eq.\ \eqref{eq:expEEpBB}.

Similarly, the leading term in a gradient expansion for the longitudinal magnetic field is
\begin{equation}
 \label{eq:ll2step4}
   \left\langle B_z^2(\tau, \mathbf{R}) \right\rangle    
   = \frac{g^6 N_c}{(N_c^2-1)}\mu_1(\mathbf{R}) \mu_2(\mathbf{R}) \frac{2}{\pi}
  \int_0^{2\tau} \frac{ dr'}{\sqrt{4\tau^2-r'^2}} \left(\frac{2}{r'}  {g_{0,1}'(r')} {g_{0,1}''(r')} \right)
  %U_1(\mathbf{R},r')U_2(\mathbf{R},r')
   \, ,
\end{equation}
while the mixed terms of longitudinal fields, $\langle E_zB_z(\tau, \mathbf{R})\rangle $, vanishes at least up to second order in gradients.
We report the integrals needed for the remaining expressions in Appendix \ref{sec:app:integrals}. We simply quote the final results for the leading terms in the gradient expansion here. For the longitudinal-transverse case we have for the leading terms in gradients
\begin{subequations}
\begin{align}
 \label{eq:tl2step2}
   \left\langle E_m B_z(\tau, \mathbf{R}) \right\rangle    
   &= \frac{g^6 N_c}{ N_c^2-1} \epsilon^{mi} \left( \mu_1(\mathbf{R})\nabla^i \mu_2(\mathbf{R}) +
   \mu_2(\mathbf{R})\nabla^i \mu_1(\mathbf{R})\right) 
   \\ \nn & \qquad \times
   \frac{1}{4 \pi\tau}
  \int_0^{2\tau}  dr'\, \sqrt{4\tau^2-r'^2} \left(\frac{2}{r'}  {g_{0,1}'(r')} {g_{0,1}''(r')} \right)
  %U_1(\mathbf{R},r')U_2(\mathbf{R},r')
  \\ 
  \left\langle B_m E_z(\tau, \mathbf{R}) \right\rangle    
   &= \frac{g^6 N_c}{ N_c^2-1} \epsilon^{im} \left( \mu_1(\mathbf{R})\nabla^i \mu_2(\mathbf{R}) +
   \mu_2(\mathbf{R})\nabla^i \mu_1(\mathbf{R})\right)
   \\ \nn & \qquad \times
   \frac{1}{4 \pi\tau}
  \int_0^{2\tau}  dr'\, \sqrt{4\tau^2-r'^2} \left(\frac{1}{r'^2}  {g_{0,1}'^2(r')} + {g_{0,1}''^2(r')} \right)
   \\ 
  \left\langle E_m E_z(\tau, \mathbf{R}) \right\rangle    
   &= \left\langle B_m B_z(\tau, \mathbf{R}) \right\rangle
\\ \nn   &= -\frac{g^6 N_c}{ N_c^2-1}  \left( \mu_1(\mathbf{R})\nabla^m \mu_2(\mathbf{R}) -
   \mu_2(\mathbf{R})\nabla^m \mu_1(\mathbf{R})\right)
   \\ \nn & \qquad \times
   \frac{1}{4 \pi\tau}
  \int_0^{2\tau}  dr'\, \frac{r'}{\sqrt{4\tau^2-r'^2}} \left(\frac{1}{r'}  {g_{0,1}'^2(r')} + g_{0,1}'(r'){g_{0,1}''(r')} \right)
   \, .
\end{align}
\end{subequations}

For transverse-transverse field expressions the leading terms are
\begin{subequations}
\begin{align}
 \label{eq:tt2step2}
   \left\langle E_m E_n(\tau, \mathbf{R}) \right\rangle    
   &= \frac{g^6 N_c}{ N_c^2-1} \delta^{mn} \mu_1(\mathbf{R}) \mu_2(\mathbf{R}) 
   \frac{1}{2 \pi\tau^2}
  \int_0^{2\tau} dr' \frac{2\tau^2- r'^2}{ \sqrt{4\tau^2-r'^2}} \left(\frac{2}{r'}  {g_{0,1}'(r')} {g_{0,1}''(r')} \right)
  %U_1(\mathbf{R},r')U_2(\mathbf{R},r')
  \\ 
  \left\langle B_m B_n(\tau, \mathbf{R}) \right\rangle    
   &= \frac{g^6 N_c}{ N_c^2-1} \delta^{mn} \mu_1(\mathbf{R}) \mu_2(\mathbf{R}) 
   \frac{1}{2 \pi\tau^2}
  \int_0^{2\tau} dr' \frac{2\tau^2- r'^2}{ \sqrt{4\tau^2-r'^2}}   \left(\frac{1}{r'^2}  {g_{0,1}'^2(r')} + {g_{0,1}''^2(r')} \right)
   \, .
\end{align}
\end{subequations}

The mixed term of the transverse fields only starts at second order in gradients of the $\mu_k$ and reads
\begin{align}
 \label{eq:tt2step2_Mix}
   \left\langle E_m B_n(\tau, \mathbf{R}) \right\rangle    
   &= \frac{g^6 N_c}{ N_c^2-1} \left( \delta^{nl}\epsilon^{mi} - \delta^{ml}\epsilon^{ni} \right) \nabla^i\left( \mu_2(\mathbf{R}) \nabla^l \mu_1(\mathbf{R})  - \mu_1(\mathbf{R})\nabla^l \mu_2(\mathbf{R})\right)  \\ \nn \qquad & \times
   \frac{1}{16\pi\tau^2}
  \int_0^{2\tau} dr' r' \sqrt{4\tau^2- r'^2}  \left(\frac{1}{r'}  {g_{0,1}'^2(r')} +{g_{0,1}''(r')}g_{0,1}(r') \right)
     \, .
\end{align}

We can now assemble the components of the energy momentum tensor from the expressions for the fields. To present our results in an organized manner, we express the gradient expansion of the energy momentum tensor in terms of coefficients with additional indices 1, 2, etc. We truncate the expansion at three or more derivatives on one $\mu_k$. For the longitudinal and transverse energy density, and the pressure anisotropy we consider the following terms 
suppressing some arguments of $\mathbf{R}$ for brevity,
\begin{align}\label{eq:T1}
\left\langle \epsilon_L(\tau,\mathbf{R})   \right\rangle &= \frac{1}{2}
 \left\langle B_z^2+E_z^2 \right\rangle    \\ \nn
 &= \mu_1\mu_2(\mathbf{R})\epsilon_{L,1}(\tau) + \nabla\mu_1\cdot\nabla\mu_2(\mathbf{R})\epsilon_{L,2}(\tau) + \lp \mu_1\triangle\mu_2 + \mu_2\triangle\mu_1 \rp(\mathbf{R})\epsilon_{L,3}(\tau) + \triangle\mu_1\triangle\mu_2(\mathbf{R})\epsilon_{L,4}(\tau)  \\ \nn
  &+ \nabla^{ij}\mu_1\nabla^{ij}\mu_2(\mathbf{R})\epsilon_{L,5}(\tau)  \, ,
  %U_1(\mathbf{R},r')U_2(\mathbf{R},r')  
  \\\label{eq:T2}
  \left\langle \epsilon_T(\tau,\mathbf{R})   \right\rangle &= \frac{1}{2}
       \left\langle B_nB_n + E_nE_n \right\rangle  \\ \nn
  &= \mu_1\mu_2(\mathbf{R})\epsilon_{T,1}(\tau) + \nabla\mu_1\cdot\nabla\mu_2(\mathbf{R})\epsilon_{T,2}(\tau) + \lp \mu_1\triangle\mu_2 + \mu_2\triangle\mu_1 \rp(\mathbf{R})\epsilon_{T,3}(\tau) + \triangle\mu_1\triangle\mu_2(\mathbf{R})\epsilon_{T,4}(\tau)  \\\nn
    &+ \nabla^{ij}\mu_1\nabla^{ij}\mu_2(\mathbf{R})\epsilon_{T,5}(\tau) \, , 
  \\ \label{eq:omegadef}
  \left\langle \omega(\tau,\mathbf{R})   \right\rangle &= \frac{1}{2}
 \left\langle B_yB_y + E_yE_y - B_xB_x - E_xE_x  \right\rangle   \\ \nn
 &= \lp \nabla^x\mu_1\nabla^x\mu_2 - \nabla^y\mu_1\nabla^y\mu_2 \rp(\mathbf{R})\omega_1(\tau) + \lp \mu_1\ls\nabla^{xx}\mu_2-\nabla^{yy}\mu_2\rs + \mu_2\ls\nabla^{xx}\mu_1-\nabla^{yy}\mu_1\rs \rp(\mathbf{R})\omega_2(\tau)\\  \nn
    &+ \lp \nabla^{xx}\mu_1\nabla^{xx}\mu_2 - \nabla^{yy}\mu_1\nabla^{yy}\mu_2 \rp(\mathbf{R})\omega_3(\tau) \, .
\end{align} 
We can build the expressions for all four diagonal elements of the energy momentum tensor from these results. The total energy density and longitudinal pressure are $\langle \epsilon\rangle =\langle \epsilon_L \rangle +\langle \epsilon_T \rangle $ and 
$\langle P_z \rangle =\langle - \epsilon_L \rangle +\langle \epsilon_T \rangle $, respectively. The transverse pressures are $\langle P_{x,y} \rangle =\langle \epsilon_L \rangle \pm\langle \omega \rangle$.
The pressure anisotropy $\omega$ is a response to gradients of flow which themselves need gradients, so it appears intuitive that its expansion starts at second order.

Similarly, one obtains the Poynting vector in transverse and longitudinal directions
\begin{align}
\left\langle S_i(\tau,\mathbf{R})   \right\rangle &= \epsilon^{ijk} 
 \left\langle E_j B_k \right\rangle   \\ \nn
  &= -\nabla^i(\mu_1\mu_2)(\mathbf{R})S_{i,1}(\tau) -\lp \nabla^i\mu_1\triangle\mu_2 + \nabla^i\mu_2\triangle\mu_1 \rp (\mathbf{R})S_{i,2}(\tau) -\nabla^i\lp\nabla\mu_1\cdot\nabla\mu_2\rp(\mathbf{R})S_{i,3}(\tau) \,  \\
  \left\langle S_z(\tau,\mathbf{R})  \right\rangle &= \epsilon^{3jk} 
 \left\langle E_j B_k \right\rangle    \\ \nn
  &= \lp \mu_1\triangle\mu_2 - \mu_2\triangle\mu_1 \rp(\mathbf{R})S_{z,1}(\tau)  \, ,
\end{align} 
respectively, describing energy flow of the gluon field. Transverse flow needs gradients in $\mu_k$. Indeed, the transverse flow vector is proportional to the negative gradient of $\mu_1\mu_2$, as anticipated in Ref.\ \cite{Chen:2015wia}. The missing components are the transverse flow of longitudinal momentum and the transverse shear component
\begin{align}
\left\langle T_{iz}(\tau,\mathbf{R})   \right\rangle &= -  \left\langle E_i E_z + B_iB_z \right\rangle    \\ \nn
   &= \lp \mu_1\nabla^i\mu_2 - \mu_2\nabla^i\mu_1 \rp(\mathbf{R})T_{iz,1}(\tau) -\lp\triangle\mu_1\nabla^i\mu_2 - \triangle\mu_2\nabla^i\mu_1 \rp(\mathbf{R})T_{iz,2}(\tau)\\ \nn
    &+ \lp\nabla\mu_1\cdot\nabla\lp\nabla^i\mu_2\rp - \nabla\mu_2\cdot\nabla\lp\nabla^i\mu_1\rp\rp(\mathbf{R})T_{iz,3}(\tau) \, ,\\ 
    \label{eq:T9}
  \left\langle T_{xy}(\tau,\mathbf{R})   \right\rangle &= -
 \left\langle B_xB_y + E_xE_y   \right\rangle   \\ \nn  &=
 \lp \nabla^x\mu_1\nabla^y\mu_2 + \nabla^y\mu_1\nabla^x\mu_2 \rp(\mathbf{R})T_{xy,1}(\tau) + \lp \mu_1\nabla^{xy}\mu_2 + \mu_2\nabla^{xy}\mu_1 \rp(\mathbf{R})T_{xy,2}(\tau) \\  \nn
    &+ \lp \triangle\mu_1\nabla^{xy}\mu_2 + \nabla^{xy}\mu_1\triangle\mu_2 \rp (\mathbf{R})T_{xy,3}(\tau)   \, .  
\end{align} 

One finds that the leading terms in the gradient expansion can be expressed in just a few functions that can be derived from the $g$-functions used to parameterize $\gamma$. For the leading terms we only need
\begin{align}
   \label{eq:newfunc}
    f_1(r) &= g^6\frac{ N_c}{N_c^2-1}{\left( \frac{1}{r}g_{01}'(r)
    + g_{01}'' \right)}^2 \, , \\
    \phi_1(r) &= g^6\frac{ N_c}{N_c^2-1}\lp\frac{1}{r^2} g_{01}'^2(r) + \frac{1}{r} g_{01}''(r) g_{01}'(r)\rp \, , \\
    f_2(r) &= g^6\frac{N_c}{N_c^2-1}\frac{1}{r^2} { g_{0,1}'}^2 \, .
\end{align}
More functions are needed if terms up to second order in the gradient expansion are computed, see Appendix \ref{sec:app:emtensor}. For the leading terms we have
\begin{subequations}\label{eq:gradient_Expansion}
\begin{align}
    \epsilon_{L,1}(\tau)  &= \frac{1}{\pi}
  \int_0^{2\tau} \frac{ dr'}{\sqrt{4\tau^2-r'^2}} f_1(r')  \, ,\\
    \epsilon_{T,1}(\tau)  &= \frac{1}{2\pi\tau^2}
  \int_0^{2\tau} dr' \frac{2\tau^2-r'^2 }{\sqrt{4\tau^2-r'^2}} f_1(r') \, ,  \\
    \omega_1(\tau) &= \frac{1}{4\pi}\int\limits_0^{2\tau} \frac{ dr'}{\sqrt{4\tau^2-r'^2}}  \lc \frac{1}{4}\lp 4\tau^2 - r'\hspace{0pt}^2 \rp f_1(r') + r'\hspace{0pt}^2f_2(r') \rc \, , \\
    S_{i,1}(\tau) 
   &= \frac{1}{4\pi\tau}
  \int_0^{2\tau} dr'\sqrt{4\tau^2-r'^2}  f_1(r')  \, ,
  \\
    S_{z,1}(\tau)
   &= - \frac{1}{8\pi\tau^2}
  \int_0^{2\tau} dr' r'^2 \sqrt{4\tau^2-r'^2} \phi_1(r')
   \, , \\
    T_{iz,1}(\tau) &=
 \frac{1}{2\pi\tau}
  \int_0^{2\tau} dr'\frac{r'^2}{\sqrt{4\tau^2-r'^2}}  \phi_1(r')  \, , \\
    T_{xy,1}(\tau) &= 
     \frac{1}{4\pi}\int\limits_0^{2\tau} \frac{ dr'}{\sqrt{4\tau^2-r'^2}} \lc \frac{1}{4}\lp 4\tau^2 - r'\hspace{0pt}^2 \rp f_1(r') + r'\hspace{0pt}^2f_2(r') \rc  \, .  
\end{align}
\end{subequations}
Recall that the depencies on transverse coordinates can be read off from Eqs.\ (\ref{eq:T1}) through (\ref{eq:T9}).
$T_{xy}$ has the same time dependence as $\omega$. This seems to be conincidental as this pattern does not hold for all orders in the gradient expansion, see Eq.\ \eqref{eq:GlasmaStressTensor}.

We now have expressions for the energy momentum tensor in the weak field limit allowing for non-constant, although reasonably smooth charge density variances $\mu_k$, and for a broad class of models for the gluon distribution. The dependence on the $\mu_k$ and their gradients is compatible with what was observed using the small time expansion \cite{Chen:2015wia}. E.g.\ the vectors $\beta^i\sim \mu_1 \nabla^i  \mu_2 -  \mu_2\nabla^i\mu_1$ and $\alpha^i\sim \nabla^i\left(\mu_1\mu_2 \right)$ defined in that reference, governing the flow of energy and longitudinal momentum in the transverse plane, agree with the direction of the corresponding quantities $T_{iz}$ and $S_i$ in our work. Another obvious test that our results should pass is energy momentum conservation, $0=\partial_{\mu} T^{\mu\nu}=\partial T^{0\nu}/\partial \tau  + \nabla^i T^{i\nu} + \tau^{-1} \partial T^{3\nu}/\partial\eta $ order by order in gradients of $\mu_k$. To compute the $\eta$-derivatives, an infinitesimal Lorentz-boost in $z$-direction can be taken first, leading to, e.g., $\partial T^{30} /\partial\eta = T^{00}+ T^{33}$ at $\eta=0$. The check of energy momentum conservation is straight forward from there on, and we leave it as an exercise for the interested reader. 

To summarize, the expressions for the energy momentum tensor derived in this subsection assume weak fields and slowly varying charge density variances. However, they are valid without specifying a concrete model for the gluon distribution in the nuclei. We will work out the last integral in each expression both for the MV-model, and in a generalized model which we introduce in the next section.

\section{An Improved Gaussian Model}\label{sec:AGaussianModel}

In Sec.\ \ref{sec:ASTE}, we will evaluate the results of the last section in the MV-model. However,  one major issue will be the susceptibility of this model to both IR and UV divergences. These singularities have to be regulated by cutoffs that are introduced ad hoc. We would like to propose an amended model for the gluon distribution of a nucleus that improves some of these issues. We will consider results obtained for this new model alongside their MV counterparts.
We will call this model the \emph{improved Gaussian model}, as it replaces $\delta$-functions in the MV-model with Gaussians. The main improvement arises from the imposition of global color neutrality which introduces a physical constraint to regulate the long-distance behavior. We also opt to introduce a UV regulator at the level of color charge correlations which makes it unnecessary to introduce additional momentum cutoffs when observables are computed. 
We will then show that the improved Gaussian model leads to a covariant gauge correlation function of the form  Eq.\  \eqref{eq:gammaf}, thus the results of the previous section apply.

\subsection{Kovchegov's Model and Its Generalization}\label{sec:AGaussianModel_ssec:KovchegovModel}

For improved IR behavior we utilize ideas outlined by Kovchegov, and by Lam and Mahlon. Kovchegov proposed a microscopic model for the the gluon distribution of a nucleus in Ref.\  \cite{Kovchegov:1996ty}. In his model nucleons are placed as spheres of equal radius inside the classical radius of the nucleus, distributed uniformly. The nucleons themselves are treated as pairs of point-like quarks and antiquarks, which are again distributed uniformly inside each nucleon. The charge volume density for $A$ nucleons in the lab frame in this model is
\begin{equation}\label{eq:NucleusCharge}
\rho_\mathrm{Kov} = g \sum\limits_{n=1}^A (t^{\underline{a}})(t^{\underline{a}}_n)\ls \delta(x^- - x_n^-)\delta^2(\mathbf{x}_\perp - \mathbf{x}_{\perp,n}) - \delta(x^- - {x'}_n^{-})\delta^2(\mathbf{x}_\perp - \mathbf{x}\hspace{1pt}'_{\perp,n}) \rs  \, .
\end{equation}
The $n$th quark is located at position $(x_n^-, \mathbf{x}_{\perp,n})$ on the light cone, and the $n$th antiquark at position $({x'}_n^{-}, \mathbf{x}\hspace{1pt}'_{\perp,n})$. $t^{\underline{a}}$ is the generator in SU($N_c$) color space and $ t^{\underline{a}}_n $ is the generator for each nucleon. 

From this ansatz, Kovchegov explicitly recovers the Gaussian distribution of color charge that is posited in the MV-model. To this end, he has to make one crucial additional assumption. When computing charge correlation functions, he is forced to drop cross terms  which correlate quarks and anti-quarks. It was pointed out by Lam and Mahlon that restoring this term leads to global color neutrality and improved IR behavior \cite{Lam:1999wu}. Here, we follow the discussion in the latter paper. 

In Kovchegov's model one can derive explicit expressions for charge correlation functions. To be precise, one finds \cite{Lam:1999wu}
${\langle \rho_{\underline{a}} \rangle}_\mathrm{Kov} = 0$ and 
\begin{align}\label{eq:rhoRhoK}
{\langle \rho_{\underline{a}}(x)\rho_{\underline{b}}(y) \rangle}_\mathrm{Kov} &= \frac{g^2}{N_c} A \delta_{\underline{a}\underline{b}}\delta(x^- - y^-)\mathcal{C}(\mathbf{x}_\perp,\mathbf{y}_\perp) \, ,
\end{align}
where $\mathcal{C}(\mathbf{x}_\perp,\mathbf{y}_\perp)$ is a transverse correlation function and $A$ is the number of nucleons in the nucleus. The correlation function has two contributions, $\mathcal{C}(\mathbf{x}_\perp, \mathbf{y}_\perp) = \mathcal{I}(\mathbf{x}_\perp, \mathbf{y}_\perp) - \bar{\mathcal{I}}(\mathbf{x}_\perp, \mathbf{y}_\perp)$, which arise from (anti)quark-(anti)quark and quark-antiquark terms respectively. They are %\cite{Lam:1999wu}
\begin{align}\label{eq:I_functions}
    \mathcal{I}(\mathbf{x}_\perp, \mathbf{y}_\perp) &= \int d^2r_\perp'dz' \int d^2r_\perp''dz'' D_N(\mathbf{r}'_\perp , z') D_q(\mathbf{r}''_\perp , z'') T_q(\mathbf{x}_\perp - \mathbf{r}'_\perp - \mathbf{r}''_\perp) T_q(\mathbf{y}_\perp - \mathbf{r}'_\perp - \mathbf{r}''_\perp)  \, , \\\nn
    \mathcal{\bar{I}}(\mathbf{x}_\perp, \mathbf{y}_\perp) &= \int d^2r_\perp'dz'\hspace{-1pt} \int d^2r_\perp''dz''\hspace{-1pt} \int d^2r_\perp''' dz''' D_N(\mathbf{r}'_\perp , z') D_q(\mathbf{r}''_\perp , z'') D_q(\mathbf{r}'''_\perp , z''') T_q(\mathbf{x}_\perp\hspace{-1pt} - \mathbf{r}'_\perp - \mathbf{r}''_\perp) T_q(\mathbf{y}_\perp\hspace{-1pt} - \mathbf{r}'_\perp\hspace{-1pt} - \mathbf{r}'''_\perp)  \, .
\end{align}
We have made the following generalizations compared to  Ref.\  \cite{Lam:1999wu}. The sharp homogeneous spheres for the nucleus and the nucleons have been replaced by more realistic distribution functions $D_N$ and $D_q$ of nucleons in a nucleus, and quarks and antiquarks in a nucleon, respectively (note the overall factor $A$ removed from $D_N$ in Eq.\ \eqref{eq:rhoRhoK}). We also replace the transverse $\delta$-functions by functions $T_q$ which we use to introduce an effective UV cutoff $l_\textrm{UV}$. This is the smallest length scale in the problem at which color charge can be resolved. The quark distributions $D_q$ in a nucleon have their own characteristic size scale $l_\textrm{conf}$. We assume that %$l_\textrm{conf}\gg l_\textrm{UV}$ and that 
$\lim_{l_\textrm{UV}\to 0} T_q(\mathbf{r}_\perp) = \delta(\mathbf{r}_\perp$). We can now be more precise and state that the assumptions of the MV-model can be recovered from this microscopic model by dropping the cross term $\bar{\mathcal{I}}$, choosing $T_q(\mathbf{r}_\perp) =\delta(\mathbf{r}_\perp$), and assuming that $D_q$ and $D_N$ represent homogeneous spheres.

We now make some choices that go beyond the MV-model but still allow us to evaluate most of the integrals in the expressions for $\mathcal{I}$ and $\bar{\mathcal{I}}$ analytically. For both $T_q$ and the quark distributions $D_q$ we choose Gaussians defined by their respective relevant scales
\begin{align}\label{eq:QuarkDist}
    T_q \left(\mathbf{r}_\perp\right) &= \frac{1}{2\pi l_\textrm{UV}^2 }e^{-\frac{\mathbf{r}_\perp^2}{2 l_\textrm{UV}^2 }} \, ,  & \int dz D_q \left(\mathbf{r}_\perp,z\right)& = \frac{1}{2\pi l_\textrm{conf}^2 }e^{-\frac{\mathbf{r}_\perp^2}{2 l_\textrm{conf}^2 }} \, .
\end{align}
One can compute that 
\begin{equation}\label{eq:I}
    \mathcal{I}(\mathbf{x}_\perp, \mathbf{y}_\perp) = \int d^2r_\perp'dz'   D_N(\mathbf{r}'_\perp , z') \, \frac{1}{\pi\lp l_\textrm{UV}^2+2 l_\textrm{conf}^2\rp}e^{-\frac{(\mathbf{R}_\perp-\mathbf{r}_\perp')^2}{l_\textrm{UV}^2+2 l_\textrm{conf}^2 }} \, \frac{1}{4\pi l_\textrm{UV}^2 }e^{-\frac{\mathbf{r}_\perp^2}{ 4 l_\textrm{UV}^2 }} 
    = \tilde \mu\left(\mathbf{R}_\perp\right)\, \frac{1}{4\pi l_\textrm{UV}^2 }e^{-\frac{\mathbf{r}_\perp^2}{ 4 l_\textrm{UV}^2 }} 
\end{equation}
after introducing center and relative coordinates $\mathbf{R}_\perp$ and $\mathbf{r}_\perp$ for $\mathbf{x}_\perp$, $\mathbf{y}_\perp$. %, and $\mathbf{y}_\perp$, and using $l_\mathrm{UV} \ll l_\mathrm{conf}$ to simplify the emerging expressions. 
We observe that the dependence on center and relative coordinates factorizes for the correlation function $\mathcal{I}$. The two factors have the same straight forward interpretations as in the MV-model. The distribution of  the relative coordinate is peaked at the UV-scale $l_\mathrm{UV}$. The dependence on $\mathbf{R}_\perp$ is given by a convolution of the nucleon distribution function in the nucleus and the profile of the nucleon, for which we introduce the notation $\tilde \mu$. Modulo normalization factors, the latter is a microscopic model for the charge density $\mu$ in Eq.\ (\ref{eq:MV}).
Furthermore, we have 
    \begin{align}\label{eq:I_Bar}
    \mathcal{\bar{I}}(\mathbf{x}_\perp, \mathbf{y}_\perp) &= \int d^2r_\perp'dz' D_N(\mathbf{r}'_\perp , z') \frac{1}{\pi\lp l_\textrm{UV}^2+l_\textrm{conf}^2\rp }e^{-\frac{(\mathbf{R}-\mathbf{r}_\perp')^2}{ \lp l_\textrm{UV}^2+l_\textrm{conf}^2\rp }} \frac{1}{4\pi \lp l_\textrm{UV}^2+l_\textrm{conf}^2\rp }e^{-\frac{\mathbf{r}_\perp^2}{ 4 \lp l_\textrm{UV}^2+l_\textrm{conf}^2\rp }}  \\\nonumber
    &\approx   \tilde \mu\left(\mathbf{R}_\perp\right) \frac{1}{4\pi \lp l_\textrm{UV}^2+l_\textrm{conf}^2\rp }e^{-\frac{\mathbf{r}_\perp^2}{ 4 \lp l_\textrm{UV}^2+l_\textrm{conf}^2\rp }} 
    \, .
    \end{align} 
The approximate sign on the second line indicates that the convolution of $D_N$ with the nucleon profile that appears here is slightly different from the definition of $\tilde \mu$. This deviation is small as long as $D_N$ is varying slowly on the length scale $\sqrt{l_\textrm{UV}^2+2 l_\textrm{conf}^2}$.
This is true in the center of a large nucleus, and we will use this approximation moving forward.
Finally, we observe that a crucial observation of Lam and Mahlon persists in our generalized calculation: The integral over the correlation function $\mathcal{C}$ disappears as long as $D_N$ is slowly varying, since
\begin{equation}
    \int d^2 r_\perp \mathcal{I}  = \tilde \mu = \int d^2 r_\perp \mathcal{\bar I}  \, .
\end{equation}
As we will see in the next subsection, this is an expression of global color neutrality.

\subsection{The Improved Gaussian Model and Color Neutrality}\label{sec:AGaussianModel_ssec:iGModel}

The factorization observed in the previous subsection justifies the following generalization of Eq.\ \eqref{eq:MV} and the MV-model
\begin{align}\label{eq:rhoRhoG}
{\langle \rho_{\underline{a}}(x)\rho_{\underline{b}}(y) \rangle}_{\mathrm{iG}} &= \frac{g^2}{N_c\hspace{0pt}^2 - 1} \delta_{\underline{a}\underline{b}}h(x^-)\delta(x^- - y^-) 
\mu(\mathbf{R}_\perp)
\mathcal{D}(\mathbf{r}_\perp) ,
\end{align}
where the new correlation function $\mathcal{D}(\mathbf{r}_\perp)$ is the difference of two functions. Following the notation of Lam and Mahlon, we write $\mathcal{D}(\mathbf{r}_\perp) = F(\mathbf{r}_\perp) - C(\mathbf{r}_\perp)$ where $F(\mathbf{r}_\perp)$ is peaked at the UV-scale and $C(\mathbf{r}_\perp)$ is spread out at the confinement scale. In addition, we demand that both functions are normalized to one, such that the overall constraint
\begin{align}\label{eq:ColorNeutrality}
\int d^2r_\perp \hspace{1pt} \mathcal{D}(\mathbf{r}_\perp) &= 0
\end{align}
holds. $\mu$  represents an average color charge density which contains all color charges considered for the creation of the classical gluon field. 

Lam and Mahlon showed in Ref.\ \cite{Lam:1999wu} that the constraint (\ref{eq:ColorNeutrality}) naturally emerges from enforcing global color neutrality for a nucleus or hadron. Global color neutrality does not only apply to ensemble averages, but also to each hadronic or nuclear state individually. Consider a single hadronic state $\lvert h \rangle$ and the expectation value of its charge density operator $\langle h \rvert \rho_{\underline{a}}(x^-, \mathbf{x}_\perp) \lvert h \rangle$. 
%For any given position, the expectation value does not have to be zero. 
If the position coordinates are integrated over all of space, then the expectation value has to be zero because all hadrons are color neutral,
\begin{align}\label{eq:hMag}
%||h||^2 &\equiv 
\langle h \rvert \int dx^- d^2x_\perp \hspace{1pt} \rho_{\underline{a}}(x^-, \mathbf{x}_\perp) \lvert h \rangle = 0   \, .
\end{align}
Lam and Mahlon then show that a similar constraint follows for the two-point function of the charge density, leading precisely to equation (\ref{eq:ColorNeutrality}).
Correlating color charges at finite distances 
has a side effect that Lam and Mahlon have noted in another paper \cite{Lam:2001ax}. The conditions for the central limit theorem are no longer satisfied %Ref.\ \cite{Lam:2001ax} demonstrates that
and the correlations cause non-Gaussian fluctuations in the weight functional $W[\rho]$, mandating higher $\rho$ statistical moments be specified. 
Since we are working in the weak field limit we will neglect any such contributions from higher moments here.

In the following, we make explicit choices for $F$ and $C$ by extracting the relative factors in \eqref{eq:I} and \eqref{eq:I_Bar}. %inspired by the microscopic model with Gaussian distributions $T_q$ and $D_q$ discussed in the previous subsection. 
Note that the functions $F$ and $C$ are rotationally invariant in the transverse plane.
Setting $l_\mathrm{UV}^2=B_q$ and $l_\mathrm{conf}^2=B_c$ in the previous calculations tells us that
%\noindent they are found to be 
%
\begin{align}
F_\mathrm{iG}(r_\perp) &= \frac{1}{4\pi B_q}\exp\ls -\frac{r_\perp^2}{4B_q} \rs & C_\mathrm{iG}(r_\perp) &= \frac{1}{4\pi \lp B_q + B_{c}\rp}\exp\ls -\frac{r_\perp^2}{4\lp B_q + B_{c}\rp} \rs .
\end{align}

Note that the choices made for the improved Gaussian model are consistent with our assumptions for the correlation function $\gamma$ in Eq.\ (\ref{eq:gamma}). We are thus prepared to evaluate the gluon correlation function (\ref{eq:TwoPointAlpha}) for this model. %Recall that we had postulated the form (\ref{eq:gammaf}). 
We can show that the $g_{i,j}(r)$ functions from Eq.\ (\ref{eq:TwoPointAlpha}) in the improved Gaussian model are
\begin{subequations}\label{eq:gFunctions}
\begin{align}\displaybreak[1]\label{eq:g01}
g_{0,1,\text{iG}}(r_\perp) &= \frac{1}{4\pi}\left( B_{q}\ls\exp\lp-\frac{r_\perp^2}{4(B_{q}+B_{c})}\rp - \exp\lp-\frac{r_\perp^2}{4B_{q}}\rp\rs - B_{c}\ls 1 - \exp\lp-\frac{r_\perp^2}{4(B_{q}+B_{c})}\rp + 2\gamma_E\rs \right. \\\nn
&\left.+ \ls\frac{r_\perp^2}{4}+B_{q}\rs E_1\lp\frac{r_\perp^2}{4B_{q}}\rp - \ls\frac{r_\perp^2}{4}+B_{q}+B_{c}\rs E_1\lp\frac{r_\perp^2}{4(B_{q}+B_{c})}\rp - 2B_{c}\ln\ls\frac{mr_\perp}{2}\rs \rp , \\\displaybreak[1]\label{eq:g02}
g_{2,1,\text{iG}}(r_\perp) &= \frac{B_{c}}{48\pi m^2} - B_{c}\frac{r_\perp^2 + 4B_{q} + 2B_{c}}{96\pi} ,\\\displaybreak[1]\label{eq:g22}
g_{2,2,\text{iG}}(r_\perp) &= \frac{1}{96\pi r_\perp^2}\lp\vphantom{\exp\lp -\frac{r_\perp^2}{4(B_{q}+B_{c})} \rp} B_{c}\ls 16B_{c}^2 + 48B_{q}^2 - r_\perp^4 - 12r_\perp^2B_{qc} - 24B_{q}\lp r_\perp^2 - 2B_{c} \rp \rs - 6r_\perp^4B_{c}\ls \gamma_E + \ln\lp\frac{mr_\perp}{2}\rp \rs  \right. \\\nn
& +\exp\lp -\frac{r_\perp^2}{4(B_{q}+B_{c})} \rp\ls B_{q}+B_{c}\rs\ls r_\perp^4 + 8r_\perp^2(B_{q}+B_{c}) - 16(B_{q}+B_{c})^2\rs - \exp\lp -\frac{r_\perp^2}{4B_{q}} \rp B_{q}  \\\nn
&\lb\times\ls r_\perp^4 + 8r_\perp^2B_{q} - 16B_{q}^2\rs  + r_\perp^4\ls \frac{r_\perp^2}{4} + 3B_{q} \rs E_1\lp\frac{r_\perp^2}{4B_{q}}\rp - r_\perp^4\ls\frac{r_\perp^2}{4}+3B_{q}+3B_{c}\rs E_1\lp\frac{r_\perp^2}{4(B_{q}+B_{c})}\rp \rp ,
\end{align}
\end{subequations}
where $E_1$ is the exponential integral $E_1(z)=\int_z^\infty e^{-t}/t \, dt$. The details of the derivation are shown in Appendix \ref{sec:app:gluoncorr}.

\section{Final Results for the Stress Tensor}\label{sec:ASTE}

To obtain final expressions for the stress tensor, the integrals over $r'$ in Eqs.\ (\ref{eq:gradient_Expansion}) need to be computed. We start by further discussing the functions $f_i$ and $\phi_i$ appearing in the integrands for both the MV- and iG-models. We will also argue that the long-time behavior of the components of the energy momentum tensor is qualitatively universal and independent of the model chosen for the sources of the color fields.

\subsection{MV-Model Stress Tensor}

In the MV-model we deal with the modified Bessel functions from Eq.\ \eqref{eq:gMV}. For example, one obtains
\begin{equation}
    f_{1,\text{MV}}(r) = \frac{g^6}{16\pi^2} \frac{N_c}{N_c^2-1} {\big( 2 K_0(mr) - mr K_1(mr)\big)}^2  \, .
    \label{eq:mvf1}
\end{equation}
It will be beneficial to turn the Bessel functions into Meijer-G functions. A definition of these special functions is given in Appendix \ref{sec:meijerg}. For that purpose we use the identity \cite{BesselK_To_MeijerG} 
\begin{align}\label{eq:KK_To_MeijerG}
    G^{4,0}_{2,4}\lp\hspace{-0pt}
\begin{array}{c}
         a, a + \frac{1}{2} \\
         b,c , 2a-c, 2a-b
\end{array} 
    \Bigg| z^2\rp &= \frac{2 z^{2a}}{\sqrt{\pi}}K_{b-c}(z)K_{b+c-2a}(z) .
\end{align}
The constants $a$, $b$, and $c$ in this equation can be chosen conveniently as long as they satisfy the indices on the right hand side.  

We now write the Meijer-G functions in terms of their definition as contour integrals, given in Eq.\ \eqref{eq:MeijerG_Def}. Then we can swap the integrations, and compute the $r'$-integral by trigonometric substitution 
\begin{align}\label{eq:SemiA_to_A}
     \int\limits_{0}^t \frac{z^{n+2s}}{\sqrt{t^2 - z^2}}dz &= \frac{\sqrt{\pi}}{2}t^{n+2s} \frac{\Gamma(\frac{n+1}{2}+s)}{\Gamma(1+\frac{n}{2}+s)}   \, .
 \end{align}
Here, 
$t = 2\tau$, $n$ is an integer and $s$ is the contour integration variable from \eqref{eq:MeijerG_Def}. The terms appearing inside the remaining contour integrals define again Meijer-G functions.
The net effect of the $r'$-integral is thus to effectively replace $mr' \to 2 m\tau$ and --- due to the presence of additional $\Gamma$-functions in the Eq.\ (\ref{eq:SemiA_to_A}) --- changed parameters of the Meijer-G functions.

We are now in a position to present analytic results in closed form for the weak field stress tensor in the MV-model. The leading terms in the gradient expansion are
\begin{subequations}\label{eq:MV_Results}
\begin{align}\label{eq:epsilonL_MV}
    \epsilon_{L,1,\text{MV}}(\tau) &= \frac{N_c}{N_c\hspace{0pt}^2 - 1}\frac{g^6}{128\pi^2 m\tau}\ls 4G^{3,1}_{2,4}\lp 
    \begin{array}{c}
         1,1 \\
         \frac{1}{2}, \frac{1}{2}, \frac{1}{2}, \frac{1}{2}
    \end{array} \Bigg| 4m^2\tau^2\rp - 4G^{3,1}_{2,4}\lp 
    \begin{array}{c}
         1,1 \\
         \frac{1}{2}, \frac{1}{2}, \frac{3}{2}, \frac{1}{2}
    \end{array} \Bigg| 4m^2\tau^2\rp  \rb \\\nn\displaybreak[2]
    &+\lb G^{3,1}_{2,4}\lp
    \begin{array}{c}
         1,2 \\
         \frac{1}{2}, \frac{3}{2}, \frac{5}{2}, \frac{1}{2}
    \end{array} \Bigg| 4m^2\tau^2\rp \rs , \\\label{eq:epsilonT_MV}
    \epsilon_{T,1,\text{MV}}(\tau) &= \frac{N_c}{N_c\hspace{0pt}^2 - 1}\frac{g^6}{128\pi^2 m\tau}\ls 4G^{3,1}_{2,4}\lp 
    \begin{array}{c}
         1,1 \\
         \frac{1}{2}, \frac{1}{2}, \frac{1}{2}, \frac{1}{2}
    \end{array} \Bigg| 4m^2\tau^2\rp - 4G^{3,1}_{2,4}\lp 
    \begin{array}{c}
         1,1 \\
         \frac{1}{2}, \frac{1}{2}, \frac{3}{2}, \frac{1}{2}
    \end{array} \Bigg| 4m^2\tau^2\rp  \rb \\\nn
    &+ G^{3,1}_{2,4}\lp
    \begin{array}{c}
         1,2 \\
         \frac{1}{2}, \frac{3}{2}, \frac{5}{2}, \frac{1}{2}
    \end{array} \Bigg| 4m^2\tau^2\rp - 8 G^{3,1}_{2,4}\lp
    \begin{array}{c}
         0,1 \\
         \frac{1}{2}, \frac{1}{2}, \frac{1}{2}, -\frac{1}{2}
    \end{array} \Bigg| 4m^2\tau^2\rp + 8 G^{3,1}_{2,4}\lp
    \begin{array}{c}
         0,1 \\
         \frac{1}{2}, \frac{1}{2}, \frac{3}{2}, -\frac{1}{2}
    \end{array} \Bigg| 4m^2\tau^2\rp \\\nn
    &- 2\lb G^{3,1}_{2,4}\lp
    \begin{array}{c}
         0,2 \\
         \frac{1}{2}, \frac{3}{2}, \frac{5}{2}, -\frac{1}{2}
    \end{array} \Bigg| 4m^2\tau^2\rp\rs , \\
    S_{i,1,\text{MV}}(\tau) &= \frac{1}{m}\frac{N_c}{N_c\hspace{0pt}^2 - 1}\frac{g^6}{256\pi^2} \ls G^{3,1}_{2,4}\lp 
    \begin{array}{c}
         1,2 \\
         \frac{1}{2}, \frac{3}{2}, \frac{5}{2}, -\frac{1}{2}
    \end{array} \Bigg| 4m^2\tau^2\rp + 4 G^{3,1}_{2,4}\lp 
    \begin{array}{c}
         1, 1 \\
         \frac{1}{2}, \frac{1}{2}, \frac{1}{2}, -\frac{1}{2}
    \end{array} \Bigg| 4m^2\tau^2\rp  \rb \\\nn
    &-4\lb G^{3,1}_{2,4}\lp
    \begin{array}{c}
         1,1 \\
         \frac{1}{2}, \frac{1}{2}, \frac{3}{2}, -\frac{1}{2}
    \end{array} \Bigg| 4m^2\tau^2\rp \rs  ,\\
    S_{z,1,\text{MV}}(\tau) &= \frac{1}{m^2}\frac{N_c}{N_c\hspace{0pt}^2 - 1}\frac{g^6 }{512\pi^2 m\tau} \ls G^{3,1}_{2,4}\lp 
    \begin{array}{c}
         1,2 \\
         \frac{3}{2}, \frac{3}{2}, \frac{5}{2}, -\frac{1}{2}
    \end{array} \Bigg| 4m^2\tau^2\rp - 2 G^{3,1}_{2,4}\lp 
    \begin{array}{c}
         1, 2 \\
         \frac{3}{2}, \frac{3}{2}, \frac{3}{2}, -\frac{1}{2}
    \end{array} \Bigg| 4m^2\tau^2\rp  \rs\\
    T_{iz,1,\text{MV}}(\tau) &= \frac{1}{m} \frac{N_c}{N_c\hspace{0pt}^2 - 1}\frac{g^6}{256\pi^2m^2 \tau^2} \ls 2 G^{3,1}_{2,4}\lp 
    \begin{array}{c}
         1,2 \\
         \frac{3}{2}, \frac{3}{2}, \frac{3}{2}, \frac{1}{2}
    \end{array} \Bigg| 4m^2\tau^2\rp - G^{3,1}_{2,4}\lp 
    \begin{array}{c}
         1, 2 \\
         \frac{3}{2}, \frac{3}{2}, \frac{5}{2}, \frac{1}{2}
    \end{array} \Bigg| 4m^2\tau^2\rp  \rs\\
    \label{eq:omega_MV}
    \omega_{1,\text{MV}}(\tau) &= T_{xy,1,\text{MV}}(\tau) = \frac{1}{m^2}\frac{N_c}{N_c\hspace{0pt}^2 - 1}\frac{g^6 m \tau}{512\pi^2} \ls 4G^{3,1}_{2,4}\lp 
    \begin{array}{c}
         1,1 \\
         \frac{1}{2}, \frac{1}{2}, \frac{1}{2}, \frac{1}{2}
    \end{array} \Bigg| 4m^2\tau^2\rp - 4 G^{3,1}_{2,4}\lp 
    \begin{array}{c}
         1, 1 \\
         \frac{1}{2}, \frac{1}{2}, \frac{3}{2}, \frac{1}{2}
    \end{array} \Bigg| 4m^2\tau^2\rp \rb \\\nn
    &+ \lb G^{3,1}_{2,4}\lp
    \begin{array}{c}
         1,2 \\
         \frac{1}{2}, \frac{3}{2}, \frac{5}{2}, \frac{1}{2}
    \end{array} \Bigg| 4m^2\tau^2\rp + 4 G^{3,1}_{2,4}\lp
    \begin{array}{c}
         0,1 \\
         \frac{1}{2}, \frac{1}{2}, \frac{3}{2}, -\frac{1}{2}
    \end{array} \Bigg| 4m^2\tau^2\rp - G^{3,1}_{2,4}\lp
    \begin{array}{c}
         0,2 \\
         \frac{1}{2}, \frac{3}{2}, \frac{5}{2}, -\frac{1}{2}
    \end{array} \Bigg| 4m^2\tau^2\rp\rs \,  .
    %
 %   \label{eq:txy_MV}
%     T_{xy,1,\text{MV}}(\tau) &= \frac{1}{m^2}\frac{N_c}{N_c\hspace{0pt}^2 - 1}\frac{g^6 m \tau}{512\pi^2} \ls 4G^{3,1}_{2,4}\lp 
%    \begin{array}{c}
%         1,1 \\
  %       \frac{1}{2}, \frac{1}{2}, \frac{1}{2}, \frac{1}{2}
  %  \end{array} \Bigg| 4m^2\tau^2\rp - 4 G^{3,1}_{2,4}\lp 
  %  \begin{array}{c}
   %      1, 1 \\
    %     \frac{1}{2}, \frac{1}{2}, \frac{3}{2}, \frac{1}{2}
   % \end{array} \Bigg| 4m^2\tau^2\rp \rb \\\nn
  %  &+ \lb G^{3,1}_{2,4}\lp
  %  \begin{array}{c}
   %      1,2 \\
   %      \frac{1}{2}, \frac{3}{2}, \frac{5}{2}, \frac{1}{2}
   % \end{array} \Bigg| 4m^2\tau^2\rp + 4 G^{3,1}_{2,4}\lp
   % \begin{array}{c}
    %     0,1 \\
     %    \frac{1}{2}, \frac{1}{2}, \frac{3}{2}, -\frac{1}{2}
   % \end{array} \Bigg| 4m^2\tau^2\rp - G^{3,1}_{2,4}\lp
    %\begin{array}{c}
     %    0,2 \\
      %   \frac{1}{2}, \frac{3}{2}, \frac{5}{2}, -\frac{1}{2}
    %\end{array} \Bigg| 4m^2\tau^2\rp\rs \, .
\end{align}
\end{subequations}

These solutions are unregularized in the UV-limit and thus exhibit logarithmic singularites at $\tau = 0$. This behavior of the MV model has been known for a while \cite{Lappi:2006hq}. %because it was implicitly assumed our models were UV safe in \eqref{eq:momentum_1} and \eqref{eq:momentum_2}. 
The single-event solution to the Yang-Mills equations itself is well-behaved; the problem arises when averaging over charge distributions $\rho$ without a UV cutoff. This allows arbitrarily large variations between color charges on arbitrarily small length scales, as encoded in Eq.\ \eqref{eq:MV}. This in turn can lead to arbitrarily large field strength.
We have constructed the improved Gaussian model with this problem in mind. However, it is worthwhile to briefly discuss how the MV-model can be UV-regulated a posteriori. The most 
natural procedure appears to be to cut off the momentum integrals in 
Eqs.\ (\ref{eq:momentum_1}) and (\ref{eq:momentum_2}). This approach, and the connection to existing solutions for the MV-model using an expansion in time $\tau$, in particular in Ref.\ \cite{Chen:2015wia}, are briefly discussed in Appendix \ref{sec:app:2015_Match}.

\begin{figure}[tb]
 \begin{center}
    \includegraphics[width=0.45\textwidth]{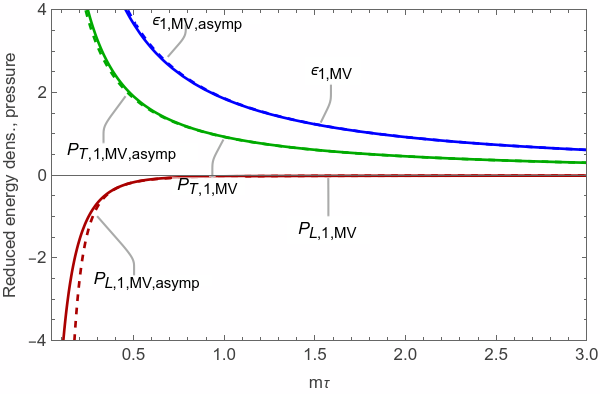}
    \includegraphics[width=0.45\textwidth]{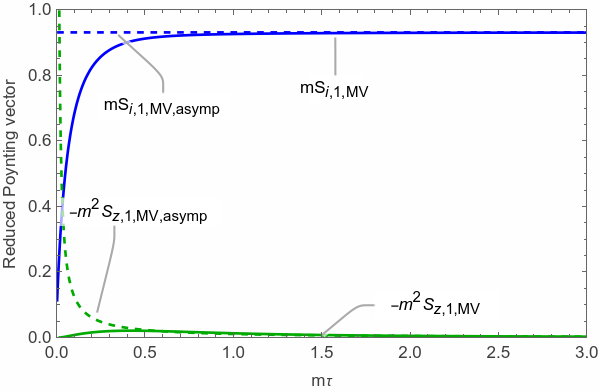}
 \end{center}
 \begin{center}
    \includegraphics[width=0.45\textwidth]{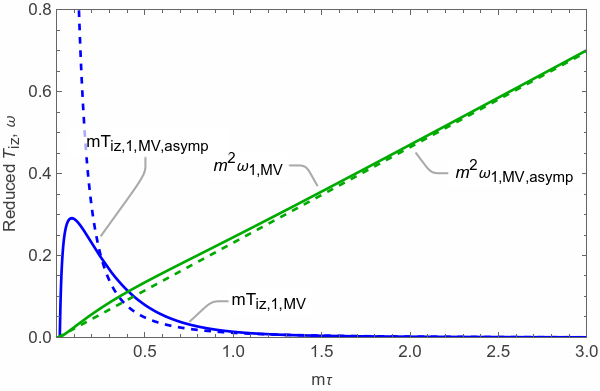}
 \end{center}
  \caption{\label{fig:mvplot1} The reduced leading order coefficients in the gradient expansion for energy density, longitudinal pressure and transverse pressure in the MV Model,  $\epsilon_{1,\text{MV}}$, $P_{L,1,\text{MV}}$, and  $P_{T,1,\text{MV}}$, resp. Reduced here means that the quantities are dimensionless and 
  a common factor $N_c g^6/(16\pi^2 (N_c^2-1))$ has been factored off (upper left panel). The reduced transverse and longitudinal Poynting vectors $m S^i_{1,\text{MV}}$ and $m^2 S^z_{1,\text{MV}}$ are plotted in the same way. In addition to the factor of color and coupling constants, factors of $m$ have been introduced to make the quantities dimensionless (upper right panel). The same is shown for $mT^{iz}_{1,\text{MV}}$ and $m^2 \omega_{1,\text{MV}}$ (lower panel). In all panels solid lines give the full results, dashed lines represent the asymptotic behavior.}
\end{figure}

The closed-form solutions to the MV model give us a clear picture of the large-time behavior of the gluon energy momentum tensor. The prefactor in the solutions above, containing a power of time $\tau$, is chosen such that the remaining factor with a linear combination of Meijer G-functions approaches a constant value for $\tau\to\infty$. We discuss in Appendix \ref{sec:meijerg} how the factoring of powers of $\tau$ from Meijer-G functions works, and how the constant asymptotic value of the remaining functions can be calculated. In short, the $\tau$-dependence of the prefactor in Eq. \eqref{eq:MV_Results} is constructed to explicitly show the asymptotic behavior for large times. We summarize the asymptotic behavior here:\comment{SR: I fixed the asymptotic Sz already.}
\begin{subequations}
\begin{align}
    \epsilon_{L,1,\text{MV},\text{asymp}}  &= \epsilon_{T,1,\text{MV},\text{asymp}} = \frac{N_c}{N_c\hspace{0pt}^2 - 1}\frac{g^6}{16\pi^2}    \frac{19 \pi}{64 m\tau}   & S_{i,1,\text{MV},\text{asymp}}  &= \frac{N_c}{N_c\hspace{0pt}^2 - 1}\frac{g^6}{16\pi^2}\frac{19 \pi}{64 m}    \\
    S_{z,1,\text{MV},\text{asymp}}  &= -\frac{N_c}{N_c\hspace{0pt}^2 - 1}\frac{g^6}{16\pi^2}    \frac{\pi}{256 m^3\tau}    & T_{iz,1,\text{MV},\text{asymp}}  &= \frac{N_c}{N_c\hspace{0pt}^2 - 1}\frac{g^6}{16\pi^2}\frac{\pi}{256 m^3 \tau^2}    \\
    \omega_{1,\text{MV},\text{asymp}}  &= T_{xy,1,\text{MV},\text{asymp}} = \frac{N_c}{N_c\hspace{0pt}^2 - 1}\frac{g^6}{16\pi^2}    \frac{19 \pi \tau}{256 m}   \, .
\end{align}
\end{subequations}

For the longitudinal pressure, the two leading terms from $\epsilon_L$ and $\epsilon_T$ cancel. For consistency it is thus advisable to analyze further terms which leads to the result that the longitudinal pressure vanishes with the cubic power of $1/\tau$, more precisely
\begin{equation}
     P_{L,1,\text{MV},\text{asymp}} = - \frac{N_c}{N_c\hspace{0pt}^2 - 1}\frac{g^6}{16\pi^2}  \frac{13 \pi}{2048 m^3 \tau^3}    \, .
\end{equation}

Fig.\ \ref{fig:mvplot1} shows the leading terms in the gradient expansion for various components of the energy momentum tensor in the MV-model as functions of time. The reduced quantities that are shown are obtained by removing a common factor $N_c g^6/(16\pi^2 (N_c^2-1))$ and by adding factors of the scale $m$ until the quantity is dimensionless. The resulting reduced quantities are independent of the transverse coordinates, and they are independent of both the coupling constant and the overall source strength $\mu_1\mu_2$. The various panels in Fig.\ \ref{fig:mvplot1}  show energy density and pressures, components of the Poynting vector, as well as the stress components $T^{iz}$ and the transverse pressure anisotropy $\omega$. For all quantities, the asymptotic expansion for large times is also shown.

Energy density and pressures display their known behaviors, with longitudinal pressure trying to reach zero from below fast. The transverse Poynting vector exhibits a sharp rise, initiated by the large transverse pressure, and then goes to a constant value, meaning that the ratio $S_i/\epsilon$ rises linearly for large times. On the other hand, the conjugated values $T_{iz}$ for longitudinal momentum flow, build up but then relax to zero quickly such that
$T_{iz}/\epsilon \sim 1/\tau$. Interestingly, the longitudinal Poynting vector decays to zero but at a rate such that
$S_z/\epsilon \sim \text{const.} $ Perhaps the most surprising result is the ``quadrupole" term $\omega$ which grows linearly, even at large times. In the central region of a nuclear collision with finite impact parameter, where $\nabla^x \mu_1 \nabla^x \mu_2 - \nabla^y \mu_1 \nabla^y \mu_2$ tends to be negative, see Eq.\ (\ref{eq:omegadef}),  the pressure $P_x$ would be driven to large negative values. The pressure asymmetry $(P_x-P_y)/(P_x+P_y)$ is consequently negative and growing as $\sim \tau^2$, trying to increase any initial spatial eccentricity of the system. It would be interesting to discuss analytically how non-abelian interactions can modify this signature behavior.
The asymptotic terms describe the full analytic results surprisingly well for almost all times $m\tau >1$. The agreement is most pronounced for the energy density and pressures, where the results are indistinguishable from the simple asymptotic behavior even down to $m\tau=0.3$.

\subsection{Improved Gaussian Model Stress Tensor}\label{sec:StressTensor_ssec:ImprovedGaussianModel}

For the improved Gaussian model, we were not able to find a completely closed expression for the last integral over $r'$. That integral must either be taken numerically,  or we have to employ a series expansion in $r'$ or $\tau$.  The latter method can provide some key insights and will be discussed next.
As an example, we treat here the function $f_1$.
We find
\begin{align}\nn
    f_{1,\text{iG}}(r) &= \frac{g^6}{16\pi^2}\frac{N_c}{N_c^2-1}\lp E_1\lp\frac{r^2}{4(B_c+B_q)}\rp - E_1\lp\frac{r^2}{4B_q}\rp \rp^2\\\label{eq:f1}
    &= \frac{g^6}{16\pi^2}\frac{N_c}{N_c^2-1}\sum\limits_{j=0}^\infty r^{2j}a_j \,\text{,} \hspace{1cm}  a_j = \sum\limits_{i=0}^j b_i b_{j-i}
    \,\text{,} 
    & b_i &= \left\{ \begin{array}{lr}
        \ln\ls\frac{B_q}{B_q + B_{c}}\rs & i = 0 \\
        \frac{(-4)^{-i}}{i!i}\lp (B_q + B_{c})^{-i} - B_q\hspace{0pt}^{-i} \rp & i \geq 1
    \end{array}\right.   .
\end{align}
It is evident %in \eqref{eq:f1} 
that the dependence on the (arbitrary) infrared cutoff $m$ has dropped out of the expression. Instead, the (physical) confinement length scale $\sqrt{B_c}$ remains as the largest length scale.
We have checked that the functions $f_2$ and $\phi_1$ are also free of $m$. This is not necessarily true for further terms needed for higher orders in the gradient expansion considered in Appendix \ref{sec:app:emtensor}. We can conclude that the global color neutrality imposed in this model makes it indeed infrared safe, as long as one does not go too far in the gradient expansion. 

The UV behavior is more subtle. The exponential integral functions $E_1$ in the expression for $f_{1,\text{iG}}$ behave logarithmically at small argument. It is the subtraction of two of them which removes the logarithmic UV singularity. Indeed, it can be shown that all terms for the energy momentum tensor considered in \eqref{eq:gradient_Expansion}  are UV-safe in the improved Gaussian model. By inspection, the $f_i$ and $\phi_j$ all behave as simple polynomials at small argument, making the integrals \eqref{eq:GlasmaStressTensor} finite at early times. Qualitatively, this can be understood as a screening of color charge by introducing a finite quark size $l_{\text{UV}}$ in \eqref{eq:QuarkDist}. However, we can not guarantee that all orders in the gradient expansion will be UV-safe as the introduction of $l_{\text{conf}}$ does not completely replace $m$ in $\epsilon_{L,3,\text{iG}}$, for example. A model that honors local color neutrality might be required to remove all singularities.

In principle, all components of the energy momentum tensor in the improved Gaussian model, corresponding to the MV results in Eq.\ \eqref{eq:MV_Results}, could now be written down as power series in $\tau$, since all integrands \eqref{eq:gFunctions} are known explicitly. However, the series expansions of $f_i$ and $\phi_i$ are not concise, with the exception of $f_1$. For that reason, we restrict ourselves to quoting the components that depend on $f_1$ only. They determine the energy density, pressures and transverse Poynting vectors to leading order in gradients,
\begin{subequations}
\begin{align}
\epsilon_{L,1,\text{iG}}(\tau) &= \frac{g^6}{32\pi^2}\frac{N_c}{N_c\hspace{0pt}^2 - 1}\sum\limits_{n=0}^\infty\frac{(2n)!}{n!\hspace{1pt}n!}a_n\tau^{2n}, & S_{i,1,\text{iG}}(\tau) &= \frac{g^6}{64\pi^2}\frac{N_c}{N_c\hspace{0pt}^2 - 1}\sum\limits_{n=0}^\infty\frac{(2n)!}{(n+1)!\hspace{1pt}n!}a_n\tau^{2n+1},\\
\epsilon_{T,1,\text{iG}}(\tau) &= -\frac{g^6}{32\pi^2}\frac{N_c}{N_c\hspace{0pt}^2 - 1}\sum\limits_{n=0}^\infty\frac{(2n)!\hspace{1pt}n}{(n+1)!\hspace{1pt}n!}a_n\tau^{2n}.
\end{align}
\end{subequations}

\begin{figure}[tb]\centering
    \includegraphics[width=0.45\textwidth]{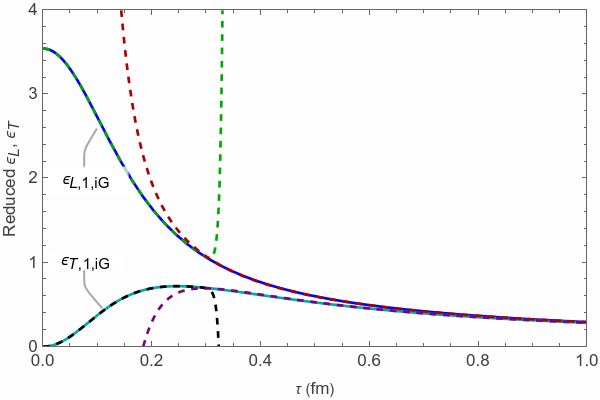}
    \caption{The reduced longitudinal and transverse energy densities $\epsilon_{L,1,\text{iG}}$ and $\epsilon_{T,1,\text{iG}}$ in the improved Gaussian model (solid lines). The small and large time expansions are also shown (dashed lines), with 35 and three terms used, respectively. Both expansions agree around $\tau = 0.3$ fm. }\label{fig:iGSeriesComparison}
\end{figure}

These can be derived using \eqref{eq:SemiA_to_A} once more. After taking the series expansion in $r'$,  $t = 2\tau$, $s=0$, and $n$ is related to the order of the expansion. The coefficients $a_n$ can be computed from Eq.\ (\ref{eq:f1}). Unlike the small-time series expansions in Ref.\ \cite{Chen:2015wia}, these series is quite well behaved and can be summed to high orders in a numerically reliable way, as can be seen in Fig.\ \ref{fig:iGSeriesComparison}.

We will now argue that the asymptotic behavior of each component of the energy momentum tensor is universal, i.e.\ the qualitative behavior is independent of the model chosen for the correlation function of color charges. To make the statement more precise, the leading powers of $\tau$ for large values of time are always the same as those found for the MV-model, although their coefficients are model dependent. To see that, one can expand Eqs.\ (\ref{eq:gradient_Expansion}) around $\tau\to\infty$. For example, for the energy densities we take a binomial expansion of the common factor $1/\sqrt{4\tau^2 - r'\hspace{1pt}^2}$ as $1/2\tau + \cdots$ and extend the integral to infinity. %This procedure works generally and is equivalent to using the asymptotic behaviour of the Meijer-G functions in the MV model. 
Likewise, the factor $1/\sqrt{4\tau^2 - r'\hspace{1pt}^2}$ can be expanded for small times.
 In either case, orders of $\tau$ decouple from orders of $r'$. 
Therefore, the orders of time are universal and models only differ in their coefficients.
Tab.\ \ref{tb:LargeTimeDependence} summarizes the leading behaviour in each regime of every term in the $\mu_i$ gradient expansion for every component of the stress tensor in the gradient expansion. The divergent behavior at large times for quantities with many derivatives in their transverse spatial dependence is indicative of the gradient expansion in $\mu$ and limits its usefulness at large times.
\begin{table}[H]
\centering
\subfloat[Small time behavior]{\begin{tabular}{|c||c c c c c||}\hline
$Q$& $Q_1(\tau)$& $Q_2(\tau)$ & $Q_3(\tau)$ & $Q_4(\tau)$ & $Q_5(\tau)$ \\ \hline
$S_z$		 & $\tau^2$ & - & - & - & - \\
$S_i$		 & $\tau$ & $\tau$ & $\tau^{3}$ & - & - \\
$T_{iz}$	 & $\tau$ & $\tau$ & $\tau^3$ & - & - \\
$\epsilon_L$ & 1 & $\tau^2$ & 1 & 1 & $\tau^{4}$ \\
$\epsilon_T$ & $\tau^2$ & $\tau^2$ & $\tau^{2}$ & $\tau^{2}$ & $\tau^{4}$ \\
$\omega$	 & $\tau^2$ & $\tau^{2}$ & $\tau^{2}$ & - & - \\
$T_{xy}$	 & $\tau^2$ & $\tau^{2}$ & $\tau^{2}$ & - & - \\
\hline
\end{tabular}}
\subfloat[Large time behavior]{\begin{tabular}{c c c c c|}\hline
$Q_1(\tau)$& $Q_2(\tau)$ & $Q_3(\tau)$ & $Q_4(\tau)$ & $Q_5(\tau)$ \\ \hline
$\tau^{-1}$ & - & - & - & - \\
1 & $\tau^{2}$ & $\tau^{2}$ & - & - \\
$\tau^{-2}$ & 1 & 1 & - & - \\
$\tau^{-1}$ & $\tau$ & $\tau$ & $\tau^{3}$ & $\tau^{3}$ \\
$\tau^{-1}$ & $\tau$ & $\tau$ & $\tau^{3}$ & $\tau^{3}$ \\
$\tau$ & $\tau$ & $\tau^{3}$ & - & - \\
$\tau$ & $\tau$ & $\tau^{3}$ & - & - \\
\hline
\end{tabular}}\caption{Asymptotic time behavior of each component of the stress tensor, generically called $Q_i$, where the index $i$ refers to the order of the gradient expansion.}\label{tb:LargeTimeDependence}
\end{table}

To find the asymptotic behavior specifically for the improved Gaussian model, the exponential integrals are replaced by Meijer-G functions \cite{ExpInt_To_MeijerG}. Writing one of them in their contour integral form allows integration with a known identity \cite{MeijerG_Integral}. For the components that depend on $f_{1,\text{iG}}$ the series are
\begin{subequations}
\begin{align}
    \epsilon_{L,1,\text{iG},\text{asymp}}  &= \frac{g^6}{16\pi^2}\frac{N_c}{N_c^2-1}\frac{1}{2\pi\tau}\sum\limits_{n=0}^{\infty}\frac{(2n)!}{n!\hspace{1pt}n!}\frac{c_n}{4^n}\tau^{-2n}, & S_{i,1,\text{iG},\text{asymp}}  &= -\frac{g^6}{16\pi^2}\frac{N_c}{N_c^2-1}\frac{1}{\pi}\sum\limits_{n=0}^{\infty}\frac{(2n-2)!n}{n!\hspace{1pt}n!}\frac{c_n}{4^n}\tau^{-2n}, \\
    \epsilon_{T,1,\text{iG},\text{asymp}}  &= -\frac{g^6}{16\pi^2}\frac{N_c}{N_c^2-1}\frac{1}{\pi\tau}\sum\limits_{n=0}^{\infty}\frac{(2n-2)!n(2n+1)}{n!\hspace{1pt}n!}\frac{c_n}{4^n}\tau^{-2n} \hspace{-2cm}
\end{align}
\end{subequations}
with coefficients
\begin{align}
    c_n &= \ls (B_q + B_c)^{n+1/2} + B_q^{n+1/2} \rs  G^{2,2}_{3,3}\lp
    \begin{array}{c}
         \frac{1}{2}- n, \frac{1}{2} - n, 1 \\
         0, 0, -\frac{1}{2} - n
    \end{array} \Bigg| 1\rp - 2B_q^{n+1/2} G^{2,2}_{3,3}\lp
    \begin{array}{c}
         \frac{1}{2}- n, \frac{1}{2} - n, 1 \\
         0, 0, -\frac{1}{2} - n
    \end{array} \Bigg| \frac{B_q}{B_q+B_c}\rp  \, .
\end{align}
Note that $(2n-2)!n$ evaluated at $n=0$ has a finite limit of $-1/2$. Indeed, the powers of the leading terms in $\tau$, written outside the sums, coincide with those of the asymptotic expressions found in the MV model.

Much like the early time expansion series, the asympotic series for the improved Gaussian model is also numerically stable. Indeed, the radii of convergence are large enough for both expansions to overlap. They agree starting around $\tau\approx 0.3$ fm for physical values of $B_c$ and $B_q$. A complete, analytic description could then be defined piecewise with relatively few terms using both series for each component. This is showcased in Fig. \ref{fig:iGSeriesComparison}. The complete numerical evaluation of $\epsilon_{L,1}$ and $\epsilon_{T,1}$, using Eq.\ (\ref{eq:gradient_Expansion}), are plotted, together with the analytic results for the small-time expansion using 35 terms, and the analytic result for the large-time expansion using the first three terms. Both expansions agree with an accuracy of 0.1\% at $\tau = 0.3$ fm. Pushing to larger orders on both sides will obviously increase the region of overlap.

\begin{figure}[tb]
 \begin{center}
    \includegraphics[width=0.47\textwidth]{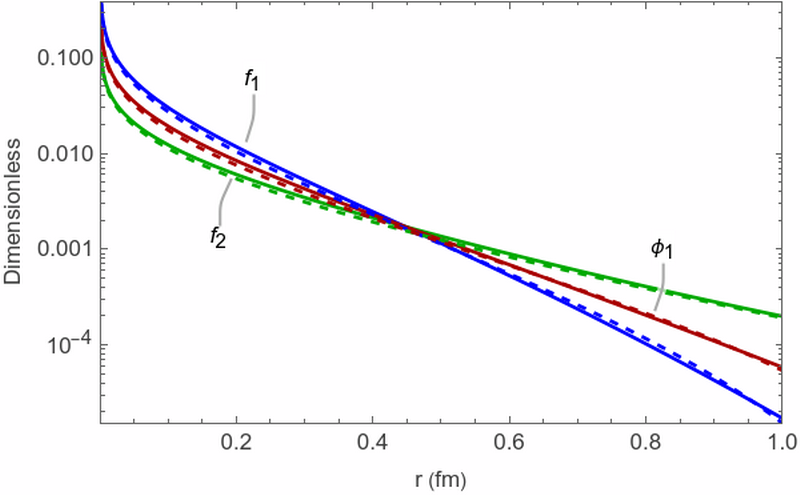}
    \includegraphics[width=0.45\textwidth]{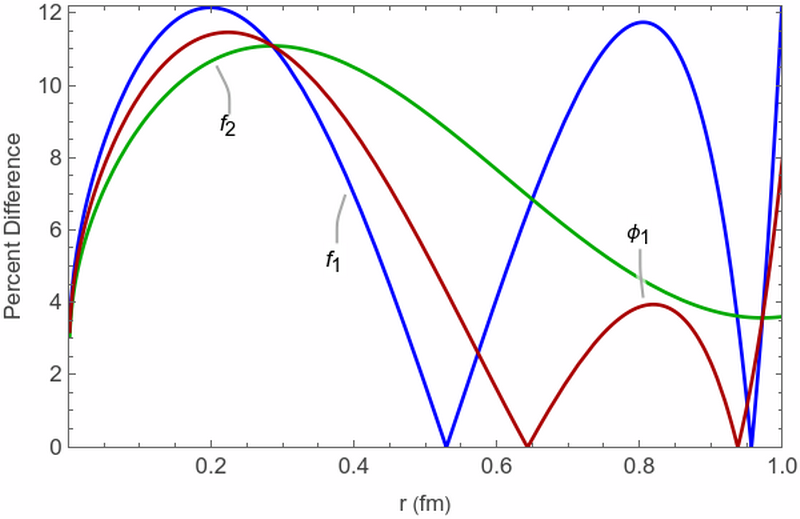}
 \end{center}
  \caption{\label{fig:MV_Comparison} Left panel: The UV-limit of the functions $f_1$, $f_2$, and $\phi_1$ in the improved Gaussian model (solid lines) and the same functions in the MV model (dashed lines). Parameter values $N_c = 3$, $g = 1$, $m = 1.28$ fm$^{-1}$, $B_q = 3\times 10^{-6}$ GeV$^{-2}$, and $B_c = 4$ GeV$^{-2}$ are used. Right panel: Corresponding absolute percent differences for the three functions in the two models.}
\end{figure}

%Feel free to re-write this. I'm just getting something down
One characteristic of the improved Gaussian model we have yet to discuss is its qualitative agreement with the MV-model at leading order gradients.
The presence of the UV-cutoff severely limits the number of available momentum modes in the improved Gaussian model. However, it has a well-defined limit for $B_q\to 0$ in which many of the qualititative characteristics of the MV model can be recovered. For this argument we notice that
\begin{equation}
   f_{1,\text{iG}}(r) \longrightarrow E_1\left( \frac{r^2}{4 B_C}\right)^2
\end{equation}
for $B_q \to 0$. This function quite closely resembles $f_{1,\text{MV}}$ numerically in Eq.\ (\ref{eq:mvf1}), if the IR regulators $m$ and $B_c$ are properly related. 
In fact, in the limit of vanishing $B_q$, the functions $f_1$, $f_2$, and $\phi_1$ can be made to agree in both models numerically to within $\sim$12\% using the relation $m \approx 0.5 B_c^{-1/2}$ as shown in Fig.\ \ref{fig:MV_Comparison}. Close agreement is not enjoyed by the remaining $f_i$ and $\phi_j$ functions. The choice of coefficient between the IR regulators minimizes the difference between the functions $f_1$, $f_2$, and $\phi_1$ across their whole domain. 
A perhaps more useful choice is a slightly smaller value $m \approx 0.46 B_c^{-1/2}$ that improves agreement at small $r'$, which dominates the integral, and sacrifices the difference at large $r'$ where the functions are approaching zero regardless.
We conclude that the improved Gaussian model can be seen as a UV regularization scheme of the MV-model as long as one does not go too far in the transverse gradient expansion.

\begin{figure}[tb]\centering
   % \hspace{1.65in}
    \includegraphics[width=0.45\textwidth]{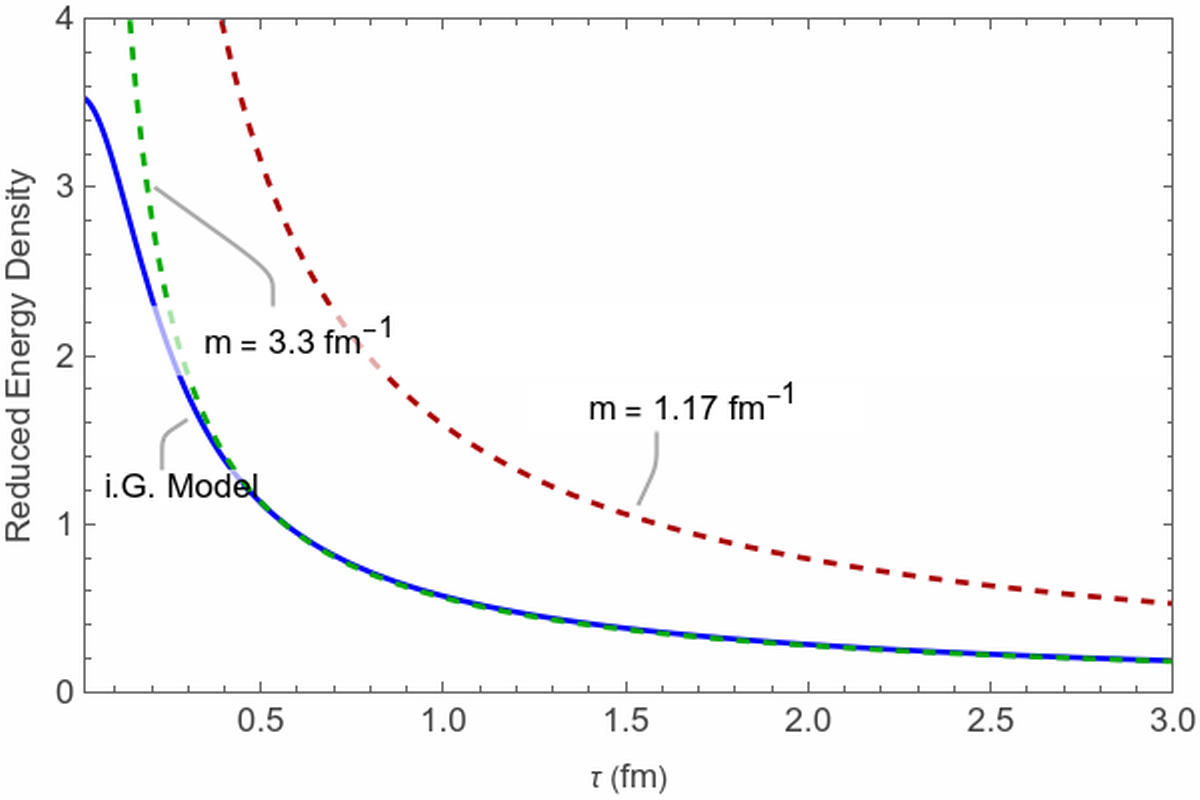}\vspace{-1.53ex}
    \caption{Reduced energy density in both the MV (dashed lines) and improved Gaussian model (solid line). The iG plot utilizes realistic values of $B_q = 0.3$ GeV$^{-2}$ and $B_c = 4$ GeV$^{-2}$. The $m = 1.17$ fm$^{-1}$ curve is the MV model which matches iG in the vanishing $B_q$ limit whereas $m = 3.3$ fm$^{-1}$ is what it would need to be for the models to agree within $\sim$2\% beyond $\tau = 0.5$ fm.}\label{fig:ig_MV_UV}
\end{figure}

It is illuminating to discuss the effects of a realistic value of $B_q$ compared to the UV-limit.  
We choose as an example the reduced energy density, shown in Fig. \ref{fig:ig_MV_UV}.
Combining a physical value of $B_q$ but keeping $m \approx 0.46 B_c^{-1/2}$, regulates the singularity at early times. In addition, the entirety of the time evolution is diminished compared to $B_q=0$ which should be qualitatively expected from energy conservation. 
%demonstrates this is not the case. While the early times are indeed regulated, the values ofe 
It is possible for the MV model to still reproduce the physical behaviour at large times, but a different relation $m \approx 1.3 B_c^{-1/2}$ should be used.

\begin{figure}[tb]\centering
%    \hspace{1.65in}
    \includegraphics[width=0.45\textwidth]{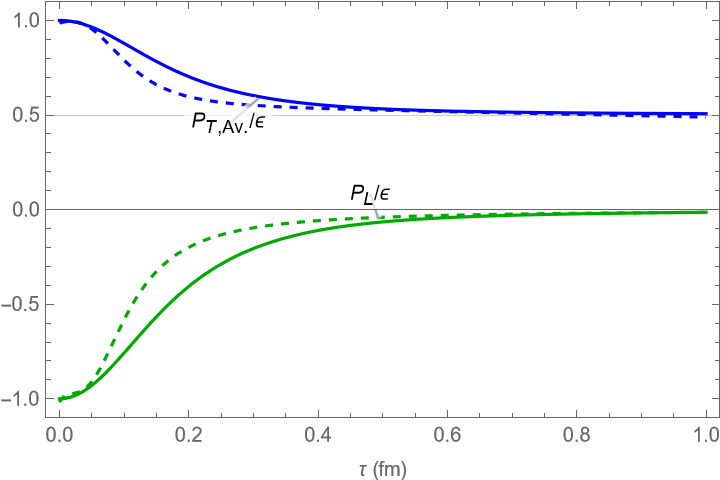}
    \includegraphics[width=0.45\textwidth]{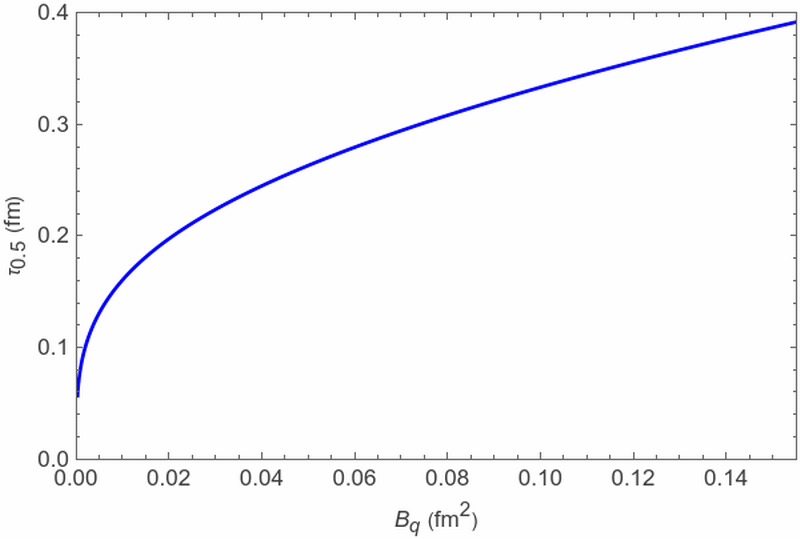}\vspace{-1.53ex}
    \caption{Left panel:Longitudinal pressure (green) and average transverse (blue) pressure normalized by the energy density for the IP-Glasma model (dashed) and the improved Gaussian model (solid lines). Right panel: Relaxation time for $P_L/\epsilon = -1/2$ as a function of $B_q$. It is found to grow approximately at $B_q^{\,\,0.4}$.}\label{fig:Pressure}
\end{figure}

Since purely numerical solutions to the MV model on discretized space-time grids exist, it is interesting to compare to our solutions. Note that numerical codes automatically impose IR and UV cutoffs through a finite grid constant and finite box size, but global color neutrality is typically not enforced. Nevertheless, we compare to our improved Gaussian model as the most physical option at our disposal.
We run the publicly available IP-Glasma code \cite{ipglasma}.
To mimic the abelian limit and to avoid complications due to gradients
we choose very small coupling constant and small constant source strengths $\mu_1=\mu_2$. In Fig.\ \ref{fig:Pressure} we plot the time dependence of the ratio of transverse and longitudinal pressure, averaged over the transverse plane, to the averaged energy density. 
These ratios are independent of $g$ and $\mu_i$ in the abelian limit.
We compare these quantities
for abelian IP-Glasma compare to our improved Gaussian results with physical values of the cutoffs. These ratios are relatively straight forward to compare,
while absolute quantities would once more require careful matching of cutoffs.
We see that the analytic and numerical results match initially and asymptotically, but display slightly different time evolutions, with the IP-Glasma model relaxing slightly more quickly to the asymptotic value. 
We quantify this relaxation rate using the time for $P_L$ to reach $-\frac{1}{2}\epsilon$ as a benchmark in the iG-model. This benchmark time is plotted numerically as a function of $B_q$ in the physical range $0\leq B_q \leq B_c = 4$ GeV$^{-2}$. A larger UV-cutoff momentum scale $\sim B_q^{-1/2}$ is found to speed up the time evolution. %This suggests that the effective UV length scale implemented in IP-Glasma is smaller than the  physical values of $B_q$.
Further analysis would require a more profound inspection of the relation between scales involved. Such studies could lead to test procedures which allow the vetting of numerical code against analytic results in the abelian limit.

\section{Angular Momentum}\label{sec:AngularMomentum}

The energy momentum tensors we have discussed can be used to easily compute the angular momentum tensors by defining a reference point $r^\mu$. 
Angular momentum in nuclear collisions has been a topic of increased interest since measurementf of the polarization of certain hadrons have become available \cite{STAR:2017ckg}. The focus of theoretical efforts has been mostly on fluid dynamic pictures with non-vanishing orbital motion, and on the mechanisms of how polarized hadrons can emerge from them \cite{Liang:2004ph,Jiang:2016woz, Becattini:2016gvu}. Here, we analyze the implications of our results for the angular momentum of weak gluon fields which offers insights into the early phase of angular momentum dynamics, which has also been discussed in Refs.\ \cite{Fries:2017ina,Carrington:2025xws}.

We choose the reference point to be the center of the collision with the two nuclei offset along the $x$-axis. %Nuclear overlap defines $t=z=0$, the impact parameter provides the $x$ axis, and $x=y=0$ at halfway along the impact vector. 
In this section we are restricting ourselves to symmetric A+A systems and define the $+z$-going nucleus to be centered around $x=b/2$ and the $-z$-going nucleus to be centered around $x=-b/2$ where $b$ is the impact parameter. The angular momentum density is then defined as usual as
\begin{align}
M^{\mu\nu\lambda} &= r^\mu T^{\nu\lambda} - r^\nu T^{\mu\lambda} & r^\mu &= \lp \tau\cosh\eta , x, y, \tau\sinh\eta\rp
\end{align}

The regular angular momentum vector emerges from integrating $M$ over a volume. For realistic spherical nuclei, the average squared charge densities $\mu_1$ and
$\mu_2$ are symmetric under reflections $x\to -x$ and $y\to -y$. 
Following Ref.\ \cite{Fries:2017ina}, a table for the parity of all components of the stress tensors with respect to reflections in both $x$ and $y$  will prove convenient to compute the necessary integrals. It can be ascertained generally from \eqref{eq:ll2}, \eqref{eq:tt2}, \eqref{eq:tl2}, \eqref{eq:gamma}, \eqref{eq:expFFAA}, and \eqref{eq:Uexp} that all terms in the gradient expansion of a particular component of the energy momentum tensor have the same parity. A subtler point in these identifications is that the sign of the impact parameter is flipped when considering $x\rightarrow -x$; swapping the positions of $\mu_1$ and $\mu_2$. Therefore field combinations that contain a factor $\delta^{ij}\epsilon^{kl}$ or $\epsilon^{ij}\delta^{kl}$ pick up an additional minus sign in the $x$ parity to swap $\mu_1$ and $\mu_2$ back to where they started. %In the case of a single nucleus, parity in $x$ is not applicable because they are shifted to one side. 
We show these parities in Tab.\ \ref{tb:ParityTable}.

\begin{table}[H]
\centering
\begin{tabular}{|c||c c c c c c c c c|}\hline
 & $S_z$ & $S_x$ & $S_y$ & $T_{xz}$ & $T_{yz}$ & $\epsilon_L$ & $\epsilon_T$ & $\omega$ & $T_{xy}$  \\ \hline
$x\to -x$ & odd & odd & even & even & odd & even & even & even & odd \\
$y\to -y$ & even & even & odd & even & odd & even & even & even & odd \\
\hline
\end{tabular}
\caption{Parity in the transverse plane for each component of the energy momentum tensor.}\label{tb:ParityTable}
\end{table}

%\subsection{Glasma}\label{sec:AngularMomentum_ssec:Glasma}

The angular momentum is then $L_i = \tfrac{1}{2}\epsilon_{ijk}\int d^3r M^{jk0}$. Because of boost-invariance, only angular momentum \emph{per unit rapidity}
is a meaningful quantity \cite{Fries:2017ina}. Using
Tab.\ \ref{tb:ParityTable} we obtain
\begin{align}\label{eq:GlasmaAngularMomentum1}
\frac{dL_y}{d\eta} &= \tau\cosh\eta\int d^2R \lp \tau\sinh^2\eta\hspace{1pt}T_{xz}(\tau,\mathbf{R}) - x\cosh(2\eta)S_z(\tau,\mathbf{R}) \rp \, , & \frac{dL_x}{d\eta}=\frac{dL_z}{d\eta}&=0 \, .
\end{align}
Here, the energy momentum tensor components are evaluated at $\eta=0$ as in the previous sections. As expected from the initial global angular momentum, only the $y$-component of angular momentum does not vanish. The two terms in $dL_y/d\eta$ can be identified with rapidity-odd directed flow in the $x$-direction, and with shear flow in the $z$-direction, respectively \cite{Fries:2017ina}.

These two terms are related which can be inferred from angular momentum conservation.
The transverse flow of angular momentum has to vanish far away from the collision, and thus the divergence theorem in the transverse plane implies
\begin{align}
0 &= \int d^2R \nabla^i M^{31i} = -\cosh\eta\int d^2R\lp T_{xz}(\tau,\mathbf{R}) + x\nabla^iT_{iz}(\tau,\mathbf{R}) \rp  \, .
\end{align}
The final integral contains three terms corresponding to the three terms of $T_{iz}$ that we consider. Because the time dependence of each of these three terms is different, the spatial integrals must cancel individually. The added divergence means that $O(\nabla^i\mu)$ terms are matched to $O(\nabla^{ij}\mu)$ terms, and $O(\nabla^{ij}\mu)$ terms to $O(\nabla^{ijk}\mu)$ terms, etc. The third order terms are not considered in this paper, so the second order terms that match to them must be considered consistent with zero in this calculation. Therefore there is only one time dependence that contributes, $\int d^2R \lp \mu_1\nabla^x\mu_2 - \mu_2\nabla^x\mu_1 \rp + x\lp \mu_1\triangle\mu_2 - \mu_2\triangle\mu_1 \rp = 0$, and we can simplify \eqref{eq:GlasmaAngularMomentum1} to
\begin{align}\label{eq:GlasmaAngularMomentum2}
\frac{dL_y}{d\eta} &= \tau\cosh\eta\lp \tau\sinh^2\eta\hspace{1pt}T_{xz,1}(\tau) + \cosh(2\eta)S_{z,1}(\tau) \rp\int d^2R \lp \mu_1\nabla^x\mu_2 - \mu_2\nabla^x\mu_1 \rp(\mathbf{R}).
\end{align}

\noindent By cross checking with Tab. \ref{tb:LargeTimeDependence} we can know that this quantity eventually becomes constant. This was conjectured to hold in the full non-abelian case in Ref. \cite{Carrington:2025xws}.

\section{Summary}

In this paper we have calculated two-point functions and the energy momentum tensor for the classical gluon field after the collision of 
two nuclei on the light cone. We present general expressions for the weak field limit that can be evaluated for rather general assumptions regarding the color charge distributions in the nuclei. 
Applying the formalism to the McLerran-Venugopalan model, we find closed analytic expressions for the components of the energy momentum tensor in the weak field limit in terms of special functions. 

We also propose the improved Gaussian model as an alternative. It enforces global color neutrality, is UV-finite to at least second order gradients, and is better behaved in the IR as long as charge densities are smooth. We present  the energy momentum tensor in the iG model both numerically and in terms of series expansions for both large and small times. We also establish that the MV model is qualitatively recovered in the UV-limit of this model.

We have shown that the qualitative asymptotic behavior of weak gluon fields at large times is universal, i.e. independent of the model used for gluon correlation functions. 
The longitudinal pressure goes to zero from below for large times, and the transverse pressure is likewise forced to approach a value of half of the energy density, while all components of the Poynting vector approach constant values.

One consistent key difference is the slower time evolution in the iG model.
It was found that early time evolution is governed by the UV-scale. 
This can also be observed when the deceleration and motion of the colliding sources in the gluon field is studied \cite{deceleration};  
when comparing results from the MV and improved Gaussian models we found comparable results on slower time scales for the latter. 

Our work could be used to develop tests for existing or future glasma event generators in the weak field limit. Interesting scenarios would involve sources which vary weakly with transverse coordinates, e.g.\ $\mu_i = \mu_{i,0} \left( 1+ \alpha_i x + \beta_i y\right)$ where $\alpha_i$ and $\beta_i$ are coefficients small enough such that $\alpha_i x, \beta_i y \lesssim 1$ on the transverse area under consideration. Analytic results, e.g.\ for the ratio $S^i/\epsilon$ could be easily computed analytically and benchmark simulations.

Future improvements to this work could be made by incorporating local color neutrality on the nucleon scale. Of course it would be desirable to extend the ideas here to expand to the non-abelian regime ($U\neq 1$ and/or beyond the weak field limit), or ideally to both. In addition, the non-Gaussian fluctuations in the weight functional $W[\rho]$ as discovered by Lam and Mahlon should be addressed.

{\bf Acknowledgements.} This work was supported by NSF awards PHY-1812431, PHY-2111568, and PHY-2413003. We are grateful for discussions with Tuomas Lappi and Dana Avramescu.

\appendix

\section{Modelling the Charge Area Density}\label{sec:app:IPGlasma}

For meaningful phenomenology, the shape of the charge density $\mu(\mathbf{R}_\perp)$ must be specified further in both the MV and improved Gaussian models. $\mu$  can be related to the saturation scale $Q_s$ \cite{Lappi:2007ku} as 
\begin{align}\label{eq:QtoMu}
\mu &= \frac{N_c^2 - 1}{n^2 g^4}Q_s^2  , 
\end{align}
where $n$ is a dimensionless parameter in the range $0.57\leq n \leq 1.15$. The value 0.8 is used in this paper. 
The saturation scale can be calculated for a particular $x$ using the IP Glasma model as discussed in Ref.\ \cite{Schenke:2012hg}. We follow the outline in that reference. The free parameters are fixed by deep inelastic scattering data from HERA \cite{Rezaeian:2012ji}. This model works on an event-by-event basis so for a given nucleus a random sample of nucleon positions is taken according to a 3D Fermi distribution\footnote{The nucleus is taken to be spherical in all cases. The weak field limit is not accurate enough for nuclear deformation to make a difference.}, fit to the classical radius and skin depth of the nucleus. Length contraction will project these positions into the transverse plane. Three quarks are assigned positions in the transverse plane around each nucleon according to a 2D Gaussian distribution \eqref{eq:QuarkDist}. Time dilation freezes the charge configuration so any virtual quarks should be considered as part of the nucleon. Event averaging makes this point moot however so $N_q$ is left at three. The IP in IP Glasma stands for impact parameter. The impact parameter in this case refers to the distance between a given position and one of the randomly sampled color charges. These positions are fed into the proton thickness function 
\begin{align}\label{eq:ProtonThickness}
T_p(\mathbf{b}_\perp) &= \frac{1}{N_q}\sum\limits_{i=1}^{N_q} T_q(\mathbf{b}_\perp - \mathbf{b}_{\perp,i}) & T_q(\mathbf{b}_\perp) &= \frac{1}{2\pi B_q}\exp\ls-\frac{b^2}{2B_q}\rs .
\end{align}

The model essentially calculates that color charge's contribution to the saturation scale at that position and sums over all charges. A cloud of gluons of a particular density is taken to be in the background around each valance quark. This computation is centered around the function
\begin{align}\label{eq:Dipole}
F(\mathbf{r}_\perp,x,\mathbf{b}_\perp) &= \frac{\pi^2}{2N_c}r^2\alpha_s(\tilde{\mu}^2)xg(x,\tilde{\mu}^2)T_p(\mathbf{b}_\perp).
\end{align}

\noindent At the saturation scale we have $\tilde{\mu}^2 = 2Q^2 + \tilde{\mu}_0^2$ with $\tilde{\mu}_0$ dictating the scale for the initial gluon density $g(x,\tilde{\mu}_0^2)$. Its form is taken by ansatz while it evolves as a function of energy according to the DGLAP evolution
\begin{align}\label{eq:GluonDensity}
xg(x,\tilde{\mu}_0^2) &= A_g x^{-\lambda_g}(1-x)^{5.6} & \frac{\partial xg(x,\tilde{\mu}^2)}{\partial\log\tilde{\mu}^2} &= \frac{\alpha_s(\tilde{\mu}^2)}{2\pi}\int\limits_x^1 dzP_{gg}(z)\frac{x}{z}g\lp\frac{x}{z},\tilde{\mu}^2\rp 
\end{align}

\noindent where the function $P_{gg}$ is the gluon splitting function. To leading order it is 
\begin{align}\label{eq:GluonSlit}
P_{gg}(z) &= 6\ls\frac{z}{(1-z)+} + \frac{1-z}{z} + z(1-z)\rs + \lp\frac{11}{2} - \frac{N_f}{3}\rp\delta(1-z).
\end{align}

\noindent In principle there is a contribution to the DGLAP evolution from annihilation of virtual quarks but this is thrown out on grounds that the quark density function is an order of magnitude smaller than that of gluons at our $x$. The function $\alpha_s(\tilde{\mu}^2)$ is the strong coupling running with the energy scale
\begin{align}\label{eq:alphaS}
\alpha_s(\tilde{\mu}^2) &= \frac{4\pi}{\lp 11-\frac{2}{3}N_f\rp\ln(\tilde{\mu}^2/\Lambda^2)}.
\end{align}

$N_f$ in the previous equations is the number of quark flavors. While there are six in reality, we only need to consider three for our relevant energies. $\Lambda$ is the length scale at which confinement emerges. The dipole radius $r$ at the saturation scale is $Q^2 = 2/r^2$. The saturation scale can then be numerically solved for by setting $F$ equal to one half\footnote{The framework code for this calculation was not available at the time. Our own code was written to reproduce the procedure.}. There are many free parameters not discussed here. Their values were constrained by HERA data as previously stated. Further discussion can be found in Refs. \cite{Schenke:2012hg, Schenke:2020mbo, Schenke:2014tga}. All values used are stated in Tab. \ref{tb:IPGlasmaParams}.
\begin{table}[H]
\centering
\begin{tabular}{|*{10}{c|}}\hline
$n$ & $\Lambda$ & $\tilde{\mu}_0^2$ & $B_q$  & $B_{c}$ & $A_g$ & $\lambda_g$ & $N_q$ & $N_c$ & $N_f$ \\ \hline
0.8 & 0.2 GeV & 1.51 GeV$^2$ & 0.3 GeV$^{-2}$ & 4 GeV$^{-2}$ & 2.308 & 0.058 & 3 & 3 & 3\\
\hline
\end{tabular}
\caption{IP Glasma parameters used in this work.}\label{tb:IPGlasmaParams}
\end{table}

The result is the initial saturation scale for a single event. This process is averaged over many events until a stable picture emerges. Given the saturation scale and \eqref{eq:alphaS}, $g$ can be calculated as well as $\mu$. A summary of the results are contained in Tab. \ref{tb:ChargeDensities}. The full spatial dependence of $\mu$ as a fitted function is left for the next appendix.
\begin{table}[H]
\centering
\subfloat[$\sqrt{s}=200$GeV]{\begin{tabular}{|c||c c c c||}\hline
& $Q_s\hspace{0pt}^2$ (GeV$^2)$ & $\alpha_s$ & $g$ & $\mu$ (fm$^{-2})$ \\ \hline
U  & 1.004 & 0.312 & 1.98 & 21.0 \\
Pb & 0.926 & 0.315 & 1.99 & 19.0 \\
Au & 0.940 & 0.315 & 1.99 & 19.4 \\
Xe & 0.826 & 0.320 & 2.00 & 16.5 \\
Ru & 0.717 & 0.325 & 2.02 & 13.8 \\
Zr & 0.732 & 0.324 & 2.02 & 14.2 \\
Cu & 0.618 & 0.330 & 2.04 & 11.6 \\
\hline
\end{tabular}}
\subfloat[$\sqrt{s}=2760$GeV]{\begin{tabular}{c c c c|}\hline
$Q_s\hspace{0pt}^2$ (GeV$^2)$ & $\alpha_s$ & $g$ & $\mu$ (fm$^{-2})$ \\ \hline
1.83 & 0.287 & 1.90 & 45.2 \\
1.69 & 0.291 & 1.91 & 40.8 \\
1.71 & 0.290 & 1.91 & 41.5 \\
1.50 & 0.296 & 1.93 & 34.9 \\
1.30 & 0.301 & 1.95 & 29.3 \\
1.34 & 0.300 & 1.94 & 30.2 \\
1.12 & 0.308 & 1.97 & 24.1 \\
\hline
\end{tabular}}\caption{Charge densities at the center of the nucleus for various ions and energies}\label{tb:ChargeDensities}
\end{table}

\section{Spatial Dependence of the Charge Density Function}

The result of the IP-Glasma initial condition is more than the value of $\mu$ at the center of the nucleus; it is a function of the center coordinate $\mathbf{R}$. Instead of working with the individual numerical function values, we choose to fit them to a function 
\begin{align}\label{eq:muFit}
\mu(\mathbf{R}) &= c\exp\ls-\frac{R^4}{d}\rs + e\exp\ls-\frac{R^4}{f}\rs .
\end{align}

\noindent The four constants are fitted parameters. A summary of the results are in Tab. \ref{tb:GlasmaFit}. Because Kovchegov, Lam, and Mahlon considered uniformly distributed nucleon, one might be curious what the spatial dependence looks like in that case. A uniform distribution must be constant over a larger area than the Fermi distribution so the value at the center is a bit lower. Quarks are actually the cause of the more substantial difference. There is a balance of Gaussian tails and crests in the bulk from the quarks. This balance is thrown off at the edge where the nucleons abruptly stop. The Gaussian crests are over represented giving the nucleus a raised edge before a Gaussian tail. A similar but less pronounced phenomenon happens with the Fermi distribution as well which is why the profile function \eqref{eq:muFit} is not chosen to be a Woods-Saxon. This is a visual explanation of what was shown analytically in \eqref{eq:I} and \eqref{eq:I_Bar}.
\begin{table}[H]
\centering
\subfloat[Set parameters\cite{Schenke:2020mbo, DEVRIES1987495}]{\begin{tabular}{|c||c c c||}\hline
   & $A$ & $R_c$ (fm) & $a$ (fm) \\\hline
U  & 238 & 6.81 & 0.55 \\
Pb & 208 & 6.62 & 0.546 \\
Au & 197 & 6.37 & 0.535 \\
Xe & 129 & 5.42 & 0.57 \\
Ru & 96 & 5.085 & 0.46 \\
Zr & 96 & 5.02 & 0.46 \\
Cu & 63 & 4.163 & 0.606 \\
\hline
\end{tabular}}
\subfloat[$\sqrt{s}=200$ GeV]{\begin{tabular}{c c c c||}\hline
$c$ (fm$^{-2}$) & $d$ (fm$^{4}$) & $e$ (fm$^{-2}$) & $f$ (fm$^{4}$) \\\hline
20.156 & 1558.814 & 0.887 & 11.457 \\
18.037 & 1411.277 & 0.949 & 12.044 \\
18.585 & 1205.154 & 0.786 & 17.372 \\
15.219 &  691.908 & 1.225 & 12.537 \\
12.918 &  515.101 & 0.895 &  7.676 \\
13.356 &  483.810 & 0.825 &  5.092 \\
9.549  &  328.404 & 1.962 & 18.904 \\
\hline
\end{tabular}}
\subfloat[$\sqrt{s}=2760$ GeV]{\begin{tabular}{c c c c|}\hline
$c$ (fm$^{-2}$) & $d$ (fm$^{4}$) & $e$ (fm$^{-2}$) & $f$ (fm$^{4}$) \\\hline
42.978 & 1576.724 & 2.135 & 19.915 \\
38.600 & 1393.442 & 2.135 & 14.935 \\
39.505 & 1211.323 & 1.910 & 17.888 \\
32.209 &  695.904 & 2.651 &  9.247 \\
27.130 &  514.579 & 2.039 &  6.174 \\
28.022 &  486.748 & 2.086 & 11.988 \\
19.874 &  326.465 & 4.172 & 24.295 \\
\hline
\end{tabular}}
\caption{Parameter list for $\mu$ fit function.}\label{tb:GlasmaFit}
\end{table}

\section{Taking the Momentum Integrals}
\label{sec:app:integrals}

Higher moments of the $d^2 q$ integrals appearing in Eq.\ (\ref{eq:ll2step}) have to be computed going from longitudinal to transverse fields, and increasing the number of gradients on $\mu_k$. All the moments needed here for complete results up to second order in gradients are presented in the following list.
\begin{subequations}\label{eq:qIntegral}
\begin{align}
\int q^i \delta(\tau-q)\delta(\tau-|\mathbf{q}-\mathbf{r}\hspace{1pt}'|) d^2q &= S r'^i \\
\int q^iq^j \delta(\tau-q)\delta(\tau-|\mathbf{q}-\mathbf{r}\hspace{1pt}'|) d^2q &= S \tau^2\lp \delta^{ij} - \frac{2\tau^2-r'^2}{2\tau^2}A^{ij} \rp\\
\int q^iq^jq^m \delta(\tau-q)\delta(\tau-|\mathbf{q}-\mathbf{r}\hspace{1pt}'|) d^2q &= S\left( \frac{6\tau^2-r'^2}{12}\ls r'^i\delta^{jm} + r'^j\delta^{im} + r'^m\delta^{ij}\rs \right.\\\nn
&\hspace{2cm}- \left.\frac{6\tau^2-2r'^2}{12}\ls r'^iA^{jm} + r'^jA^{im} + r'^mA^{ij}\rs\rp \displaybreak[2]\\
\int q^iq^jq^mq^n \delta(\tau-q)\delta(\tau-|\mathbf{q}-\mathbf{r}\hspace{1pt}'|) d^2q &= S\lp\frac{4\tau^4+4\tau^2r'^2-r'^4}{24}\ls\delta^{ij}\delta^{mn} + \delta^{im}\delta^{jn} + \delta^{in}\delta^{jm}\rs \right. \displaybreak[2]\\\nn
&\hspace{0cm}- \frac{2\tau^4-\tau^2r'^2}{12}\ls\delta^{ij}A^{mn} + \delta^{im}A^{jn} + \delta^{in}A^{jm} + A^{ij}\delta^{mn} + A^{im}\delta^{jn} + A^{in}\delta^{jm}\rs \displaybreak[2]\\\nn
&+\left. \frac{2\tau^4 - 4\tau^2r'^2 + r'^4}{12}\ls A^{ij}A^{mn} + A^{im}A^{jn} + A^{in}A^{jm}\rs\rp \displaybreak[2]\\
\int q^i q^j q^m q^n q^k \delta(\tau-q)\delta(\tau-|\mathbf{q}-\mathbf{r}\hspace{1pt}'|) d^2q &= S\lp \frac{20\tau^4 - r'^4}{240}\ls r'^k\lp\delta^{ij}\delta^{mn} + \delta^{im}\delta^{jn} + \delta^{in}\delta^{jm}\rp + \text{4 sym. terms} \rs \right.\hspace{-5pt} \displaybreak[2]\\\nn
&\hspace{-5cm}-  \frac{20\tau^4 - 10r'^2\tau^2 + r'^4}{240}\ls r'^k\lp \delta^{ij}A^{mn} + \delta^{im}A^{jn} + \delta^{in}A^{jm} + A^{ij}\delta^{mn} + A^{im}\delta^{jn} + A^{in}\delta^{jm} \rp + \text{4 sym. terms} \rs \displaybreak[2]\\\nn
&\hspace{-5cm}+ \left. \frac{5\tau^4 - 5r'^2\tau^2 + r'^4}{60}\ls r'^k\lp A^{ij}A^{mn} + A^{im}A^{jn} + A^{in}A^{jm}\rp + \text{4 sym. terms} \rs \rp \displaybreak[2]\\
\int q^i q^j q^m q^n q^k q^l \delta(\tau-q)\delta(\tau-|\mathbf{q}-\mathbf{r}\hspace{1pt}'|) d^2q &= S\lp \frac{4\tau^6 + 12r'^2\tau^4 - 3r'^4\tau^2}{240}\ls\delta^{ij}\delta^{mn}\delta^{kl} + \text{14 sym. terms}\rs\right. \displaybreak[2]\\\nn
&\hspace{-6.5cm}-   \frac{8\tau^6 + 4r'^2\tau^4 - 6r'^4\tau^2 + r'^6}{480}\ls A^{ij}\delta^{mn}\delta^{kl} + \text{44 sym. terms} \rs + \frac{2\tau^6 - 4r'^2\tau^4 + r'^4\tau^2}{120}\ls A^{ij}A^{mn}\delta^{kl} + \text{44 sym. terms} \rs \displaybreak[2]\\\nn
&\hspace{-6.5cm}- \left. \frac{2\tau^6 - 9r'^2\tau^4 + 6r'^4\tau^2 - r'^6}{120}\ls A^{ij}A^{mn}A^{kl} + \text{14 sym. terms} \rs \right)  \, .\displaybreak[2]
\end{align}
\end{subequations}
We have defined the short-hand notations
\begin{align}
S &= 2\tau^2\lp 4\tau^2 - r'^2 \rp^{-1/2} \frac{\Theta(2\tau-r')}{r'} \, , & A^{ij} &= \lp\frac{r'^ir'^j}{r'^2}-\epsilon^{ik}\epsilon^{jl}\frac{r'^kr'^l}{r'^2}\rp  
\, .
\end{align}
Subsequently, the azimuthal angle part of the $d^2 r'$-integral needs to be carried out, as demonstrated in Eq.\ (\ref{eq:ll2step3}). Once more, going to transverse fields and higher order gradients introduce factors of $r'^i$. Denoting the azimuthal angle by $\alpha$, the integrals needed up to second order in gradients are 
\begin{subequations}\label{eq:rpIntegral}
\begin{align}
\frac{1}{2\pi}\int\limits_0^{2\pi} r'\hspace{0pt}^i  d\alpha &= 0 & \frac{1}{2\pi}\int\limits_0^{2\pi} r'\hspace{0pt}^ir'\hspace{0pt}^j  d\alpha &= \frac{r'\hspace{0pt}^2}{2}\delta^{ij}\\
\frac{1}{2\pi}\int\limits_0^{2\pi} r'\hspace{0pt}^ir'\hspace{0pt}^jr'\hspace{0pt}^k  d\alpha &= 0 & \frac{1}{2\pi}\int\limits_0^{2\pi} r'\hspace{0pt}^ir'\hspace{0pt}^jr'\hspace{0pt}^kr'\hspace{0pt}^l  d\alpha &= \frac{r'\hspace{0pt}^4}{8}\ls\delta^{ij}\delta^{kl} + \delta^{ik}\delta^{jl} + \delta^{il}\delta^{jk}\rs \\
\frac{1}{2\pi}\int\limits_0^{2\pi} r'\hspace{0pt}^ir'\hspace{0pt}^jr'\hspace{0pt}^kr'\hspace{0pt}^lr'\hspace{0pt}^m  d\alpha &= 0 & \frac{1}{2\pi}\int\limits_0^{2\pi} r'\hspace{0pt}^ir'\hspace{0pt}^jr'\hspace{0pt}^kr'\hspace{0pt}^lr'\hspace{0pt}^mr'\hspace{0pt}^n  d\alpha &= \frac{r'\hspace{0pt}^6}{48}\ls\delta^{ij}\delta^{kl}\delta^{mn} + \text{14 sym. terms}\rs
\, .
\end{align}
\end{subequations}

\section{Energy Momentum Tensor to Higher Orders in Gradients}
\label{sec:app:emtensor}

In this section we report results for various components of the energy momentum tensor discussed in Sec.\ \ref{sec:emtensor} that are complete up to second order in gradients of either $\mu_k$, and are still independent of the model chosen for the gluon distribution. They can be obtained using the integrals in the previous appendix. These components are
%T_{iz,2}(\tau), T_{iz,3}(\tau), \omega_3(\tau), and T_{xy,3}(\tau) are unconstrained
\begin{subequations}\label{eq:GlasmaStressTensor}
\begin{align}\displaybreak[2]
\langle\epsilon_L\rangle &= \frac{1}{2\pi}\int\limits_0^{2\tau} dr' \lp 4\tau^2 - r'\hspace{0pt}^2 \rp^{-1/2} \ls 2\mu_1\mu_2(\mathbf{R})f_1(r') + \nabla\mu_1\cdot\nabla\mu_2(\mathbf{R})\lc \frac{1}{4}\lp 4\tau^2 - r'\hspace{0pt}^2 \rp f_1(r') + r'\hspace{0pt}^2f_2(r') \rc \rb \\\nn
& +\lp \mu_1\triangle\mu_2 + \mu_2\triangle\mu_1 \rp(\mathbf{R})\lc \frac{4\tau^2 - r'^2}{8}f_1(r') + 2f_3(r') + f_4(r') \rc + \triangle\mu_1\triangle\mu_2(\mathbf{R})\lc \frac{(4\tau^2 - r'^2)^2}{256}f_1(r') \rb\\\nn\displaybreak[2]
&\left.+ \frac{4\tau^2r'^2 - r'^4}{16}f_2(r') + \frac{4\tau^2 - r'^2}{4}f_3(r') + \frac{12\tau^2 - 3r'^2}{16}f_4(r') + 2f_5(r') + 2f_6(r') + \frac{1}{4}f_9(r') \rc \\\nn\displaybreak[2]
&\left. + \nabla^{ij}\mu_1\nabla^{ij}\mu_2(\mathbf{R})\lc \frac{(4\tau^2 - r'^2)^2}{128}f_1(r') + \frac{4\tau^2r'^2 - r'^4}{8}f_2(r') - \frac{4\tau^2 - r'^2}{8}f_4(r') + 2f_7(r') + f_8(r') + \frac{1}{2}f_9(r') \rc \rs \\\displaybreak[2]
\langle\epsilon_T\rangle &= \frac{1}{4\pi\tau^2}\int\limits_0^{2\tau} dr'\frac{2\tau^2-r'\hspace{0pt}^2}{\lp 4\tau^2 - r'\hspace{0pt}^2 \rp^{1/2}}\ls 2\mu_1\mu_2(\mathbf{R})f_1(r') + \nabla\mu_1\cdot\nabla\mu_2(\mathbf{R})\lc \frac{1}{4}\lp 4\tau^2 - r'\hspace{0pt}^2 \rp f_1(r') + r'\hspace{0pt}^2f_2(r') \rc \right. \\\nn
& +\lp \mu_1\triangle\mu_2 + \mu_2\triangle\mu_1 \rp(\mathbf{R})\lc \frac{4\tau^2 - r'^2}{8}f_1(r') + 2f_3(r') + f_4(r') \rc + \triangle\mu_1\triangle\mu_2(\mathbf{R})\lc \frac{(4\tau^2 - r'^2)^2}{256}f_1(r') \right.\\\nn
&\left.+ \frac{4\tau^2r'^2 - r'^4}{16}f_2(r') + \frac{4\tau^2 - r'^2}{4}f_3(r') + \frac{12\tau^2 - 3r'^2}{16}f_4(r') + 2f_5(r') + 2f_6(r') + \frac{1}{4}f_9(r') \right\} \\\nn\displaybreak[2]
&\left. + \nabla^{ij}\mu_1\nabla^{ij}\mu_2(\mathbf{R})\lc \frac{(4\tau^2 - r'^2)^2}{128}f_1(r') + \frac{4\tau^2r'^2 - r'^4}{8}f_2(r') - \frac{4\tau^2 - r'^2}{8}f_4(r') + 2f_7(r') + f_8(r') + \frac{1}{2}f_9(r') \rc \rs \\\displaybreak[2]
\langle\omega\rangle &= \frac{1}{2\pi\tau^2}\int\limits_0^{2\tau}dr'\lp 4\tau^2 - r'\hspace{0pt}^2 \rp^{-1/2}\frac{\tau^2}{2}\ls\lp \nabla^x\mu_1\nabla^x\mu_2 - \nabla^y\mu_1\nabla^y\mu_2 \rp(\mathbf{R})\lc \frac{1}{4}\lp 4\tau^2 - r'\hspace{0pt}^2 \rp f_1(r') + r'\hspace{0pt}^2f_2(r') \rc\right. \\\nn
&+ \lp \mu_1\ls\nabla^{xx}\mu_2-\nabla^{yy}\mu_2\rs + \mu_2\ls\nabla^{xx}\mu_1-\nabla^{yy}\mu_1\rs \rp\lc \frac{4\tau^2 - r'^2}{8}f_1(r') - f_4(r') \rc + \lp \nabla^{xx}\mu_1\nabla^{xx}\mu_2 - \nabla^{yy}\mu_1\nabla^{yy}\mu_2 \rp  \\\nn\displaybreak[2]
&\lb \times\lc \frac{(4\tau^2 - r'^2)^2}{64}f_1(r') + \frac{4\tau^2r'^2 - r'^4}{4}f_2(r') + \frac{4\tau^2 - r'^2}{4}f_3(r') - 2f_6(r') - f_8(r') - f_9(r') \rc\rs \\\displaybreak[2]
\langle S_i \rangle &= \frac{1}{\pi\tau}\int\limits_0^{2\tau} dr'\lp 4\tau^2 - r'^2 \rp^{1/2}\ls -\nabla^i(\mu_1\mu_2)(\mathbf{R}) \frac{f_1(r')}{4} -\lp \nabla^i\mu_1\triangle\mu_2 + \nabla^i\mu_2\triangle\mu_1 \rp (\mathbf{R})\lc\frac{4\tau^2 - r'^2}{128}f_1(r') \rb\rb \\\nn\displaybreak[2]
&\lb\lb+ \frac{1}{16}\ls r'^2f_2(r') + 4f_3(r') + 3f_4(r')\rs\rc -\nabla^i\lp\nabla\mu_1\cdot\nabla\mu_2\rp(\mathbf{R}) \lc\frac{4\tau^2 - r'^2}{64}f_1(r') + \frac{1}{8}\ls r'^2f_2(r') - f_4(r')\rs\rc\rs\\\displaybreak[2]
\langle S_z \rangle &= \frac{1}{2\pi\tau^2}\int\limits_0^{2\tau} dr'\left( 4\tau^2 - r'^2 \rp^{-1/2}r'^2\left( \mu_1\triangle\mu_2 - \mu_2\triangle\mu_1 \rp(\mathbf{R})\lc -\frac{4\tau^2 - r'^2}{4}\phi_1(r') \rc \\\displaybreak[2]
\langle T_{iz} \rangle &= \frac{1}{2\pi\tau}\int\limits_0^{2\tau}dr' \frac{r'\hspace{0pt}^2}{\lp 4\tau^2 - r'^2 \rp^{1/2}}\ls \lp \mu_1\nabla^i\mu_2 - \mu_2\nabla^i\mu_1 \rp(\mathbf{R}) \phi_1(r') -\lp\triangle\mu_1\nabla^i\mu_2 - \triangle\mu_2\nabla^i\mu_1 \rp(\mathbf{R})\lc\hspace{-2pt}\frac{4\tau^2 - r'^2}{32} \right.\right.\\\nn\displaybreak[2]
&\lb\lb\times\phi_1(r') + \phi_2(r') - \phi_3(r') - \frac{3}{4}\phi_4(r') \rc +\lp\nabla\mu_1\cdot\nabla\lp\nabla^i\mu_2\rp - \nabla\mu_2\cdot\nabla\lp\nabla^i\mu_1\rp\rp\lc\frac{4\tau^2 - r'^2}{16}\phi_1(r') + \frac{1}{2}\phi_4(r')\rc \rs \\
\langle T_{xy}\rangle &= \frac{1}{2\pi\tau^2}\int\limits_0^{2\tau}dr'\lp 4\tau^2 - r'\hspace{0pt}^2 \rp^{-1/2}\frac{\tau^2}{2}\ls\lp \nabla^x\mu_1\nabla^y\mu_2 + \nabla^y\mu_1\nabla^x\mu_2 \rp(\mathbf{R})\lc \frac{1}{4}\lp 4\tau^2 - r'\hspace{0pt}^2 \rp f_1(r') + r'\hspace{0pt}^2f_2(r') \rc\rb \\\nn
&+ \lp \mu_1\nabla^{xy}\mu_2 + \mu_2\nabla^{xy}\mu_1 \rp(\mathbf{R})\lc \frac{4\tau^2 - r'^2}{4}f_1(r') - 2f_4(r') \rc + \lp \triangle\mu_1\nabla^{xy}\mu_2 + \nabla^{xy}\mu_1\triangle\mu_2 \rp (\mathbf{R}) \\\nn
&\left. \times\lc \frac{(4\tau^2 - r'^2)^2}{64}f_1(r') + \frac{4\tau^2r'^2 - r'^4}{4}f_2(r') + \frac{4\tau^2 - r'^2}{2}f_3(r') - 2f_6(r') - f_8(r') - f_9(r') \rc\right]  \, .
\end{align}
\end{subequations}
For brevity, we have introduced new functions in addition to the ones already defined in (\ref{eq:newfunc}), namely
\begin{align}
    f_3(r) & = g^6\frac{N_c}{N_c^2-1}\left\{ \frac{1}{r} g_{0,1}' + g_{0,1}'' \right\} 
    \left\{ -\frac{1}{4} g_{0,1} + \frac{1}{r} g_{2,1}' +g_{2,1}'' + \frac{2}{r^2} g_{2,2}' 
    -\frac{1}{2r} g_{2,2}' -\frac{1}{2} g_{2,2}'' \right\}  \displaybreak[2]\\
    f_4(r) & = g^6\frac{N_c}{N_c^2-1}\left\{ \frac{1}{r} g_{0,1}' + g_{0,1}'' \right\} 
    \left\{ -\frac{4}{r^2} g_{2,2} + \frac{1}{r} g_{2,2}' +g_{2,2}''  \right\} \displaybreak[2]\\ \nn
    f_5(r) &= g^6\frac{N_c}{N_c^2-1}\ls{\lp - \frac{1}{4}g_{0,1} + \frac{1}{r}g_{2,1}' + g_{2,1}'' - \frac{1}{2}g_{2,2}'' 
    \rp}^2 
     + \frac{4g_{2,2}}{r^2} \lp \frac{1}{2}g_{0,1} + \frac{1}{r}g_{2,1}' + g_{2,1}'' - \frac{1}{r^2}g_{2,2} - \frac{1}{2r}g_{2,2}' - \frac{1}{2}g_{2,2}'' \rp\rb \displaybreak[2]\\  
       & \qquad \lb+\frac{1}{2r}g_{2,2}' \lp \frac{1}{2}g_{0,1} + \frac{2}{r}g_{2,1}' - 2g_{2,1}'' + \frac{1}{2r}g_{2,2}' + g_{2,2}''\rp\vphantom{{\lp - \frac{1}{4}g_{0,1} + \frac{1}{r}g_{2,1}' + g_{2,1}'' - \frac{1}{2}g_{2,2}'' \rp}^2 }\rs   \displaybreak[2]\\
     f_6(r) &=  g^6\frac{N_c}{N_c^2-1}\lp\frac{g_{2,2}'}{r}+g_{2,2}''\rp\lp-\frac{g_{0,1}}{4}+\frac{g_{2,1}'}{r}+g_{2,1}''+\frac{4g_{2,2}}{r^2}-\frac{g_{2,2}'}{2r}-\frac{g_{2,2}''}{2}\rp-\frac{4g_{2,2}}{r^2}\lp\frac{g_{0,1}}{2}+\frac{g_{2,1}'}{r}+g_{2,1}''\rp
     \displaybreak[2]\\
     f_7(r) &= g^6\frac{N_c}{N_c^2-1}\frac{4g_{2,2}}{r^2}\lc-\frac{3g_{0,1}}{4} + \frac{2g_{2,2}}{r^2}\rc 
     \displaybreak[2]\\
     f_8(r) &= g^6\frac{N_c}{N_c^2-1}\frac{8g_{2,2}}{r^2}\lc\frac{3g_{0,1}}{4} - \frac{2g_{2,2}}{r^2}\rc  
     \displaybreak[2]\\
     f_9(r) &= g^6\frac{N_c}{N_c^2-1}{\lp-\frac{4g_{2,2}}{r^2} + \frac{g_{2,2}'}{r} + g_{2,2}''\rp}^2  
     \displaybreak[2]\\
     \phi_2(r) &= g^6\frac{N_c}{N_c^2-1}\frac{g_{0,1}'}{r}\lc\frac{g_{0,1}}{4} - \frac{2g_{2,2}}{r^2}\rc 
      \displaybreak[2]\\
     \phi_3(r) &= g^6\frac{N_c}{N_c^2-1}\frac{g_{0,1}'}{r}\lc\frac{g_{2,1}'}{r} + g_{2,1}'' - \frac{g_{2,2}'}{2r} - \frac{g_{2,2}''}{2}\rc
     \displaybreak[2]\\
     \phi_4(r) &= g^6\frac{N_c}{N_c^2-1}\frac{g_{0,1}'}{r}\lc - \frac{4g_{2,2}}{r^2} + \frac{g_{2,2}'}{r} + g_{2,2}''\rc  \, .
\end{align}

\section{Gluon Correlation Function in the Improved Gaussian Model}
\label{sec:app:gluoncorr}

We would like to evaluate \eqref{eq:gamma}
\begin{align}~\nonumber
\gamma(\mathbf{x}_\perp, \mathbf{y}_\perp) &= \int d^2u_\perp d^2v_\perp G(\mathbf{u}_\perp-\mathbf{v}_\perp/2)G(\mathbf{u}_\perp+\mathbf{v}_\perp/2)\mu( \mathbf{R}_\perp - \mathbf{u}_\perp) \mathcal{D}(\mathbf{v}_\perp-\mathbf{r}_\perp)
\end{align}
for some Gaussian function $\mathcal{D}$ using the Taylor expansion of $\mu$ introduced in Sec.\ \ref{sec:gluecorr}, keeping up to second order in gradients. The result of the $\mathbf{u}_\perp$ integral is simply the convolution of $\gamma_\mathrm{MV}$ from the MV model and the function $\mathcal{D}$. Explicit vector components $\mathbf{v}_\perp^i$  appear in the integrand and can be evaluated by using the identities \cite{Arfken}
\begin{align}\label{eq:IntBesselI}
\int\limits_{0}^{2\pi}d\phi \cos(n\phi)e^{a\cos(\phi)} &= 2\pi I_n(a), & \int\limits_{0}^{2\pi}d\phi \sin(n\phi)e^{a\cos(\phi)} &= 0 \, .
\end{align}
The remaining integral takes the general form
\begin{align}\label{eq:qMagInt}
&\int\limits_0^\infty dv\hspace{1pt}v^\rho e^{-v^2/4B} K_\ell(mv)I_n\lp \frac{rv}{2B}\rp 
\end{align}
for some indices $\rho$, $\ell$, and $n$. 
Since the improved Gaussian model is much less sensitive to the IR regulator $m$ we take the freedom to expand the modified Bessel functions of the second kind $K_\ell$ and omit all terms of order $m$, $m\ln m$ or higher. %Be aware of the additional $1/m$ factor in each $g_{i,j,\text{MV}}$ for this expansion. 
Next we utilize the fact that the modified Bessel functions of the first kind can be written in terms of the Meijer-G functions, see Appendix \ref{sec:meijerg}. To be precise, it holds that \cite{BesselI_To_MeijerG}
\begin{align}
    I_n(z) &= \lp\frac{z}{2}\rp^n
    G^{1,0}_{0,2}\lp\hspace{-3pt}
    \begin{array}{c}
         - \\
         0 , -n
    \end{array} \Bigg| -\frac{z^2}{4}\rp .
\end{align}

\noindent Using this identity together with the definition in Appendix \ref{sec:meijerg}, and using $g_{2,1}(r)$ as an example, the integral becomes
\begin{align*}\numberthis
    &\frac{1}{96\pi B}e^{-\frac{r^2}{4B}}\frac{1}{2\pi i}\int_L ds \frac{\Gamma(-s)}{\Gamma(1+s)}\lp-\frac{r^2}{16B^2}\rp^s \int\limits_0^\infty dv \hspace{1pt} v^{2s}e^{-\frac{v^2}{4B}}\ls\frac{2}{m^4}v - \frac{1}{4m^2}v^3 + \frac{1}{32}v^5 \rs \\
    &=\frac{1}{96\pi B}e^{-\frac{r^2}{4B}}\frac{1}{2\pi i}\int_L ds \frac{\Gamma(-s)}{\Gamma(1+s)}\lp-\frac{r^2}{16B^2}\rp^s \ls\frac{1}{m^4}(4B)^{1+s}\Gamma(1+s) - \frac{1}{8m^2}(4B)^{2+s}\Gamma(2+s) + \frac{1}{64}(4B)^{3+s}\Gamma(3+s) \rs \\
    &= \frac{1}{24\pi m^4}e^{-\frac{r^2}{4B}}G^{1,0}_{0,1}\lp\hspace{-3pt}
    \begin{array}{c}
         - \\
         0 
    \end{array} \Bigg| -\frac{r^2}{4B}\rp - \frac{B}{48\pi m^2}e^{-\frac{r^2}{4B}}G^{1,1}_{1,2}\lp\hspace{-3pt}
    \begin{array}{c}
         -1 \\
         0, 0 
    \end{array} \Bigg| -\frac{r^2}{4B}\rp + \frac{B^2}{96\pi}e^{-\frac{r^2}{4B}}G^{1,1}_{1,2}\lp\hspace{-3pt}
    \begin{array}{c}
         -2 \\
         0, 0 
    \end{array} \Bigg| -\frac{r^2}{4B}\rp ,
\end{align*}

\noindent where we have used a Gaussian integral identity \cite{GaussianIntegral} in the second line. The final line can be recognized as a simple (Laguerre) polynomial after consulting the Meijer-G documentation \cite{MeijerG}. Setting $B = B_q , B_q+B_{c}$ and taking the difference completes the derivation for $g_{2,1}(r)$. Many terms in the derivation for $g_{0,1}(r)$ and $g_{2,2}(r)$ can also be computed with this procedure with the exception of terms like $v^\rho\ln(v)$. For these terms we use a differentiation under the integral trick, $v^\rho\ln(v) = \partial_\rho(v^\rho)$, and proceed as before. For $g_{0,1}(r)$ one reaches
\begin{align}
    \frac{1}{4\pi m^2} + \frac{1}{16\pi}\lp \gamma_E + \ln(m^2B) \rp\lp r^2 + 4B \rp + \frac{B}{4\pi}\partial_\rho\lp L^0_\rho\lp-\frac{r^2}{4B}\rp\rp_{\rho=1} .
\end{align}

For brevity define $\partial_\rho\lp L^\alpha_\rho\lp x\rp\rp = L^\alpha_{\rho'}\lp x\rp$. The differential equation it satisfies,
\begin{align}
    x {L^\alpha_{\rho'}}''\lp x\rp + (\alpha + 1 - x){L^\alpha_{\rho'}}'\lp x\rp + \rho L^\alpha_{\rho'}\lp x\rp = - L^\alpha_{\rho}\lp x\rp ,
\end{align}

\noindent can be derived from taking the derivative of the Laguerre equation with respect to $\rho$. It is of the same form as the original equation with an in-homogeneous term added. It can be solved with a Frobenius solution and is fixed up to the initial value $L^\alpha_{\rho'}\lp 0\rp$. To be consistent with the Laguerre polynomials however \cite{Laguerre_To_1F1, 1F1_Derivative}, it is completely determined in terms of harmonic numbers $L^\alpha_{\rho'}\lp 0\rp = \binom{\rho + \alpha}{\rho}\lp H_{\rho + \alpha} - H_{\rho}\rp$. The general result, though mundane, doesn't appear to be in the literature so we present it here:
\begin{align}
    L_{\rho'}^\alpha(x) &= (\alpha+\rho)!\sum\limits_{i=0}^{\rho} \frac{(-1)^{i}}{i!}\frac{H_{\alpha+\rho} - H_{\rho - i}}{(\alpha+i)!(\rho-i)!}x^{i} + (-1)^{\rho+1}(\alpha+\rho)!\sum\limits_{i=\rho+1}^\infty \frac{(i-\rho-1)!}{i!(i+\alpha)!}x^{i}.
\end{align}

For the $g_{0,1}(r)$ derivation specifically one needs 
\begin{align}
    L^0_{1'}\lp x\rp = - x + \sum\limits_{i=2}^\infty \frac{x^i}{i(i-1)i!} = x + 1 - e^x + (1-x)\ls E_1(-x) + \ln(-x) + \gamma_E\rs .
\end{align}

\noindent The final equality follows after one line of algebra and one identity for the exponential integral
\begin{align}
     E_1(-x) + \ln(-x) + \gamma_E &= - \sum\limits_{i = 1}^\infty \frac{x^i}{i(i!)}.
\end{align}

\noindent The derivation of $g_{2,2}(r)$ uses the same procedure, but this time needs
\begin{align}
    L^2_{1'}\lp x\rp = \frac{5}{2} -\frac{11}{6}x + \sum\limits_{i=2}^\infty \frac{x^i}{i(i-1)(i+2)!} = &x + 1 - e^x + \frac{1}{x}\lp\frac{1}{x}+2\rp e^x - \frac{1}{x}\lp\frac{1}{x}+3\rp \\\nn
    &+ (3-x)\ls E_1(-x) + \ln(-x) + \gamma_E\rs .
\end{align}

\noindent The $g_{i,j}(r)$ functions are then \eqref{eq:gFunctions}.

\section{The Meijer-G Function}
\label{sec:meijerg}

The Meijer-G function is defined as
\begin{align}\label{eq:MeijerG_Def}
    G^{m,n}_{p,q}\lp\hspace{-3pt}
    \begin{array}{c}
         a_1,\cdots , a_p \\
         b_1,\cdots , b_q
    \end{array} \Bigg| z\rp &= \frac{1}{2\pi i}\int_L \frac{\prod\limits_{j=1}^m \Gamma(b_j-s) \prod\limits_{j=1}^n \Gamma(1-a_j+s)}{\prod\limits_{j=m+1}^q \Gamma(1-b_j+s) \prod\limits_{j=n+1}^p \Gamma(a_j-s)}z^s ds 
\end{align}
where $L$ is a suitable integration path in the complex plane. For all Meijer-G functions appearing explicitly in this work, an integration path parallel to the imaginary $s$-axis from $-i\infty$ to $+i\infty$ can be assumed, with all poles of the terms $\Gamma(b_j-s)$, $j=1,\ldots,m$ to the right of the integration path and all poles of the terms $\Gamma(1-a_j+s)$, $j=1,\ldots ,n$ to its left.

An important identity which follows immediately from the definition  is
\begin{align}
    z^\rho G^{m,n}_{p,q}\lp\hspace{-3pt}
    \begin{array}{c}
         \mathbf{a} \\
         \mathbf{b}
    \end{array} \Bigg| z\rp &= G^{m,n}_{p,q}\lp\hspace{-3pt}
    \begin{array}{c}
         \mathbf{a} + \rho \\
         \mathbf{b} + \rho
    \end{array} \Bigg| z\rp   \, ,
\end{align}
where $\mathbf{a}$, $\mathbf{b}$ are tuples of coefficients. It is useful in simplifying relations.

The asymptotic behavior of the Meijer-G functions most relevant here, for large real arguments $z\to\infty$, is given by Theorem 2 in Sec. 5.7 of Ref. \cite{luke1969special}
\begin{align}
    G^{3,1}_{2,4}\lp\hspace{-3pt}
    \begin{array}{c}
         a_1, a_2 \\
         b_1,b_2,b_3,b_4
    \end{array} \Bigg| z\rp &\sim z^{a_1 - 1}\frac{}{} \frac{\Gamma(b_1 - a_1 + 1)\Gamma(b_2 - a_1 + 1)\Gamma(b_3 - a_1 + 1)}{\Gamma(a_2 - a_1 + 1)\Gamma(a_1 - b_4)} {}_4F_{1}\lp\hspace{-3pt}
    \begin{array}{c}
        1-a_1+\mathbf{b} \\
        1-a_1+a_2 
    \end{array} \Bigg| \frac{1}{z}\rp
\end{align}

\noindent which holds on the conditions $1 = n \leq p < q$, $1 \leq m \leq q$, and $a_j - b_k \neq 1, 2, 3,\cdots$ for any of the $a_j$, $b_k$. In our asymptotic values of the stress tensor, we evaluate the ordinary hypergeometric-F function at zero for simplicity.

\section{UV Regularized Series Expansion at Early Times}\label{sec:app:2015_Match}

Models that are not UV-safe, like the MV model, require high momentum modes in the transverse plane to be cut off to regularize the result. The most obvious choice of regularization is to place a momentum cutoff $Q'$ into Eqs.\ (\ref{eq:momentum_1}) and (\ref{eq:momentum_2}). The resulting integrals can be carried out after a power series expansion and lead to
\begin{subequations}\label{eq:momentum_upper_bound}
\begin{align}\label{eq:momentum_Q}
    \int_0^{Q'}  \frac{ d p_\perp}{2\pi } p_\perp J_0\left (p_\perp |\mathbf{x}_\perp-\mathbf{x}'_\perp|\right) J_0(p_\perp \tau)
    = \frac{Q'\hspace{0pt}^2}{4\pi} \sum\limits_{j=0}^\infty \frac{1}{j!j!}\lp -\frac{\tau^2Q'\hspace{0pt}^2}{4}\rp^j  &G^{1,1}_{1,3}\lp\hspace{-3pt}
    \begin{array}{c}
         -j \\
         0, 0, -1-j 
    \end{array} \Bigg| \frac{1}{4}(\mathbf{x}_\perp-\mathbf{x}'_\perp)^2Q'\hspace{0pt}^2\rp,  \\\label{eq:momentum_Q2}
    \int_0^{Q'}  \frac{ d p_\perp}{2\pi } J_0\left (p_\perp |\mathbf{x}_\perp-\mathbf{x}'_\perp|\right) J_1(p_\perp \tau)
    = \frac{Q'\hspace{0pt}^2\tau}{8\pi} \sum\limits_{j=0}^\infty \frac{1}{j!(j+1)!}\lp -\frac{\tau^2 Q'\hspace{0pt}^2}{4}\rp^{j}  &G^{1,1}_{1,3}\lp\hspace{-3pt}
    \begin{array}{c}
         -j \\
         0, 0, -1-j 
    \end{array} \Bigg| \frac{1}{4}(\mathbf{x}_\perp-\mathbf{x}'_\perp)^2Q'\hspace{0pt}^2\rp ,
\end{align}
\end{subequations}

\noindent after using results from Refs.\ \cite{Q_Identity1, Q_Identity2}. This procedure naturally leads to a power series in time $\tau$, but note that the series is different from the result in Ref. \cite{Chen:2015wia}. A prime is placed on the UV cutoff to indicate that the two regularization schemes are different. In Ref.\ \cite{Chen:2015wia} all integrals are evaluated order-by-order in an expansion in $\tau$ while the relative coordinate $r$ of the two fields in the expectation value is kept finite. As a last step, the limit $r\to 0$ is taken which would lead to the UV singularities. These are then regularized by fixing $r=1/Q$. Therefore, the coarse graining is introduced for the coordinate $r$ rather than for $r'$.

One can nevertheless match the two schemes relatively well numerically.
Both the UV and IR cutoffs must be rescaled slightly to maintain consistency at for selective orders in time. That is, if $m$ and $Q$ are the regulators in Ref.\ \cite{Chen:2015wia}, then the regulators after using \eqref{eq:momentum_upper_bound}  should be $m' = 0.95\sqrt{2}m$ and $Q' = 0.95\sqrt{2} Q$ to obtain the same result at second order in time. For fourth order in time they should be $m' = 0.97\sqrt{2}m$ and $Q' = 0.97\sqrt{2} Q$. The overall scale factor drifting with the order in time indicates that there is no consistent mapping between this regularization scheme and the one used in Ref.\ \cite{Chen:2015wia}. The scale factors were found by performing the $\mathbf{r}'$ and $\mathbf{q}$ integrals numerically and minimizing the percent difference with the coefficients $\delta$, and $\rho$ from Ref.\ \cite{Chen:2015wia}, respectively. There is a constant 8\% difference in the leading term, $\epsilon_0$, because it relies on the ratio $Q/m$.

\bibliographystyle{apsrev4-1}
\bibliography{weakfield}

\end{document}